\documentclass[10pt, onecolumn, a4paper, fleqn]{article}

\usepackage{STY/PACKAGES}
\usepackage{STY/DEFINITIONS}
\usepackage{STY/SETTINGS}
\usepackage{STY/TIKZ}
\usepackage{STY/TITLEPAGE}

\usepackage{mdframed} 

\begin{document}
\edef\myindent{\the\parindent}
\author{Kronberg, Vi \and Anthonissen, Martijn \and ten Thije Boonkkamp, Jan \and IJzerman, Wilbert}
\title{Three-Dimensional Freeform Reflector Design with a Scattering Surface}

\thispagestyle{empty} 
\pagenumbering{gobble} 

\newgeometry{margin=19mm,top=30mm}

\begin{figure}[t]
	\centering
	\includegraphics[width=0.4\textwidth]{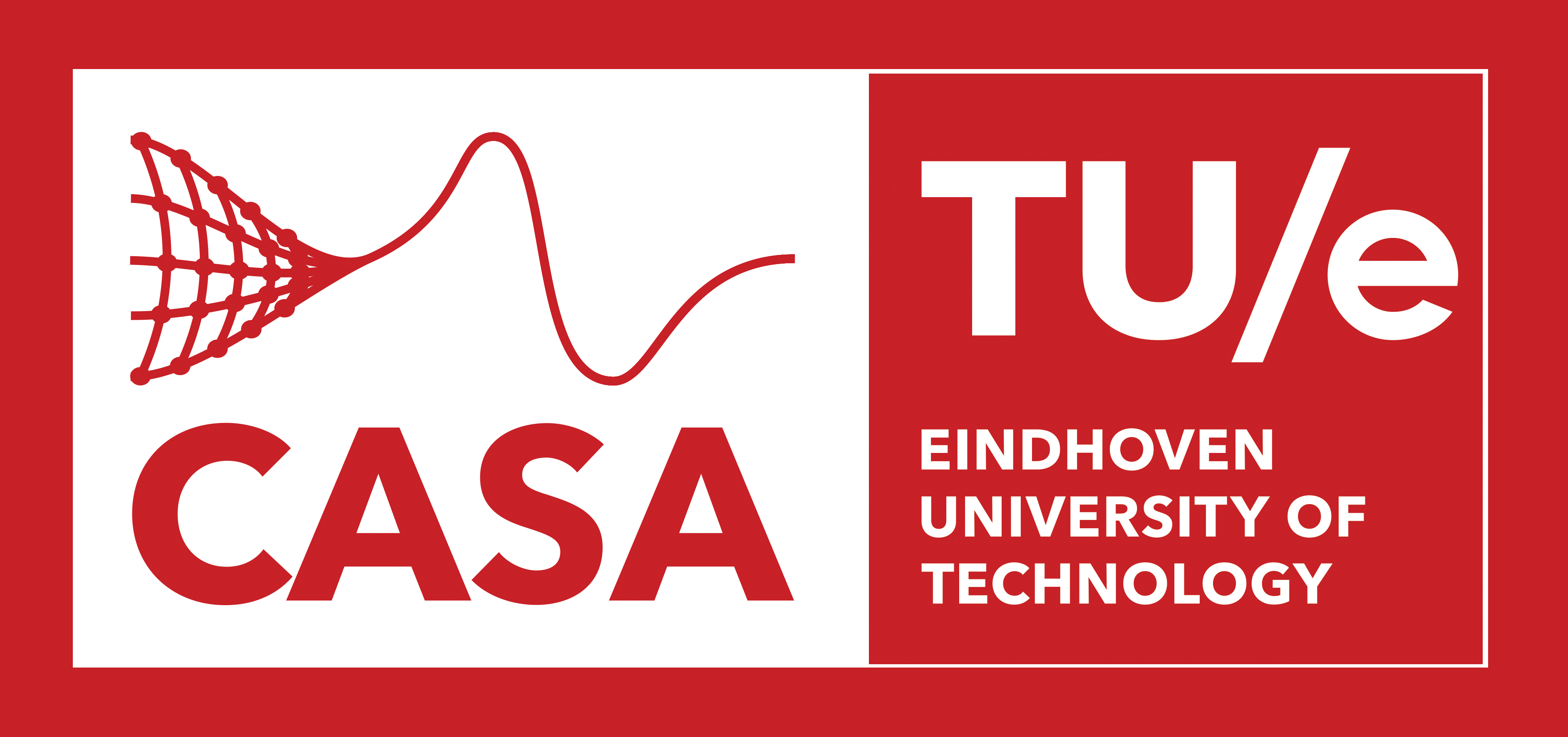}
\end{figure}

\vspace*{-0.7cm}
\begin{center}
	\LARGE \textbf{Three-Dimensional Freeform Reflector Design with a Scattering Surface}\\[0.5cm]
\end{center}

\begin{center}
	\large
	{\itshape V\`{i}}~\textsc{Kronberg},\textsuperscript{1,*} {\itshape Martijn}~\textsc{Anthonissen},\textsuperscript{1}\\
	{\itshape Jan}~\textsc{ten Thije Boonkkamp},\textsuperscript{1} and {\itshape Wilbert}~\textsc{IJzerman}\textsuperscript{1,2}
\end{center}

\begin{changemargin}{1.5cm}{1.5cm}
\noindent
\textsuperscript{1}{\itshape Department of Mathematics and Computer Science, Eindhoven University of Technology},\\PO Box 513, 5600 MB Eindhoven, The Netherlands\\
\textsuperscript{2}{\itshape Signify Research}, High Tech Campus 7, 5656 AE Eindhoven, The Netherlands\\
\textsuperscript{*}{\color{blue}\href{mailto:v.c.e.kronberg@tue.nl}{\texttt{v.c.e.kronberg@tue.nl}}}\\
\url{https://www.win.tue.nl/~martijna/Optics/}\\
\end{changemargin}

\hrule
\begin{changemargin}{1.5cm}{1.5cm}
\textsc{\textbf{Keywords:}} Surface scattering $\cdot$ Reflector design $\cdot$ Inverse problem\\[1mm]
\textsc{\textbf{PACS:}} 02.30.Z $\cdot$ 42.15.-i $\cdot$ z42.25.Fx $\cdot$ 42.79.Fm\\[1mm]
\textsc{\textbf{AMS:}} 78A05 $\cdot$ 78A45 $\cdot$ 78A46
\end{changemargin}
\hrule
\begin{changemargin}{1.5cm}{1.5cm}
	\textsc{\textbf{Abstract:}}
	We introduce a novel approach to calculating three-dimensional freeform reflectors with a scattering surface.
	Our method is based on optimal transport and utilizes a Fredholm integral equation to express scattering.
	By solving this integral equation through a process similar to deconvolution, which we call `unfolding,' we can recover a typical specular design problem.
	Consequently, we consider freeform reflector design with a scattering surface as a two-step process wherein the target distribution is first altered to account for scattering, and then the resulting specular problem is solved.
	We verify our approach using a custom raytracer that implements the surface scattering model we used to derive the Fredholm integral.
\end{changemargin}
\hrule

\clearpage

\pagestyle{main}
\pagenumbering{arabic}
\setcounter{page}{1}
\restoregeometry

\section{Introduction}
Lighting plays a crucial role in our current society, and since the introduction of light-emitting diodes (LEDs), the prevalence of beam-shaping optical elements has increased.
This is partly because the sharp, point-like light from a bare LED package is typically considered undesirable and partly due to the increasing demand for aesthetic and personalized lighting, such as RGB LED lights.
These optical elements are typically designed in an iterative and largely manual process, requiring significant experience and knowledge on the part of the optical designer, as well as considerable time \cite[Ch.~1.9]{koshelIlluminationEngineeringDesign2013}.
While a specular reflector can shape the light into a desired light distribution, it cannot necessarily reduce the sharpness of the light source since the mirrored surface may result in undesirable glare.
Scattering elements may help address the glare, such as rough reflector surfaces with scattering or transmissive scattering elements combined with a specular reflector \cite[Ch.~1.8.4]{koshelIlluminationEngineeringDesign2013}.
Introducing scattering in the system generally means that some light control is lost, i.e., achieving the specified target cannot be guaranteed \textit{a priori}.
This work includes surface scattering in a consistent way into the existing framework for computing specular reflectors in the context of inverse freeform design to regain control over the light.

More precisely, the problem of directly computing an optical system given source and target distributions is often referred to as the \textit{inverse problem of illumination optics}.
Many methods of solving the inverse specular problem for reflectors and lenses have been developed over the last few decades, such as by solving a system of coupled ordinary differential equations (ODEs) in the case of rotationally or cylindrically symmetric systems \cite{maesMathematicalMethodsReflector1997}.
For three-dimensional freeform optical surfaces --- i.e., surfaces without any overall symmetry --- a method that has proven successful is based on solving a Monge-Amp{\`e}re equation \cite{prinsInverseMethodsIllumination2014,yadavMongeAmpereProblemsNonquadratic2018,romijnGeneratedJacobianEquations2021}.

While the specular inverse design problem is well-researched, literature concerning the direct computation of scattering optical surfaces is scarce.
The best reference we have found is Lin \textit{et al.}~\cite{linNovelOpticalLens2015}, who designed a lens with a freeform scattering inner surface and a spherical outer one.
Their approach represented the freeform surface by B{\'e}zier curves.
The initial shape was iteratively modified to take into account the differences between the prescribed target distribution and the resulting raytraced distribution.

As we showed in \cite{kronbergModellingSurfaceLight2023,kronbergTwodimensionalFreeformReflector2023}, the problem of computing two-dimensional --- i.e., rotationally or cylindrically symmetric --- reflectors with a scattering surface reduces to computing a deconvolution, followed by solving a specular reflector design problem.
This manuscript will extend these results to compute three-dimensional freeform reflectors with scattering surfaces.
To do so, we shall first find a mathematical relation between the light reflected from a perfectly smooth reflector with a specific shape and the scattered light from the same reflector made from a scattering material.
We will show that this relation takes the shape of a Fredholm integral equation of the first kind, and we will then show how we solved this integral relation to gain a suitable target function to use in the specular design problem.
This approach is thus analogous to the two-dimensional one we presented in \cite{kronbergModellingSurfaceLight2023}.

The manuscript is structured as follows.
The scattering model is first derived in Sec.~\ref{sec:scatteringModel} based on ideas from optimal transport theory.
Next, the freeform specular design problem is discussed in some detail in Sec.~\ref{sec:specularReflectorDesign}, followed by an outline of how we verified the aforementioned model in Sec.~\ref{sec:verification}.
Two numerical examples are shown in Sec.~\ref{sec:examples} --- the first showcases how we propose to use our model in a typical workflow, and the second shows how varying the amount of scattering influences the reflector shape.

\section{Scattering Model}\label{sec:scatteringModel}
\noindent This section treats the theoretical aspects needed to develop the scattering model and to apply it in the context of freeform reflector design.

\subsection{Key Assumptions}
We shall make several assumptions throughout our derivation of the Fredholm integral equation governing scattering in our model.
The key assumptions are discussed here; additional assumptions will be introduced when they become relevant.
The first assumption is that light scattering can be described using geometric optics by considering incoming, specularly reflected, and scattered light rays.
Statistically, this is equivalent to the more physical notion of scattering whereby one incident direction yields multiple outgoing directions.
Furthermore, light scattering is assumed to be fully elastic, i.e., the incident energy is scattered without absorption or other losses.
The medium surrounding the reflector is also assumed to be lossless, and light is assumed to be scattered exclusively at the reflector surface.
Finally, in this manuscript, we shall only design reflectors illuminated by zero-{\'e}tendue parallel light with far-field targets.
Note that the derivation of the scattering model is also valid for zero-{\'e}tendue point sources since the scattering event occurs at the surface, irrespective of the system's symmetry (or lack thereof).

\clearpage
\subsection{Geometry}

\hspace{1pt} 
\vspace{-\fsize} 

\begin{wrapfigure}{r}{0.3\linewidth}
	\centering
	\includegraphics[width=\linewidth]{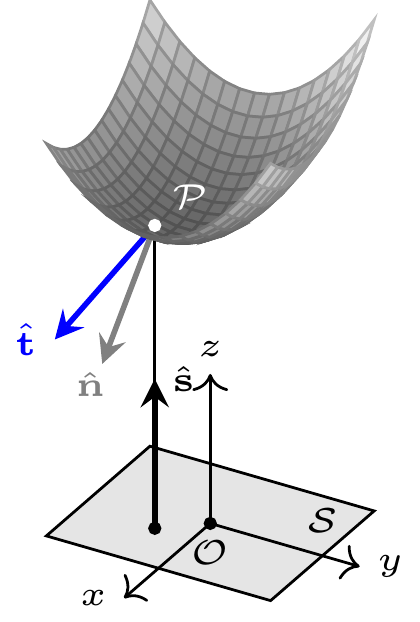}
	\captionsetup{width=\linewidth}
	\caption{Specular reflection from a smooth reflector.}
	\label{fig:geometry}
\end{wrapfigure}

\noindent Suppose we have a Cartesian $xyz$-coordinate system in $\mathbb{R}^3$, with a parallel-ray source (henceforth \textit{parallel source}) on a rectangular domain $\mathcal{S} = [a_1,a_2] \times [b_1,b_2] \subset \mathbb{R}^2$ in the $xy$-plane, centred around the origin $\mathcal{O}$ --- see Fig.~\ref{fig:geometry}.
Rays emitted from the source (so-called \textit{source rays} or \textit{incident rays}) propagate in a fixed `upwards' direction, i.e., with a positive $z$-component, given by the unit vector $\us$ (hats ($\hat{\ }$) denote unit vectors throughout this manuscript).
For simplicity, we shall align the source rays with the $z$-axis, i.e., $\us \equiv \ue_z = (0,0,1)^\intercal$.

A specularly reflected ray, i.e., one abiding by the familiar law of specular reflection (a so-called \textit{specular ray} or \textit{reflected ray}), propagates along the unit vector
\begin{equation}
	\ut := \ut(\psi,\chi) := \big(\! \sin(\psi)\cos(\chi),\sin(\psi)\sin(\chi),\cos(\psi) \big)^\intercal,
\end{equation}
where $\psi \in [0,\pi]$ and $\chi \in [0,2\pi]$.
Note that $\ut$ is given by the vectorial law of reflection (LoR):
\begin{equation}\label{eq:LoR}
	\ut = \us - 2(\us \cdot \un)\un,
\end{equation}
where $\un$ is the surface normal at the point of intersection, $\mathcal{P}$.
By convention, we choose $\us \cdot \un < 0$, i.e., the normal pointing towards the light source.
Note that $\us$, $\un$ and $\ut$ are coplanar; they span the so-called \textit{plane of incidence}.


An off-specular ray leaving the surface (a so-called \textit{scattered ray}) propagates along the unit vector
\begin{equation}\label{eq:uu_gammanu}
	\uu := \uu(\gamma,\nu) := \big(\! \sin(\gamma)\cos(\nu),\sin(\gamma)\sin(\nu),\cos(\gamma) \big)^\intercal,
\end{equation}
where $\gamma \in [0,\pi]$ and $\nu \in [0,2\pi]$, respectively.
The following section concerns how the scattered direction relates to the specular direction.

\subsection{Model Derivation}\label{subsec:model}
\noindent We shall now consider the scattering model in detail.
It is based on Monge's formulation of the optimal transport problem in mathematics.
The next few sections will show how this setup and subsequent analysis yield an expression for the scattered light in the form of a Fredholm integral equation of the first kind.

\subsubsection{Mappings}
Speaking in general terms, it can be shown that the optical map, i.e., the mapping that gives the specular direction $(\psi,\chi)$ corresponding to a given incident direction $(\th,\ph)$ parametrizing $\us(\th,\ph)$, is injective for strictly convex mirrors (perfect specular reflectors) \cite{romijnInverseReflectorDesign2020}.
That is, the specular direction $(\psi,\chi)$ is unique for any given incident direction.
Let us denote this map by $\m$, so that $(\psi,\chi) = \m(\th,\ph)$.

Consider now a rough reflector, i.e., one where the resulting light is scattered into a direction $(\gamma,\nu)$, typically different from $(\psi,\chi)$.
To relate the two directions, let us first return to a static $xyz$-coordinate system, starting with defining the elemental rotation matrices
\begin{equation}
	\R_y(\theta) :=
	\begin{pmatrix}
		\cos(\theta) & 0 &\sin(\theta)\\
		0 & 1 & 0\\
		-\sin(\theta) & 0 & \cos(\theta)
	\end{pmatrix}
\end{equation}
and
\begin{equation}
	\R_z(\phi) :=
	\begin{pmatrix}
		\cos(\phi) & -\sin(\phi) & 0\\
		\sin(\phi) & \cos(\phi) & 0\\
		0 & 0 & 1
	\end{pmatrix},
\end{equation}
corresponding to rotations around the $y$-axis and $z$-axis by the right-hand rule, i.e., counter-clockwise in the $zx$-plane and counter-clockwise in the $xy$-plane, respectively.
Thus, by construction, $\ut(\psi,\chi) = \R_z(\chi) \R_y(\psi) \ue_z$, and $\uu(\gamma,\nu) = \R_z(\nu) \R_y(\gamma) \ue_z$.

\begin{figure}[htb!]
	\centering
	\includegraphics[width=0.4\linewidth]{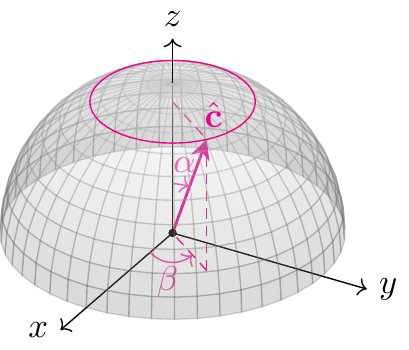}
	\captionsetup{width=\linewidth}
	\caption{The cone vector $\uc$ traces a circle on the unit sphere for fixed $\alpha$.}
	\label{fig:coneVector}
\end{figure}

Let $\alpha \in [0,\pi/2]$ and $\beta \in [0,2\pi]$ be the polar (incline) and azimuthal angles of the so-called \textit{cone vector},
\begin{equation}
	\uc := \uc(\alpha,\beta) := \big(\! \sin(\alpha)\cos(\beta),\sin(\alpha)\sin(\beta),\cos(\alpha) \big)^\intercal.
\end{equation}
As for the origin of the name, notice that for a fixed $\alpha \in [0,\pi/2]$, the parametric curve traced by $\uc(\alpha, \beta),\ 0 \leq \beta \leq 2\pi$, is a circle on the unit sphere, centered around $\big(0,0,\cos(\alpha)\big)$ with radius $\sin(\alpha)$ --- see Fig.~\ref{fig:coneVector}.
In other words, the vector $\uc$ is located on a cone coaxial with the $z$-axis with base radius $\sin(\alpha)$ and height $\cos(\alpha)$.
This is, of course, true for \textit{any} unit vector parametrized by polar and azimuthal angles, but this observation will become relevant later when we construct the scattered ray direction $\uu$.

\begin{wrapfigure}{r}{0.3\linewidth}
	\centering
	\includegraphics[width=\linewidth]{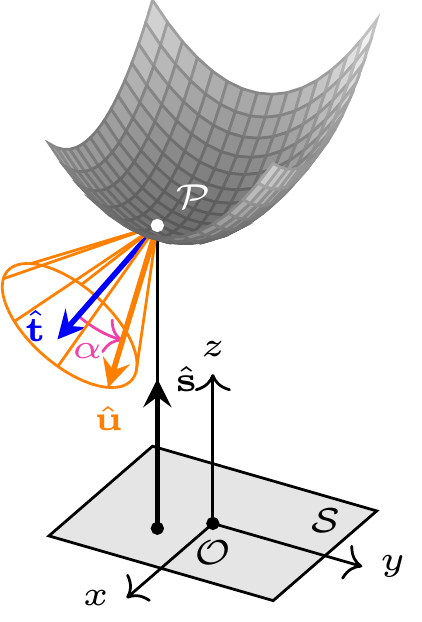}
	\captionsetup{width=\linewidth}
	\caption{Scattering by a rough reflector.}
	\label{fig:geometry_scattering}
\end{wrapfigure}

Note that, by definition, $\uc(\alpha,\beta) = \R_z(\beta)\R_y(\alpha) \ue_z$.
Suppose we apply the rotation matrix $\R_y(\psi)$ followed by $\R_z(\chi)$ to $\uc$.
Then, for a fixed $\alpha \in [0,\pi/2]$, the resulting vector would trace a tilted cone coaxial with $\ut$ by letting $0 \leq \beta \leq 2\pi$, with base radius $\sin(\alpha)$ and height $\cos(\alpha)$.
This is what we want for our scattered vector $\uu$, i.e.,
\begin{equation}\label{eq:u_rot}
	\uu = \R_z(\chi)\R_y(\psi) \uc(\alpha,\beta) = \R_z(\chi)\R_y(\psi) \R_z(\beta)\R_y(\alpha) \ue_z.
\end{equation}
Fig.~\ref{fig:geometry_scattering} shows the scattering geometry for a fixed $\alpha$.
The circle traced by $\uu$ is achieved by letting $\beta$ vary from $0$ to $2\pi$.
By sampling $\alpha$ and $\beta$, we can thus control the direction of $\uu$ with respect to $\ut$.
Note that for $\beta = 0$ or $\beta=2\pi$, $\uu$ lies in the plane of incidence spanned by $\us$, $\ut$ and $\un$, since $\R_z(0) = \R_z(2\pi) = \mathbb{I}$, the identity matrix, so that $\uu = 
\R_z(\chi)\R_y(\psi + \alpha) \ue_z.$
Similarly, when $\beta = \pi$, $\uu$ also lies in the plane of incidence, since $\R_z(\pi)$ has nonvanishing elements $\{-1,-1,1\}$ on the diagonal.
Also notice that $\uu = \R_z(\chi)\R_y(\psi) \R_z(\beta)\R_y(\alpha - \psi)\R_z(-\chi) \ut(\psi,\chi)$, since, by construction, $\ue_z = \R_y(-\psi)\R_z(-\chi) \ut(\psi,\chi)$.
This gives a direct, albeit cumbersome, relation between the specular direction $\ut$ and scattered direction $\uu$ for fixed $\alpha$ and $\beta$.

\begin{figure*}[ht!]
	\centering
	\includegraphics[width=0.8\linewidth]{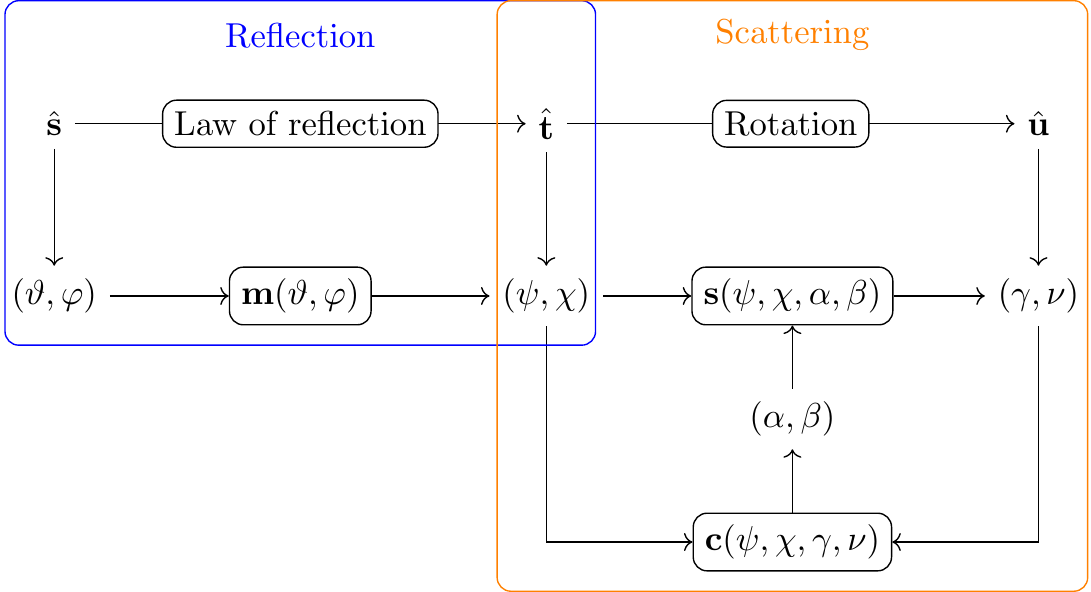}
	\captionsetup{width=\linewidth}
	\caption{Relations between unit vectors and corresponding spherical coordinates.}
	\label{fig:mappings}
\end{figure*}

Equating the representation of $\uu$ in Eq.~\eqref{eq:u_rot} with $\uu(\gamma,\nu)$ in Eq.~\eqref{eq:uu_gammanu} allows us to solve for $(\gamma,\nu)$ for any given specular direction $(\psi,\chi)$ and cone vector direction $(\alpha,\beta)$.
That is, we may find a so-called \textit{scattering map}, $\s$, such that  $\s(\psi,\chi,\alpha,\beta) = (\gamma,\nu)$.
Suppose we instead want $(\alpha,\beta)$ for some known $(\psi,\chi)$ and $(\gamma,\nu)$ pairs.
This yields a third map, the so-called \textit{cone map}, say $\c$, such that $\c(\psi,\chi,\gamma,\nu) = (\alpha,\beta)$.
These relations are summarised in Fig.~\ref{fig:mappings}.
We shall focus on the scattering part, i.e., finding $\s(\psi,\chi,\alpha,\beta)$ and $\c(\psi,\chi,\gamma,\nu)$.

\paragraph{Finding the scattering map}
We shall first find the scattering map, $\s$, which returns the scattered direction $(\gamma,\nu)$ of some specular direction $(\psi,\chi)$ and cone direction $(\alpha,\beta)$.
Starting with $\gamma$, note that, by construction, $\cos(\gamma) = u_3$, where $u_3$ is the third component of $\uu$.
Thus, by evaluating Eq.~\eqref{eq:u_rot}, we find that
\begin{equation}\label{eq:gammaMap}
	\gamma(\psi,\chi,\alpha,\beta) = \arccos\!\big(\! \cos(\psi)\cos(\alpha) - \sin(\psi)\sin(\alpha)\cos(\beta) \big).
\end{equation}
Next, $\nu$ can be found by noticing that $\tan(\nu) = u_2/u_1$.
Computing the components of $\uu$ in Eq.~\eqref{eq:u_rot} yields
\begin{equation}\label{eq:nuMap}
	\nu(\psi,\chi,\alpha,\beta) = \arctan\!\big( x_\nu(\psi,\chi,\alpha,\beta), y_\nu(\psi,\chi,\alpha,\beta) \big),
\end{equation}
where $\arctan(x,y)$ is the inverse tangent of $y/x$, taking into account the quadrant of the point $(x,y)$, and where
\begin{equation}
	\begin{split}
		x_\nu &= \cos (\chi ) \big(\! \sin (\alpha ) \cos (\beta )  \cos (\psi )+\cos (\alpha ) \sin (\psi )\big)\\
		&\quad - \sin (\alpha ) \sin (\beta ) \sin (\chi ),\\
		y_\nu &= \sin (\chi ) \big(\! \sin (\alpha ) \cos (\beta ) \cos (\psi )+\cos (\alpha ) \sin (\psi ) \big)\\
		&\quad + \sin (\alpha ) \sin (\beta ) \cos (\chi ).
	\end{split}
\end{equation}
Thus, the scattering map is
\begin{equation}\label{eq:scatterMap}
	\s(\psi,\chi,\alpha,\beta) = \big( \gamma(\psi,\chi,\alpha,\beta), \nu(\psi,\chi,\alpha,\beta) \big),
\end{equation}
with $\gamma$ and $\nu$ given by Eqs.~\eqref{eq:gammaMap} and \eqref{eq:nuMap}, respectively.

\paragraph{Finding the cone map}
We shall now find the cone map, $\c$, yielding $(\alpha,\beta)$ for given directions $(\psi,\chi)$ and $(\gamma,\nu)$.
Starting with $\alpha$, note that by construction (recall Fig.~\ref{fig:geometry_scattering}), $\cos(\alpha) = \ut \cdot \uu$, so that,
\begin{equation}\label{eq:alphaMap}
	\alpha(\psi,\chi,\gamma,\nu) = \arccos\!\big(\!\cos(\psi)\cos(\gamma) + \sin(\psi)\sin(\gamma)\cos(\nu - \chi)\big).
\end{equation}

Finding $\beta$ requires significantly more effort.
Theoretically, one could equate the two representations of $\uu$ in Eqs.~\eqref{eq:uu_gammanu} and \eqref{eq:u_rot} and solve for $\beta$; in practice, however, this turns out to be very difficult.
Since we shall assume rotational symmetry later in this manuscript, meaning the explicit expression for $\beta$ is no longer necessary, we only briefly summarise how it was derived below.
We first considered a representation of $\uu$ in terms of the stereographic components of the reflected vector $\ut$ and the components of $\uc$, i.e., $\y(\ut)$, $c_1$, $c_2$, and $c_3$, where $\y$ is the 2-tuple associated with the unit vector $\ut$ via stereographic projection from the north pole:
\begin{equation}\label{eq:sterNP}
	\y(\ut) =
	\begin{pmatrix}
		y_1\\
		y_2
	\end{pmatrix}
	=
	\frac{1}{1-t_3}
	\begin{pmatrix}
		t_1\\
		t_2
	\end{pmatrix}
	=
	\frac{\sin(\psi)}{1-\cos(\psi)}
	\begin{pmatrix}
		\cos(\chi)\\
		\sin(\chi)
	\end{pmatrix}.
\end{equation}

This allowed us to solve for $c_1$ and $c_2$ in terms of $y_1$ and $y_2$, or, via the stereographic projection, in terms of $\psi$ and $\chi$, as well as $\gamma$ and $\nu$.
Then, we used the fact that, by construction, $\tan(\beta) = c_2/c_1$, and the scattering map in Eq.~\eqref{eq:scatterMap} to conclude that
\begin{equation}\label{eq:betaMap}
	\beta(\psi,\chi,\gamma,\nu) = \arctan\big( x_\beta(\psi,\chi,\gamma,\nu) , y_\beta(\chi,\gamma,\nu) \big),
\end{equation}
where
\begin{equation}
	\begin{split}
		x_\beta &= \cos(\psi)\sin(\gamma)\cos(\nu-\chi) - \sin(\psi)\cos(\gamma),\\
		y_\beta &= \sin(\gamma)\sin(\nu-\chi).
	\end{split}
\end{equation}
Thus, the cone map is
\begin{equation}\label{eq:coneMap}
	\c(\psi,\chi,\gamma,\nu) = \big(\alpha(\psi,\chi,\gamma,\nu), \beta(\psi,\chi,\gamma,\nu)\big),
\end{equation}
with $\alpha$ and $\beta$ given by Eqs.~\eqref{eq:alphaMap} and \eqref{eq:betaMap}, respectively.

\subsubsection{Energy Balances}
Let us introduce the light distributions associated with the source, the specular and scattered light.
The light source is parallel, meaning it is prescribed in the form of an exitance $[\mathrm{W} \cdot \mathrm{m}^{-2}]$ denoted by $f(x,y) > 0$, $(x,y) \in \mathcal{S} \subset \mathbb{R}^2$, where $\mathcal{S} = [a_1,a_2] \times [b_1,b_2]$.
Both the reflected light and the scattered light may be described using intensity distributions [$\mathrm{W} \cdot \mathrm{sr}^{-1}$] in the far field.
We have:
\begin{itemize}
	\item virtual specular target intensity distribution $g(\psi,\chi) > 0$, $(\psi,\chi) \in \T$,
	\item scattered target intensity distribution $h(\gamma,\nu) > 0$, $(\gamma,\nu) \in \U$,
\end{itemize}
where $\T$ and $\U$ are angular domains such that $\ut$ and $\uu$ are on $\SS^2$.
They form the supports of the intensity distributions, i.e., $g = h = 0$ outside of these domains.
The addition of \textit{virtual} to the specular target intensity distribution comes from the fact that $g$ is never observed from a rough reflector.
Instead, $f$ and $h$ are prescribed, and $g$ is computed in some manner we are yet to describe, which in turn allows the shape of the rough freeform reflector to be calculated by solving a specular design problem.

Before discussing the freeform design problem, we shall formulate the relation between $g$ and $h$.
To do so, let us first note that our assumptions regarding the absence of losses in the system lead to the following global energy balances:
\begin{equation}\label{eq:energyBalance}
	\begin{split}
		\int_{\S} f(x,y) \, \dd x \dd y &=
		\int_{\T} g(\psi,\chi) \sin(\psi) \, \dd\psi \dd\chi\\
		&= \int_{\U} h(\gamma,\nu) \sin(\gamma) \, \dd\gamma \dd\nu,
	\end{split}
\end{equation}
i.e., all the energy of the source distribution $f$ is contained in the specular light distribution $g$ and the scattered light distribution $h$.

\begin{figure}[htb!]
	\centering
	\includegraphics[width=0.4\linewidth]{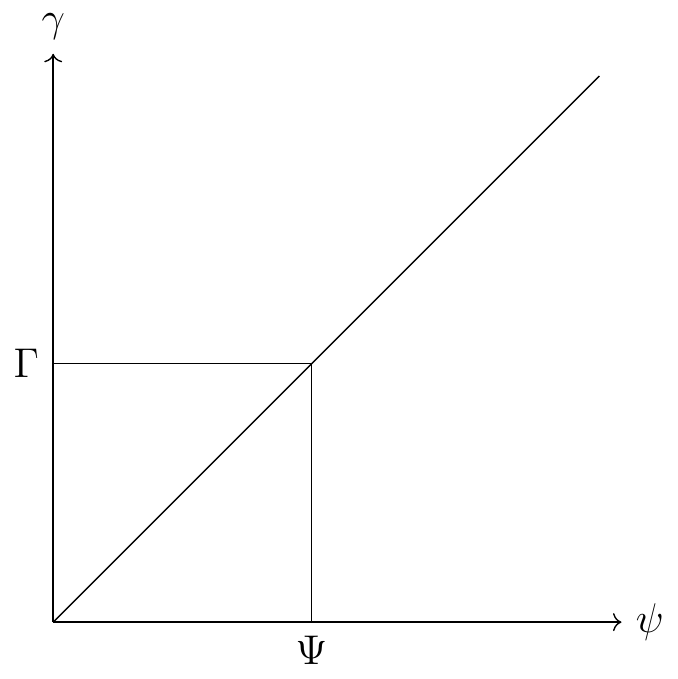}%
	\includegraphics[width=0.4\linewidth]{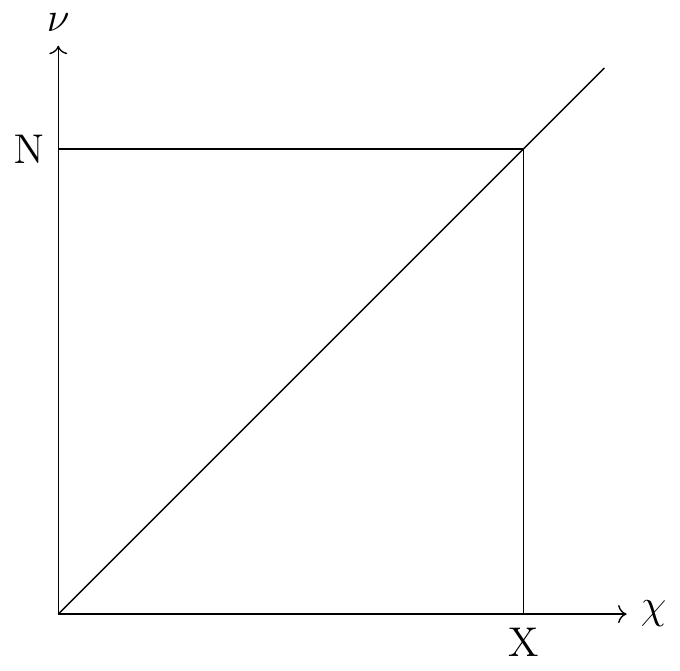}
	\captionsetup{width=\linewidth}
	\caption{Specular maps $\psi \to \gamma$ and $\chi \to \nu$ with fixed points $\Psi \to \Gamma$ and $\Chi \to \Nu$.}
	\label{fig:specularMaps}
\end{figure}

\paragraph{Optimal transport}
Suppose we fix $\psi = \Psi$ and $\chi = \Chi$ such that $(\Psi,\Chi) \in \T$.
Consider perfect specular reflection, i.e., reflection from a perfectly mirrored surface.
In that case, $\alpha$ vanishes and $\beta$ is irrelevant, so that Eq.~\eqref{eq:u_rot} gives $\uu \equiv \ut$.
Then, $\gamma = \Gamma \equiv \Psi$ and $\nu = \Nu \equiv \Chi$, i.e., $\s(\Psi,\Chi,\alpha,\beta)$ is the identity map for all $\alpha \in [0,\pi/2]$ and $\beta \in [0,2\pi]$.
This is true for any $\Psi$ and $\Chi$ so that we get the plots in Fig.~\ref{fig:specularMaps}.

\begin{figure}[htb!]
	\centering
	\includegraphics[width=0.4\linewidth]{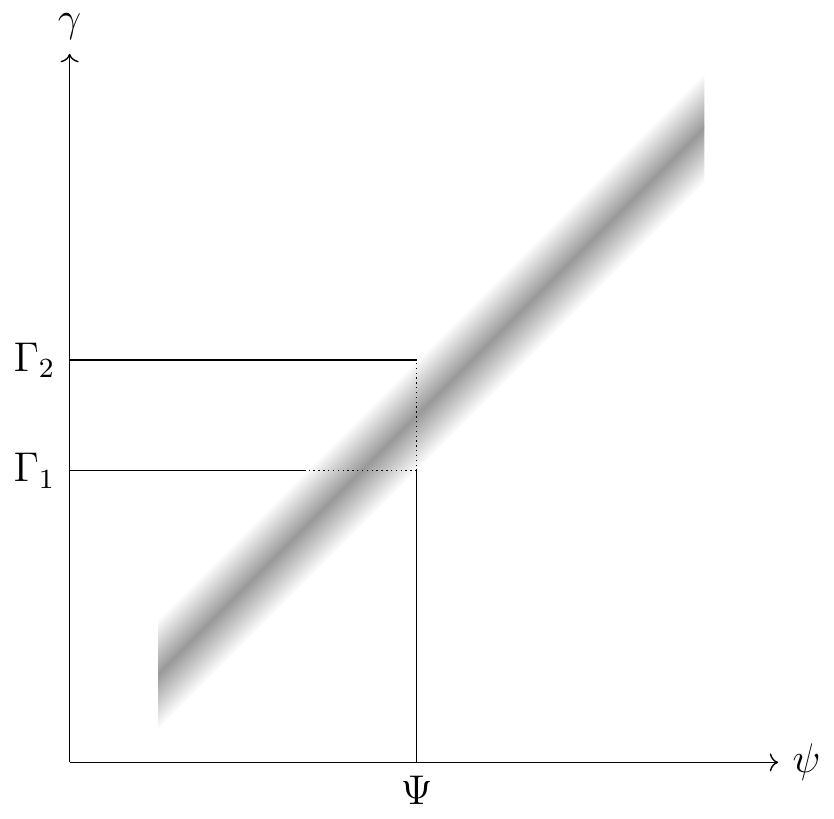}%
	\includegraphics[width=0.4\linewidth]{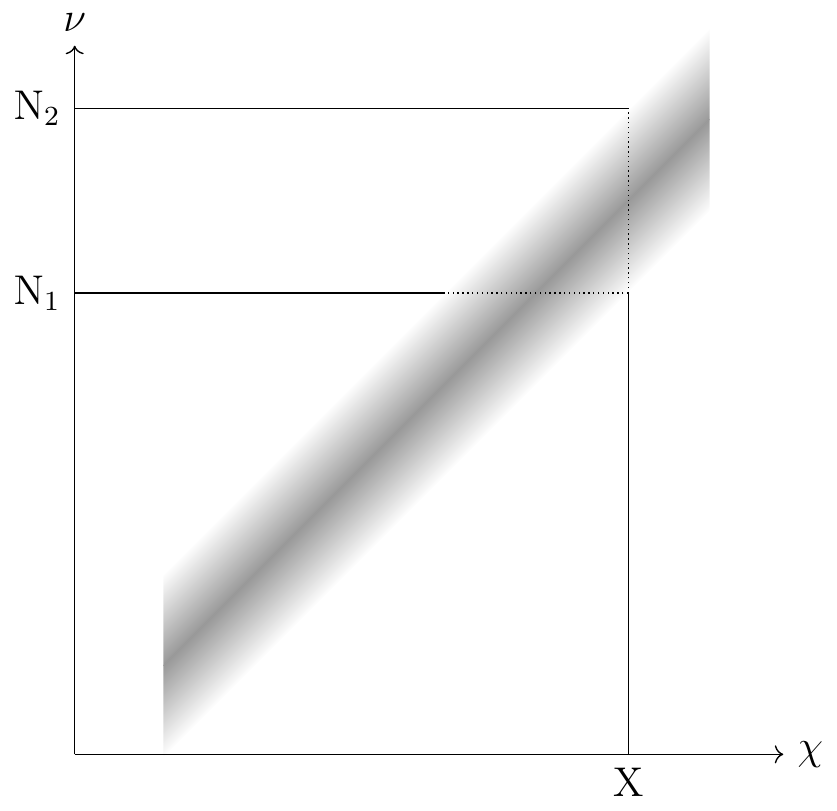}
	\captionsetup{width=\linewidth}
	\caption{Schematic scattering maps $\psi \to \gamma$ and $\chi \to \nu$ with fixed points $\Psi \to [\Gamma_1, \Gamma_2]$ and $\Chi \to [\Nu_1, \Nu_2]$.}
	\label{fig:scatteringMaps}
\end{figure}

Suppose instead we have scattering from a rough surface, then $\alpha$ and $\beta$ are nonvanishing, and the simple one-to-one relationship schematically shown in Fig.~\ref{fig:specularMaps} is replaced by a richer relationship.
Schematically, we can imagine a broadening of the lines, indicating a probability to go in that direction --- see Fig.~\ref{fig:scatteringMaps}.
Here, $[\Gamma_1,\Gamma_2] \times [\Nu_1,\Nu_2] \subseteq \U$ indicates the nonzero region where the direction $(\Psi,\Chi) \in \T$ is mapped.
The relationship is more complex, but this is an intuitive starting point.

Let us now explore the connection between our approach of modeling scattering and optimal transport, particularly so-called Monge-Kantorovich problems \cite[Ch.~1]{villaniTopicsOptimalTransportation2003}.
Let $\rho(\psi,\chi,\gamma,\nu) > 0$, $(\psi,\chi) \in \T$, $(\gamma,\nu) \in \U$ represent a \textit{density} with properties
\begin{equation}\label{eq:rhoIntegrals}
	\begin{split}
		\int_{\U} \rho(\psi,\chi,\gamma,\nu) \sin(\gamma) \, \dd\gamma \dd\nu &= g(\psi,\chi),\\
		\int_{\T} \rho(\psi,\chi,\gamma,\nu) \sin(\psi) \, \dd\psi \dd\chi &= h(\gamma,\nu).
	\end{split}
\end{equation}
If we have a direction $(\psi,\chi)$, integrating over the domain $\U$ will provide us with the specularly reflected light in that direction.
Similarly, integrating over the domain $\T$ for a direction $(\gamma,\nu)$ will give us the scattered light in that direction.
The second energy balance in Eq.~\eqref{eq:energyBalance} is fulfilled by direct substitution of the relations in Eq.~\eqref{eq:rhoIntegrals} after a change of order of integration (note that $\rho$ has finite support so that the change of integration order is always allowed):
\begin{equation}
 	\begin{split}
 		&\int_{\T} \int_{\U} \rho(\psi,\chi,\gamma,\nu) \sin(\gamma) \, \dd\gamma \dd\nu \, \sin(\psi) \, \dd\psi \dd\chi\\
 		= &\int_{\U} \int_{\T} \rho(\psi,\chi,\gamma,\nu) \sin(\psi) \, \dd\psi \dd\chi \, \sin(\gamma) \, \dd\gamma \dd\nu.
 	\end{split}
\end{equation}

Returning to the schematic scattering maps in Fig.~\ref{fig:scatteringMaps}, it seems reasonable to make the following ansatz.
Let $p$ be a probability density function on the unit sphere depicting the broadening of the lines.
Then, the density $\rho$ that we shall pick is given by the product
\begin{equation}\label{eq:rhoChoice}
	\rho(\psi,\chi,\gamma,\nu) = p\big(\c(\psi,\chi,\gamma,\nu)\big)g(\psi,\chi),
\end{equation}
since this encapsulates the smearing out of the light from direction $(\psi,\chi)$ due to scattering.
Inserting this density into the second relation in Eq.~\eqref{eq:rhoIntegrals} and noting that $\T$ and $\U$ constitute the finite support of $\rho$ so that we may readily extend the integrations to the whole unit sphere, we get
\begin{equation}\label{eq:scatteringEq}
	h(\gamma,\nu) = \int_{0}^{2\pi} \int_{0}^{\pi} p\big(\c(\psi,\chi,\gamma,\nu)\big) g(\psi,\chi) \sin(\psi) \, \dd\psi \dd\chi,
\end{equation}
which further motivates our choice of density.
In particular, notice that this Fredholm integral equation of the first kind reduces to a two-dimensional convolution integral if the kernel depends on the shift between the variables, i.e., if $\c(\psi,\chi,\gamma,\nu) = \tilde{\c}(\gamma - \psi, \nu - \chi)$.
Thus, we can reasonably expect it to act similarly, i.e., that the kernel $p$ will `smear out' the function $g$.
This is consistent with the blurring of an image when light is scattered from rough surfaces versus perfect mirrors \cite[Sec.~1.8.4]{koshelIlluminationEngineeringDesign2013}, \cite[Ch.~10]{stoverOpticalScatteringMeasurement2012}.

If we insert our choice of $\rho$ from Eq.~\eqref{eq:rhoChoice} into the first relation of Eq.~\eqref{eq:rhoIntegrals}, meanwhile, and extend the limits to those of the unit sphere, we get, for all $\psi \in [0,\pi]$ and $\chi \in [0,2\pi]$,
\begin{equation}
	\int_{0}^{2\pi} \int_{0}^{\pi} p\big(\c(\psi,\chi,\gamma,\nu)\big) \sin(\gamma) \, \dd\gamma \dd\nu = 1.
\end{equation}
Transforming the integrations over $\gamma$ and $\nu$ to $\alpha$ and $\beta$ gives (recall the relations summarised in Fig.~\ref{fig:mappings})
\begin{equation}
	\int_{0}^{2\pi} \int_{0}^{\pi} p(\alpha,\beta) \sin\!\big(\gamma(\psi,\chi,\alpha,\beta)\big) \, \abs{\pdv{\s(\psi,\chi,\alpha,\beta)}{(\alpha,\beta)}} \, \dd\alpha \dd\beta = 1.
\end{equation}
The Jacobian, $\abs{\partial \s/\partial(\alpha,\beta)} = 1$, can be directly computed from Eq.~\eqref{eq:scatterMap}, and $\sin(\gamma)$ can be evaluated using Eq.~\eqref{eq:gammaMap}.
Doing so yields the integral
\begin{equation}
	\int_{0}^{2\pi} \int_{0}^{\pi} p(\alpha,\beta) \sin(\alpha) \, \dd\alpha \dd\beta = 1,
\end{equation}
i.e., we see that $p$ is a probability density function (PDF) on the unit sphere, as required.
Physically, it is clear that (at least for a flat reflector surface) $\alpha \in [0,\pi/2)$ and $\beta \in [0,2\pi]$, so we can safely integrate over the upper hemisphere and maintain energy conservation.
Note that $\alpha$ is typically much smaller than $\pi/2$.
We shall return to this point when considering the examples in Sec.~\ref{sec:examples}.

\paragraph{Rotationally symmetric scattering}
Suppose the PDF $p$ in Eq.~\eqref{eq:scatteringEq} is rotationally symmetric, i.e., $p(\alpha,\beta) = p(\alpha)$ for all $\beta \in [0,2\pi]$.
In that case, Eq.~\eqref{eq:scatteringEq} reduces to
\begin{equation}\label{eq:scatteringEq_rotSym}
	h(\gamma,\nu) = \int_{0}^{2\pi} \int_{0}^{\pi} p\big(\alpha(\psi,\chi,\gamma,\nu)\big) g(\psi,\chi) \sin(\psi) \, \dd\psi \dd\chi,
\end{equation}
where $\alpha(\psi,\chi,\gamma,\nu)$ is given by Eq.~\eqref{eq:alphaMap}, and $p$ is subject to the normalisation
\begin{equation}
	2\pi \int_{0}^{\pi/2} p(\alpha) \sin(\alpha) \, \dd\alpha = 1.
\end{equation}
Note that Eq.~\eqref{eq:scatteringEq_rotSym} is still a Fredholm integral equation.
For simplicity, the forthcoming section with numerical examples, Sec.~\ref{sec:examples}, will focus on PDFs that fulfill $p(\alpha,\beta) = p(\alpha)$.
Physically, this means that any rotation of the scattered ray around the specular direction is equally likely, and only the deviation in polar angle is modulated --- recall Fig.~\ref{fig:geometry_scattering}.
Specifically, we shall choose $p$ such that the most likely value of $\alpha$ drops off from its peak at $\alpha = 0$, similar to what is observed from so-called \textit{glossy} reflections \cite[Ch.~18]{fernandoGPUGemsProgramming2004}.

\begin{figure*}[htb!]
	\centering
	\includegraphics[width=0.33\linewidth]{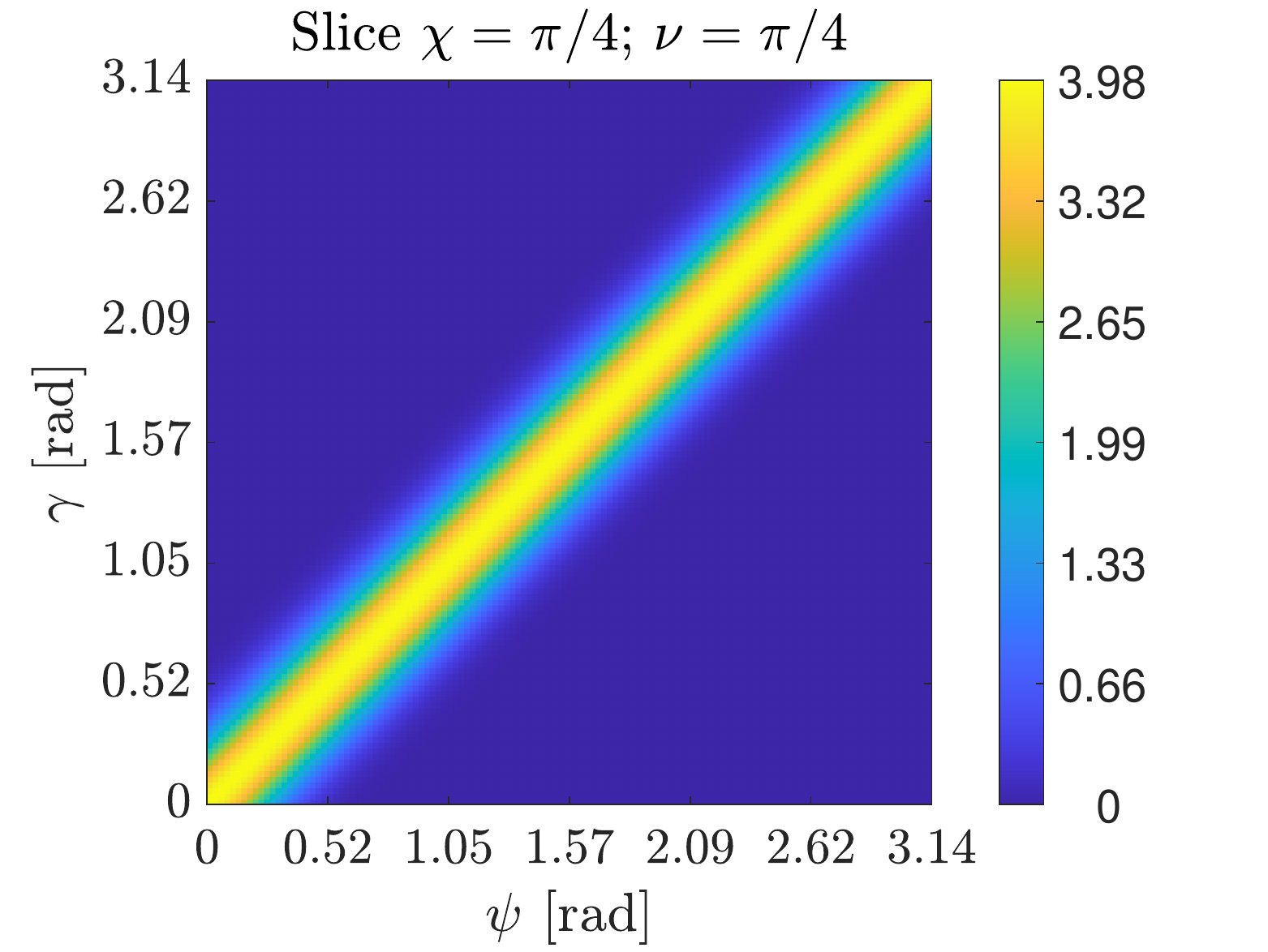}%
	\includegraphics[width=0.33\linewidth]{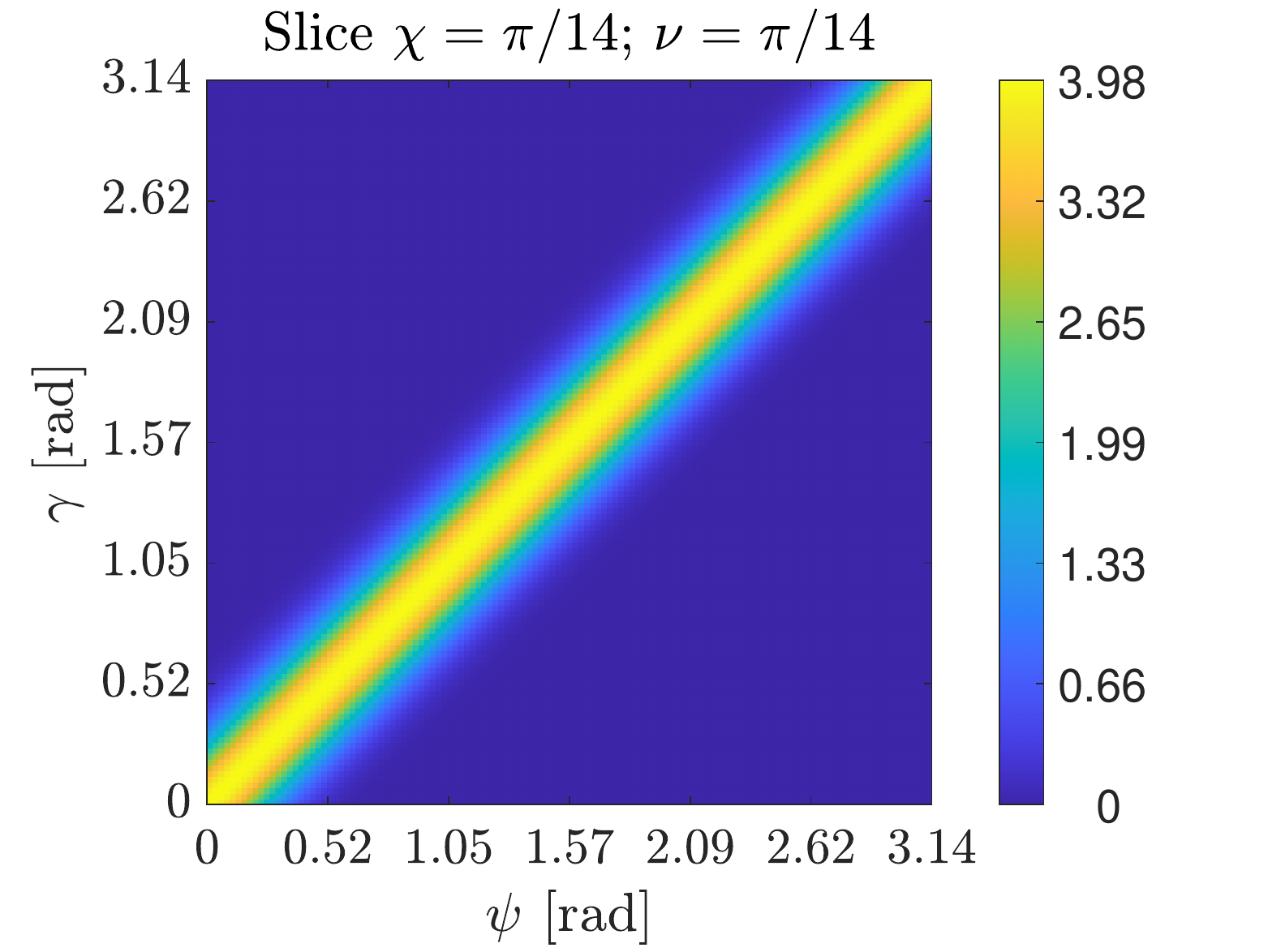}%
	\includegraphics[width=0.33\linewidth]{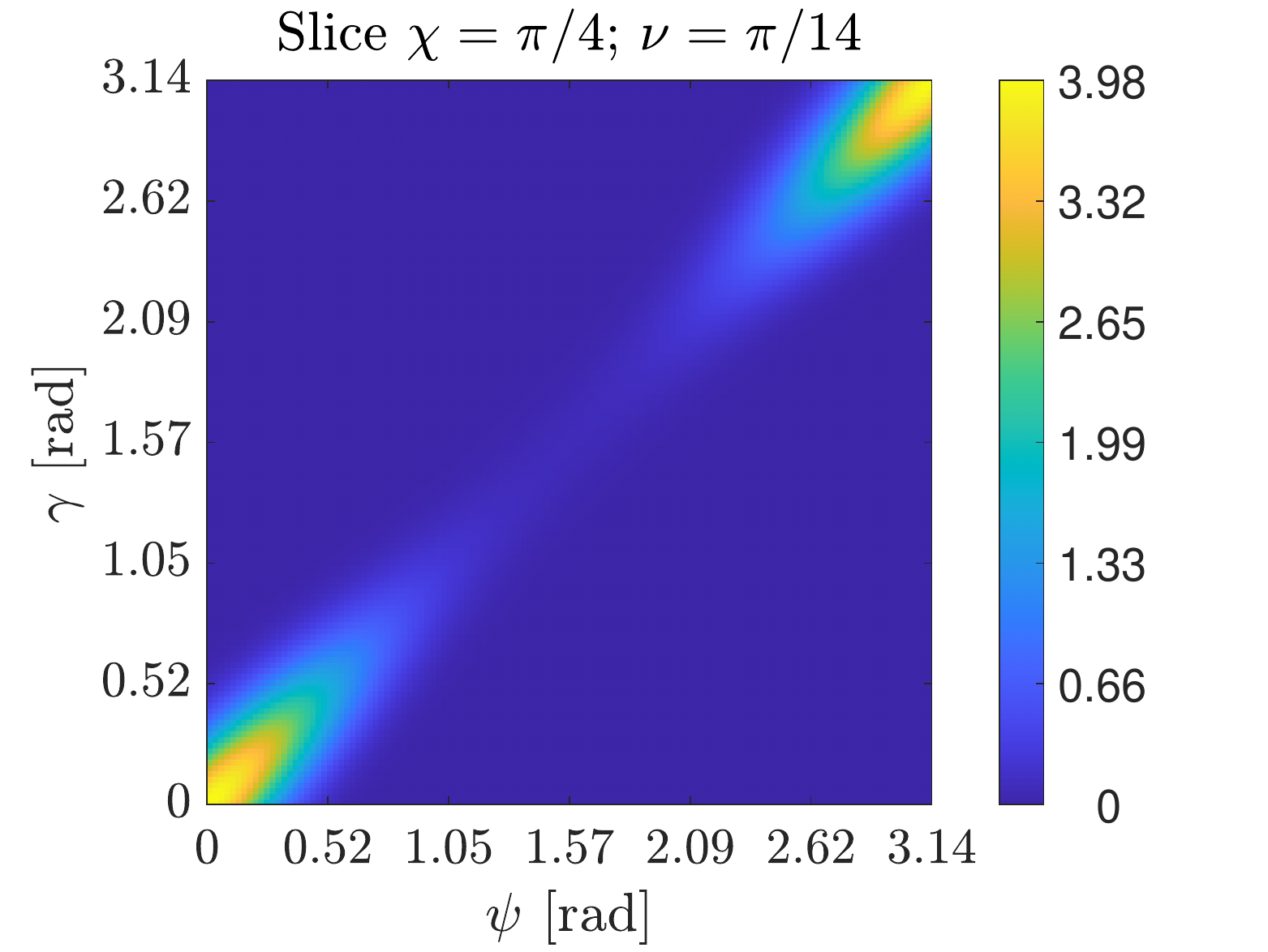}\\[5pt]
	\includegraphics[width=0.33\linewidth]{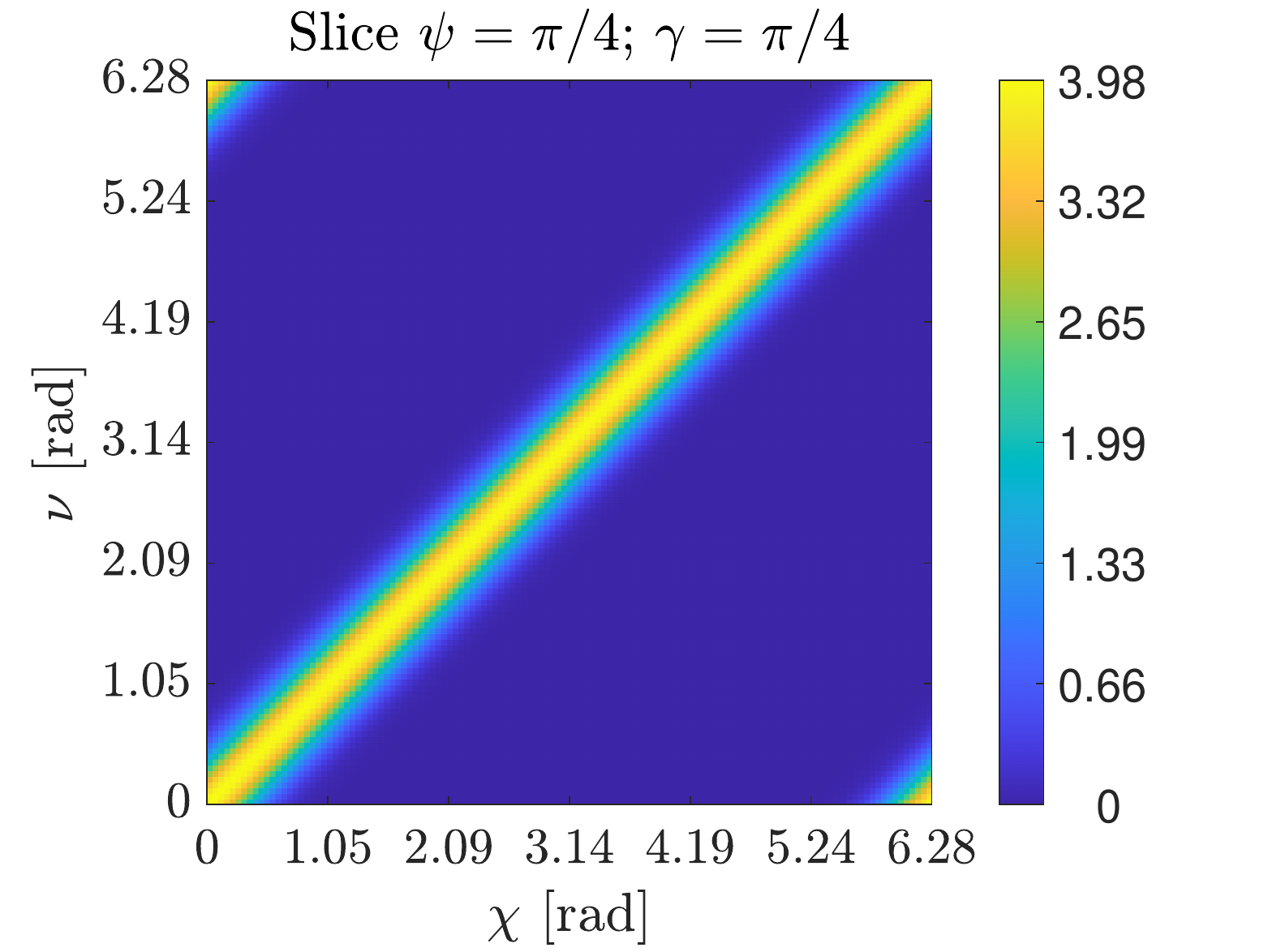}%
	\includegraphics[width=0.33\linewidth]{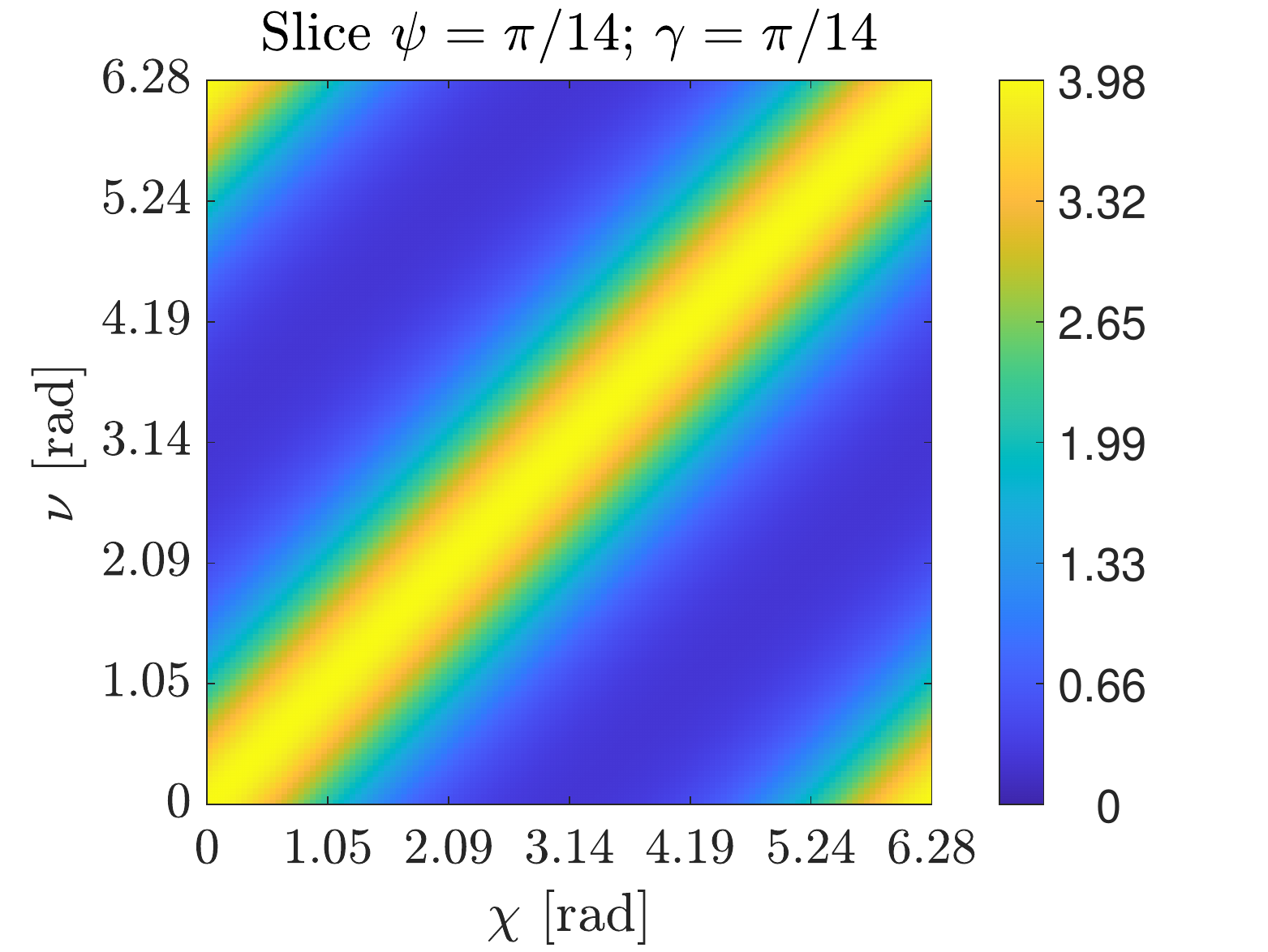}%
	\includegraphics[width=0.33\linewidth]{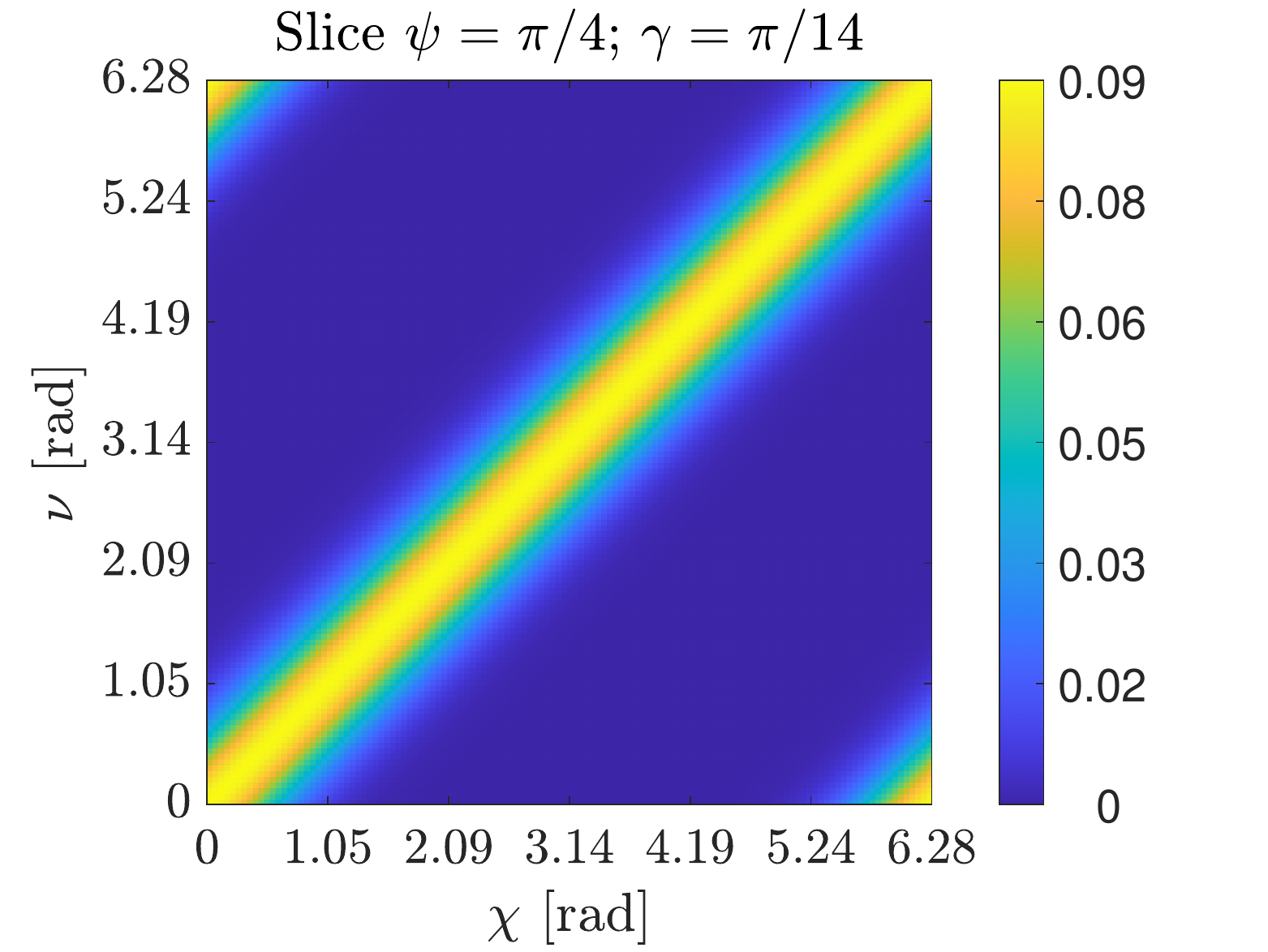}%
	\captionsetup{width=\linewidth}
	\caption{Scattering maps $\psi \to \gamma$ and $\chi \to \nu$ for various fixed values of $\chi$ and $\nu$ or $\psi$ and $\gamma$ from Example \#1 in Sec.~\ref{sec:examples}.}
	\label{fig:scatteringMaps_real}
\end{figure*}

Let us now return to the schematic scattering map in Fig.~\ref{fig:scatteringMaps}.
In particular, compare the schematic versions to Fig.~\ref{fig:scatteringMaps_real}, which depicts the kernel $p$ from Example \#1 in Sec.~\ref{sec:examples}.
Since the kernel is a four-dimensional quantity via the mapping $\alpha(\psi,\chi,\gamma,\nu)$, we only consider slices with two fixed angles --- $\chi$ and $\nu$ or $\psi$ and $\gamma$.
Focusing on the top row, it is clear that the mapping $\psi \to \chi$ does not change for slices where the azimuthal angles $\chi$ and $\nu$ are equal.
However, the mapping significantly differs when $\chi \neq \nu$.
In particular, values close to the poles are more likely to remain close to the poles and cannot readily reach, e.g., the equator.
Meanwhile, the situation is quite different for the mapping $\chi \to \nu$ (bottom row).
First, notice that the map is naturally periodic at $\chi = 2\pi$ and $\nu = 2\pi$ since this is the period of the sphere.
Next, note that these mappings \textit{do} change for slices where the polar angles $\psi$ and $\gamma$ are equal.
This is consistent with what is expected from the rotationally symmetric scattering probability density function since we chose $p(\alpha)$ such that it is most significant close to $\alpha = 0$ (i.e., $\psi = \gamma$) and then drops relatively rapidly to near-vanishing values --- see Example \#1 in Sec.~\ref{sec:examples}.

\subsubsection{Unfolding the Fredholm Integral Equation}
Suppose we want to solve the inverse problem; given a target intensity distribution $h$ and a scattering function $p$, can we compute the virtual specular distribution $g$?
If the scattering equation had been a convolution integral, solving the inverse problem would be known as \textit{deconvolution}. 
As they are Fredholm integrals, we shall refer to the process as \textit{unfolding} for historical reasons --- cf., e.g., \cite{dicolaAnalysisNumericalMethods1967}.
Formally, there are constrained situations for which unique, closed-form, analytical solutions can be constructed when unfolding Fredholm integral equations \cite[Ch.~12]{polianinHandbookIntegralEquations2008}, but we are interested in more general, numerical methods for obtaining an approximation of $g$.

We shall apply Richardson-Lucy deconvolution to the Fredholm integral problem.
This method is based on maximum likelihood arguments; Richardson and Lucy independently developed a ratio deconvolution method due to a need for deblurring images from telescopes and in the context of fluorescence microscopy \cite{richardsonBayesianBasedIterativeMethod1972,lucyIterativeTechniqueRectification1974}.
Specifically, Richardson used Bayesian statistics and assumed a conditional probability caused the blurring, while Lucy considered maximizing the likelihood of the observed sample within the solution space.
We shall not formally show that Richardson-Lucy deconvolution applies to unfolding our Fredholm integral equation. 
Still, the derivation by Lucy in \cite{lucyIterativeTechniqueRectification1974} is so general that it is enough to assume that the blurring occurs via a Poisson process.

Before stating the final Richardson-Lucy expression, let us discretize the Fredholm integral equation in Eq.~\eqref{eq:scatteringEq} (analogously for the rotationally symmetric case in Eq.~\eqref{eq:scatteringEq_rotSym}).
Fix a rectangular grid of $N_1 \times N_2$ points, and let $\h$ be the matrix representation of $h$ with components $h_{ij} = h(\gamma_i, \nu_j)$, where $i \in [1,N_1]$ and $j \in [1,N_2]$.
Similarly, let $\g$ be the matrix representation of $g$ such that $g_{kl} = g(\psi_k,\chi_l)\sin(\psi_k)$, where $k \in [1,N_1]$ and $l \in [1,N_2]$.
Finally, let $\p$ be the tensor representation of $p$ such that $p_{ij}^{kl} = p(\c(\psi_k,\chi_l,\gamma_i, \nu_j))$, where $i,k \in [1,N_1]$ and $j,l \in [1,N_2]$.
Then, the Fredholm integral can be written as
\begin{equation}\label{eq:scatteringEq_disc}
	\h = \p \g,
\end{equation}
or, element-wise as
\begin{equation}\label{eq:scatteringEq_discComp}
	h_{ij} = p_{ij}^{kl} g_{kl},
\end{equation}
where Einstein summation is implied.

Let us now denote \textit{element-wise multiplication} (i.e., the Hadamard product) of two square matrices $\mathbf{A}$ and $\mathbf{B}$, fulfilling $\dim(\mathbf{A}) = \dim(\mathbf{B})$, as $\mathbf{A} \odot \mathbf{B}$.
The resulting matrix has elements
\begin{equation}
	(\mathbf{A} \odot \mathbf{B})_{ij} := A_{ij}B_{ij}.
\end{equation}
Analogously, \textit{element-wise division} (i.e., Hadamard division) is denoted $\mathbf{A} \oslash \mathbf{B}$, with matrix elements
\begin{equation}
	(\mathbf{A} \oslash \mathbf{B})_{ij} := \frac{A_{ij}}{B_{ij}},
\end{equation}
where $B_{ij} \neq 0$.
The Richardson-Lucy method in the context of Fredholm integral equations may then be written as
\begin{equation}\label{eq:RL_unfolding}
	\g_\uf^{(n+1)} = \g_\uf^{(n)} \odot \Bigg[\p \bigg( \h \oslash \Big(\p \g_\uf^{(n)}\Big) \bigg) \Bigg],
\end{equation}
where $n \in \mathbb{N}$ is the iteration variable, and $\g_\uf$ is the discretized matrix representation of the approximation of $g$.
As a starting point for the iteration, we take $\g_\uf^{(0)} = \h$, and we do not use a formal stopping criterion, such as convergence of the solution --- instead, we stop after a fixed number of iterations, deemed large for the method to have converged.
The tensor multiplications involving $\p$ are carried out like in Eq.~\eqref{eq:scatteringEq_discComp}.

It is worth noting that unfolding a Fredholm integral, much like deconvolution, is an ill-posed problem \cite[Ch.~12.12]{polianinHandbookIntegralEquations2008}.
In certain situations, one can show that deconvolution methods converge \cite[Ch.~5]{janssonDeconvolutionImagesSpectra2012}, but they are too restrictive to be of use to us and certainly do not apply to unfolding Fredholm integral equations using iterative deconvolution methods. 
We shall return to how well this approach works when discussing the numerical examples in Sec.~\ref{sec:examples}.

\section{Freeform Specular Reflector Design}\label{sec:specularReflectorDesign}
We shall now discuss how we computed the reflector surfaces that yield a desired target light distribution when light scattering is accounted for.
First, Eq.~\eqref{eq:scatteringEq_rotSym} was discretized, then unfolded using the Richardson-Lucy method in Eq.~\eqref{eq:RL_unfolding}, thus yielding a virtual specular target distribution $\g_\uf$.
Next, we computed a reflector that fulfills the resulting \textit{specular} problem.
To this end, we used the numerical Monge-Amp{\`e}re solver first introduced in our group by Prins \cite{prinsInverseMethodsIllumination2014} and later expanded by Yadav \cite{yadavMongeAmpereProblemsNonquadratic2018} and Romijn \cite{romijnGeneratedJacobianEquations2021}.
The details are outside the scope of this manuscript, but a summary is given below.

\subsection{Stereographic Coordinates}
Because some equations become simpler to work with in stereographic coordinates, these are used in the code for computing the reflector surfaces.
In particular, we must transform our specular target intensity $\g_\uf$ [$\mathrm{W} \cdot \mathrm{sr}^{-1}$] into one defined in stereographic coordinates.
Let us consider the continuous case with $g_\uf(\psi,\chi)$.
This section will omit the subscript ${}_\mathrm{uf}$ on $g$ to simplify the notation.

Let $f(\x) > 0$, $\x \in \mathcal{S}$ be the source exitance distribution, and let $g(\psi,\chi)$ be the specular target intensity (we shall return to the domain momentarily).
Recall the geometry from Fig.~\ref{fig:geometry}, i.e., parallel rays leave the domain $\mathcal{S}$ in the $xy$-plane parallel to the positive $z$-axis and strike a reflector parametrized using some height function $z = u(\x) > 0$, $\x \in \mathcal{S}$.
Solving the inverse problem thus reduces to finding $u(\x)$ such that $f$ is transformed into $g$.

Recall that $\ut(\psi,\chi)$ is the direction of the specular ray, and that $\y$ is the 2-tuple stereographic representation of $\ut$ defined in Eq.~\eqref{eq:sterNP},
where we have chosen stereographic projection from the north pole since the reflector surface is positioned above the parallel source, meaning the reflected rays typically travel `downwards,' i.e., in negative $z$-direction.
Note that the stereographic projection from the north pole is undefined at the north pole itself, i.e., we consider $t_3 \neq 1$ and $\psi \neq 0$.
The corresponding inverse stereographic projection is
\begin{equation}\label{eq:sterNPInverse}
	\ut(\y) = \frac{1}{1 + \norm{y}^2}
	\begin{pmatrix}
		2 y_1\\
		2 y_2\\
		\norm{y}^2 - 1
	\end{pmatrix}.
\end{equation}

\subsection{Energy Conservation}
Let $\tilde{g}$ be the stereographic representation of $g(\psi,\chi)$ such that $\tilde{g}(\y) = g\big(\psi(\y),\chi(\y)\big)$.
Let $\mathcal{A} \subseteq \mathcal{S}$ be a (sub)set of the source domain.
Local energy conservation in the far-field approximation then states
\begin{equation}
	\int_\A f(\x) \, \dd\x = \int_{\ut(\A)} g(\psi,\chi) \, \dd\mathbf{S}(\psi,\chi),
\end{equation}
where $\dd\mathbf{S}(\psi,\chi) = \sin(\psi) \, \dd\psi\dd\chi$ and $\ut(\A) \subset \SS^2$ is the so-called \textit{image set} of $\A$ on the unit sphere.
For local energy conservation, $\mathcal{A} \subset \mathcal{S}$, whilst for global energy conservation, $\mathcal{A} \equiv \mathcal{S}$, so that $\ut(\A) \equiv \ut(\S) = \T$.
Transforming the integration over part of the unit sphere into an integration over the corresponding stereographic domain and recalling the definition of $\tilde{g}(\y)$ gives
\begin{equation}\label{eq:localEnergyConservation}
	\int_\A f(\x) \, \dd\x = \int_{\y\big(\ut(\A)\big)} \tilde{g}(\y) \, \abs{\pdv{\ut}{y_1} \cross \pdv{\ut}{y_2}} \, \dd\y,
\end{equation}
where $\y\big(\ut(\A)\big)$ constitutes the stereographic projection of the image set $\ut(\A)$.
The Jacobian may readily be evaluated using Eq.~\eqref{eq:sterNPInverse}:
\begin{equation}
	\abs{\pdv{\ut}{y_1} \cross \pdv{\ut}{y_2}} = \frac{4}{\big(1 + \norm{\y}^2\big)^2}.
\end{equation}

Let the optical map in stereographic coordinates be $\y = \widetilde{\m}(\x)$ (recall that $(\psi,\chi) = \m(\x)$ is the optical map in spherical coordinates).
Then, Eq.~\eqref{eq:localEnergyConservation} becomes (after substitution and transformation to integration over $\x$):
\begin{equation}
	\int_\A f(\x) \, \dd\x = \int_\A \tilde{g}\big( \widetilde{\m}(\x) \big) \frac{4}{\big(1 + \norm{\widetilde{\m}(\x)}^2\big)^2} \det\!\big(\D\widetilde{\m}(\x)\big) \, \dd\x,
\end{equation}
where the omission of absolute values around the determinant means that we restrict ourselves to a positive Jacobian $\det\!\big(\D\widetilde{\m}(\x)\big)$, and where $\D\widetilde{\m}(\x)$ signifies the Jacobian matrix with respect to $\x$.
Since the above relation holds for every $\A \subseteq \S$, it follows that, pointwise,
\begin{equation}
	\det\!\big(\D\widetilde{\m}(\x)\big) = \frac{1}{4} \big(1 + \norm{\widetilde{\m}(\x)}^2\big)^2 \frac{f(\x)}{\tilde{g}\big( \widetilde{\m}(\x) \big)}.
\end{equation}
Finally, the mapping $\widetilde{\m} = \grad u$ for the case of parallel incoming light and a far-field target \cite[Sec.~3.2]{romijnGeneratedJacobianEquations2021}.
Whence, we recover the so-called \textit{standard Monge-Amp{\`e}re equation}
\begin{equation}\label{eq:MongeAmpere}
	\det(\D^2 u(\x)) = \frac{1}{4} \big(1 + \norm{\grad u(\x)}^2\big)^2 \frac{f(\x)}{\tilde{g}\big( \grad u(\x) \big)},
\end{equation}
where $\D^2 u(\x)$ denotes the Hessian matrix.
To find the reflector, one must solve this nonlinear PDE for the height function $u$.

\subsection{Numerical Solution to \texorpdfstring{Monge-Amp{\`e}re}{Monge–Ampère} Equation}
Solving the standard Monge-Amp{\`e}re equation in Eq.~\eqref{eq:MongeAmpere} is nontrivial, and thus a numerical least-squares approach was chosen.
We need to venture further outside the scope of this manuscript to describe the method in detail.
Thus, we point the reader to the works of Prins, Yadav, and Romijn \cite{prinsInverseMethodsIllumination2014,yadavMongeAmpereProblemsNonquadratic2018,romijnGeneratedJacobianEquations2021}.
Note, however, that we shall always compute the strictly convex solution, such that $\m$ and $\widetilde{\m}$ are injective mappings.

\section{Verification}\label{sec:verification}
This section shows how we numerically verified our model in Sec.~\ref{sec:examples}.
In particular, once the reflectors have been computed in the manner described in Sec.~\ref{sec:specularReflectorDesign}, they were raytraced, and the resulting distributions were then compared to the ones predicted by our model.

\subsection{Raytracing}
To this end, we wrote a custom raytracer that directly implements the model of scattering presented in Sec.~\ref{sec:scatteringModel}.
This approach was chosen instead of using pre-existing raytracing software such as LightTools to completely control the scattering behavior so that the model could be reliably verified.

\subsubsection{Implementation}
The raytracer was implemented in Matlab, and it works as follows.
First, the normals of the reflector are computed for each sampling point on the rectangular grid using Matlab's \texttt{surfnorm} routine.
Next, a ray in direction $\us \equiv (0,0,1)^{\intercal}$ is sampled from the source distribution using Matlab's \texttt{rand} command (for simplicity, we always use a constant source).
The normal $\un$ at the point of intersection is then found using Matlab's \texttt{interp2} routine with piecewise linear interpolation, i.e., each component of the normal vector is assumed to change linearly between the closest known normals on the initial rectangular grid.
The reflected direction $\ut(\psi,\chi)$ associated with this source ray is computed using the vectorial law of reflection, Eq.~\eqref{eq:LoR}.
Next, the scattered ray is computed by applying Eq.~\eqref{eq:u_rot} using $\psi$ and $\chi$ from $\ut$ and $\alpha$ and $\beta$ by sampling from the appropriate PDF $p$ --- see the next paragraph for details on this sampling.

We discretized the domains $\S$, $\T$, and $\U$ to collect the source, specular and scattered rays, forming so-called `bins.'
Next, we applied Matlab's \texttt{dsearchn} nearest-point search routine and incremented the number of rays in the returned bins.

Once the desired number of rays has been traced, the ray count per bin is converted into an exitance or intensity, such that it may be compared to $f$, $g$, or $h$.
For instance, suppose we have an $N_1 \times N_2$ grid of bins.
Then, the exitance of the source is estimated using ($i \in [1,N_1]$, $j \in[1,N_2]$)
\begin{equation}\label{eq:estimateF}
	\begin{split}
		E_{ij} = &\frac{\mathrm{Pr}(x_{i-1}\leq x < x_i\ \&\ y_{j-1}\leq y < y_j)}{\Delta x \, \Delta y}\\
		&\times \int_\S f(x,y) \, \dd x \dd y,
	\end{split}
\end{equation}
where $\mathrm{Pr}(x_{i-1}\leq x < x_i\ \&\ y_{j-1}\leq y < y_j)$ is the number of rays in the $ij$th bin divided by the total number of rays traced, i.e., the probability of falling in the $ij$th bin.
The integral in Eq.~\eqref{eq:estimateF} represents the total flux of the source, and $\Delta x \, \Delta y$ is the size of the bins.
Similarly, the specular intensity distribution is estimated using
\begin{equation}\label{eq:estimateG}
	\begin{split}
		I_{ij} = &\frac{\mathrm{Pr}(\psi_{i-1}\leq \psi < \psi_i\ \&\ \chi_{j-1}\leq \chi < \chi_j)}{\sin(\psi_i) \, \Delta \psi \, \Delta \chi}\\
		&\times \int_\S f(x,y) \, \dd x \dd y,
	\end{split}
\end{equation}
where the symbols have similar meanings to before.
The scattered intensity is also estimated using Eq.~\eqref{eq:estimateG}, with $\psi$ and $\chi$ replaced by $\gamma$ and $\nu$, respectively.
More details regarding this approach to raytracing can be found in \cite[p.~34]{filosaPhaseSpaceRay2018}.

\paragraph{Sampling of \texorpdfstring{$\alpha$}{α} and \texorpdfstring{$\beta$}{β}}
We shall now consider the problem of sampling $\alpha$ and $\beta$ in our raytracer.
We have already mentioned that $\beta$ will be sampled uniformly on $[0,2\pi]$, so we only need to consider how $\alpha$ is sampled.
For instance, suppose we would like a `rotationally symmetric Gaussian on the sphere.'
This is a vague definition that can be interpreted in many ways.
For example, we could pick $\alpha$ from a regular one-dimensional Gaussian via a process known as inverse transform sampling and then pick $\beta$ uniformly, or we could use a generalized distribution for picking normally distributed points on a sphere, like the Kent distribution used in geology and bioinformatics \cite{kentFisherBinghamDistributionSphere1982}.

We have instead chosen the following approach.
First, we pick two independent normally distributed variables $q_1 \sim \N(0,\sigma)$ and $q_2 \sim \N(0,\sigma)$, where $\N(\mu,\sigma)$ is the normal distribution with mean $\mu$ and standard deviation $\sigma$.
The point $\q := (q_1,q_2)^\intercal \in \mathbb{R}^2$ in the $xy$ plane is picked from a rotationally symmetric two-dimensional normal distribution.
Applying inverse stereographic projection from the south pole to $\q$ thus yields a point on the unit sphere, representing the direction of the cone vector $\uc$.
That is,
\begin{equation}
	\uc = \frac{1}{1+\norm{\q}^2}
	\begin{pmatrix}
		2q_1\\
		2q_2\\
		1-\norm{\q}^2\\
	\end{pmatrix}.
\end{equation}
This allows us to find closed expressions for $\alpha$ and $\beta$ in terms of $\q$:
\begin{equation}\label{eq:sampleRT}
	\begin{split}
		\alpha(q_1,q_2) &= \arccos\left( \frac{\norm{\q}-1}{\norm{\q}+1} \right),\\
		\beta(q_1,q_2) &= \arctan(q_1,q_2).
	\end{split}
\end{equation}
It can be shown --- see the Appendix, Sec.~\ref{sec:appendix} --- that the PDF, $p(\alpha; \sigma)$, associated with this approach of picking $\alpha$ and $\beta$ is given by
\begin{equation}\label{eq:pAlpha}
	p(\alpha; \sigma) = \frac{1}{8\pi \sigma^2} \, \sec^4\!\bigg(\frac{\alpha}{2}\bigg) \exp\!\Bigg(-\frac{1}{2\sigma^2} \tan^2\!\bigg(\frac{\alpha}{2}\bigg)\Bigg).
\end{equation}
Whence, the raytracer picks $\alpha$ and $\beta$ by sampling $q_1 \sim \N(0,\sigma)$ and $q_2 \sim \N(0,\sigma)$ followed by applying Eq.~\eqref{eq:sampleRT}.
The predicted scattered light distribution, meanwhile, is computed by inserting $p$ from Eq.~\eqref{eq:pAlpha} into Eq.~\eqref{eq:scatteringEq_rotSym} and computing the discretized version in Eq.~\eqref{eq:scatteringEq_discComp}.

\subsection{RMS Error}
The root mean square (RMS) error will quantify the error between the raytraced and exact distributions.
For a discretization grid of $N_1 \times N_2$, i.e., $N_1$ polar angles and $N_2$ azimuthal angles, the RMS error between $\h$ and the raytraced $\h^*$ is defined as
\begin{equation}
	\varepsilon(\h,\h^*) := \sqrt{ \frac{1}{N_1 N_2} \sum_{j=1}^{N_2} \sum_{i=1}^{N_1} \abs{h_{ij} - h^*_{ij}}^2 }.
\end{equation}
Note that an upper-index asterisk $({}^{*})$ denotes a raytraced distribution henceforth.

\section{Numerical Examples}\label{sec:examples}
This section discusses two numerical examples to showcase the design procedure outlined in this manuscript and the effect varying amounts of surface scattering have on the shape of the computed freeform reflectors.
Let
\begin{equation}\label{eq:2DGaussian}
	\begin{split}
		&\N(x,y;\mu_x,\mu_y,\sigma_x,\sigma_y) :=\\
		&\frac{1}{2\pi \sigma_x \sigma_y} \mathrm{exp}\Bigg(-\frac{1}{2}\bigg(\frac{x - \mu_x}{\sigma_x}\bigg)^2 - \frac{1}{2}\bigg(\frac{y - \mu_y}{\sigma_y}\bigg)^2\Bigg)
	\end{split}
\end{equation}
be the two-dimensional normal distribution (Gaussian) with means $\mu_x$ and $\mu_y$ and standard deviations $\sigma_x$ and $\sigma_y$.

\subsection{Example \#1: Overlapping Gaussians}
The first example we considered is outlined below, where $p(\alpha;\sigma)$ can be found in Eq.~\eqref{eq:pAlpha}:
\begin{equation*}
	\begin{split}
			&\text{Source domain:}\ \S = [-1,1] \times [-1,1]\\
			&\text{Target domain:}\ \text{see text}\\
			&\text{Source distribution:}\ f(x,y) = \frac{1}{4}\\
			&\text{Scattered target distribution:}\\
			&h(\gamma,\nu) = \frac{1}{1.8202} \Bigg[\N\bigg(\gamma,\nu;\frac{41\pi}{60},\frac{5\pi}{6},0.3,0.4\bigg)\\
			&+ \N\bigg(\gamma,\nu;\frac{2\pi}{3},\frac{7\pi}{6},0.25,0.45\bigg) + 0.3\, \N\bigg(\gamma,\nu;\frac{19\pi}{24},\pi,0.2,0.4\bigg)\Bigg]\\
			&\text{Surface scattering function:}\ p(\alpha;0.1)
	\end{split}
\end{equation*}

\begin{figure*}[htb!]
	\centering
	\includegraphics[width=0.33\linewidth]{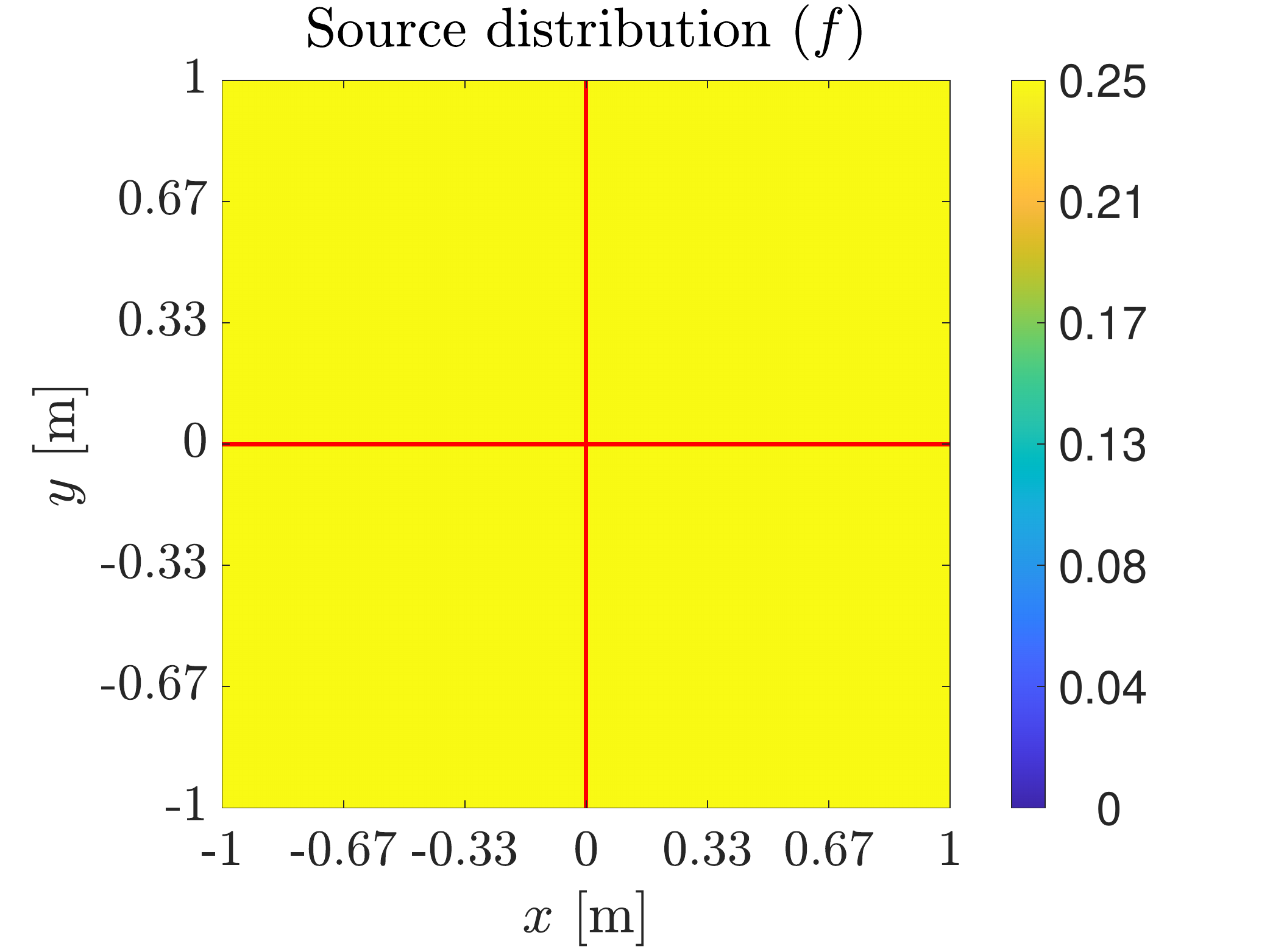}%
	\includegraphics[width=0.33\linewidth]{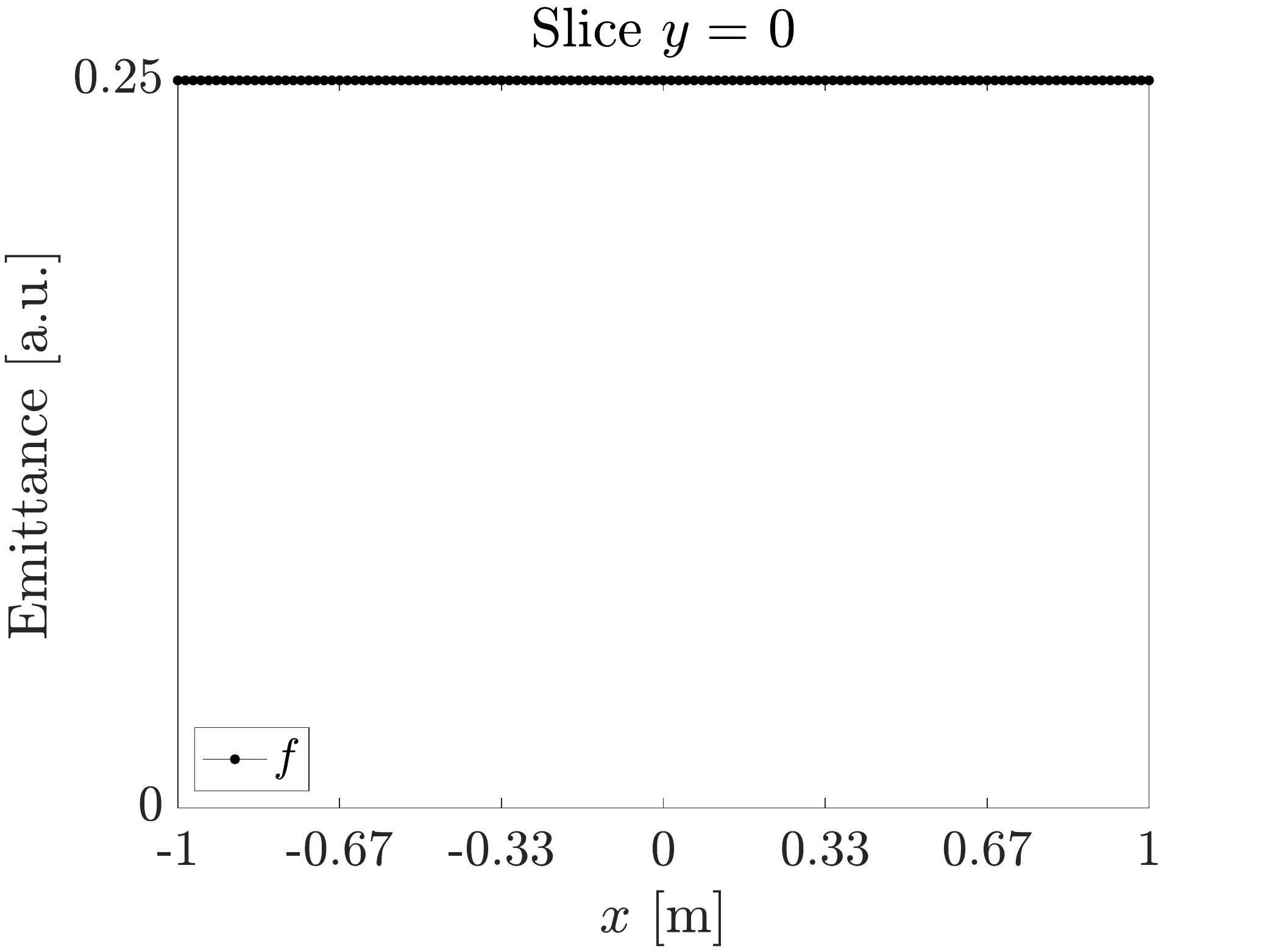}%
	\includegraphics[width=0.33\linewidth]{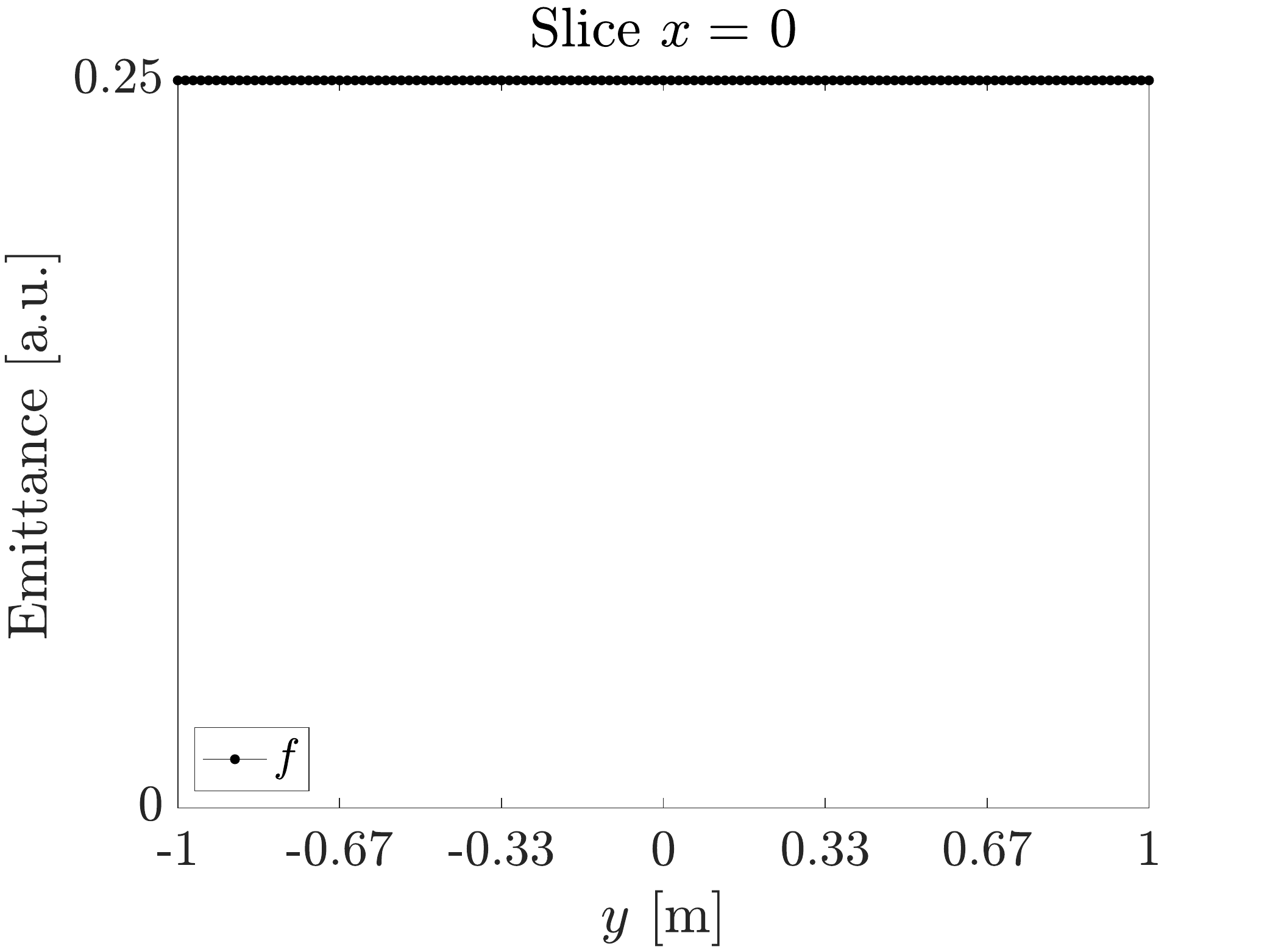}\\[10pt]
	\includegraphics[width=0.33\linewidth]{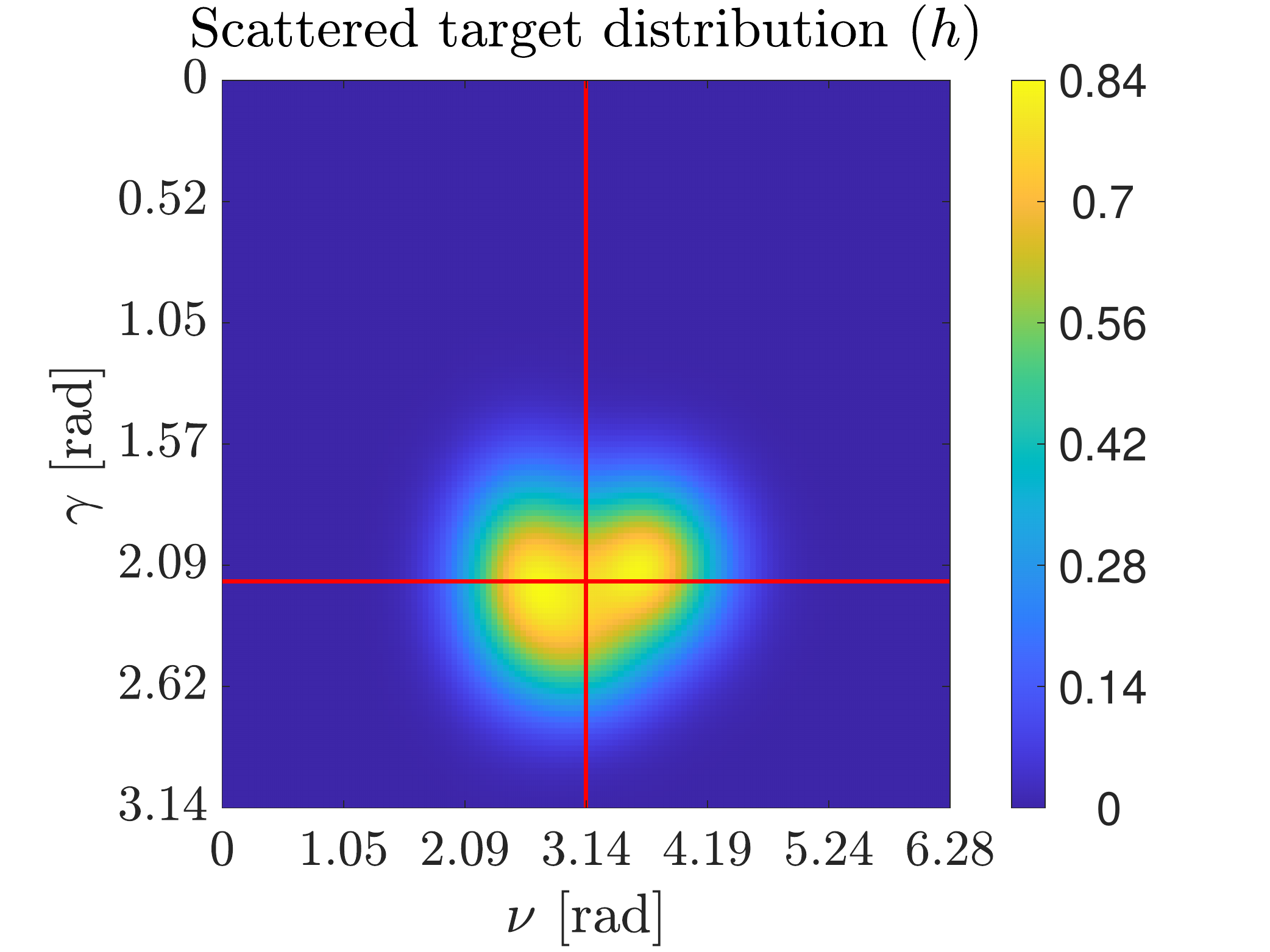}%
	\includegraphics[width=0.33\linewidth]{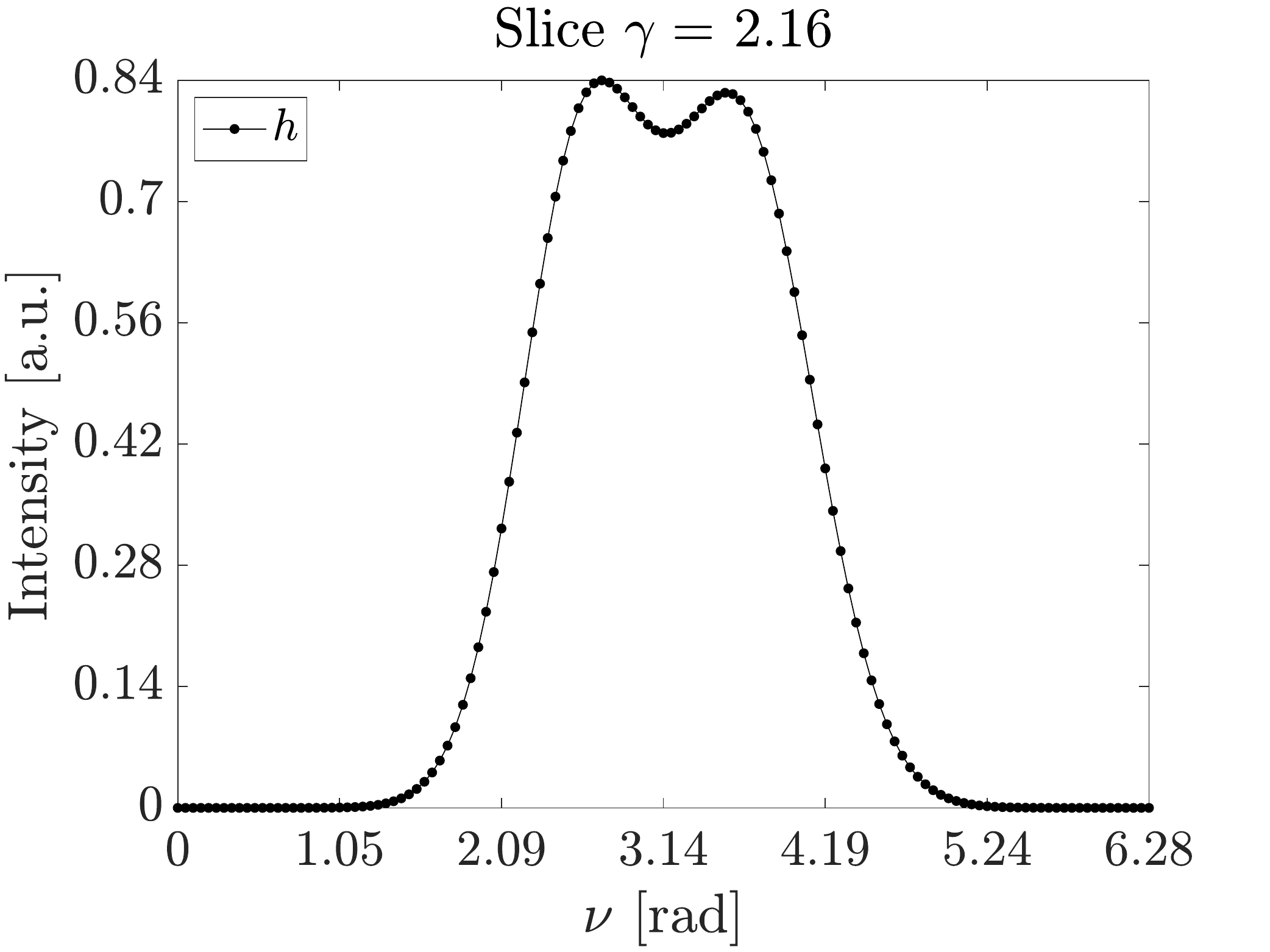}%
	\includegraphics[width=0.33\linewidth]{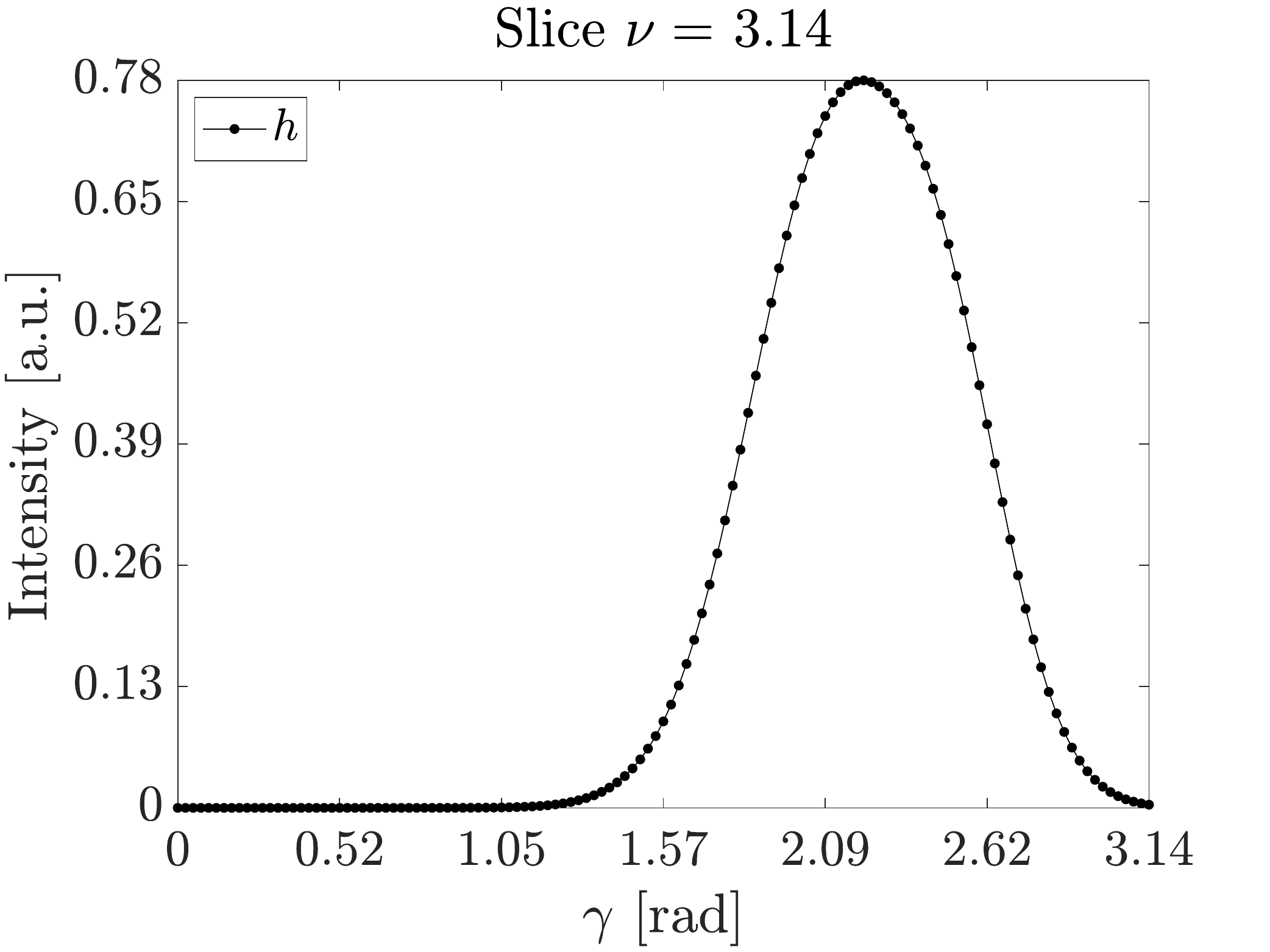}\\[10pt]
	\includegraphics[width=0.33\linewidth]{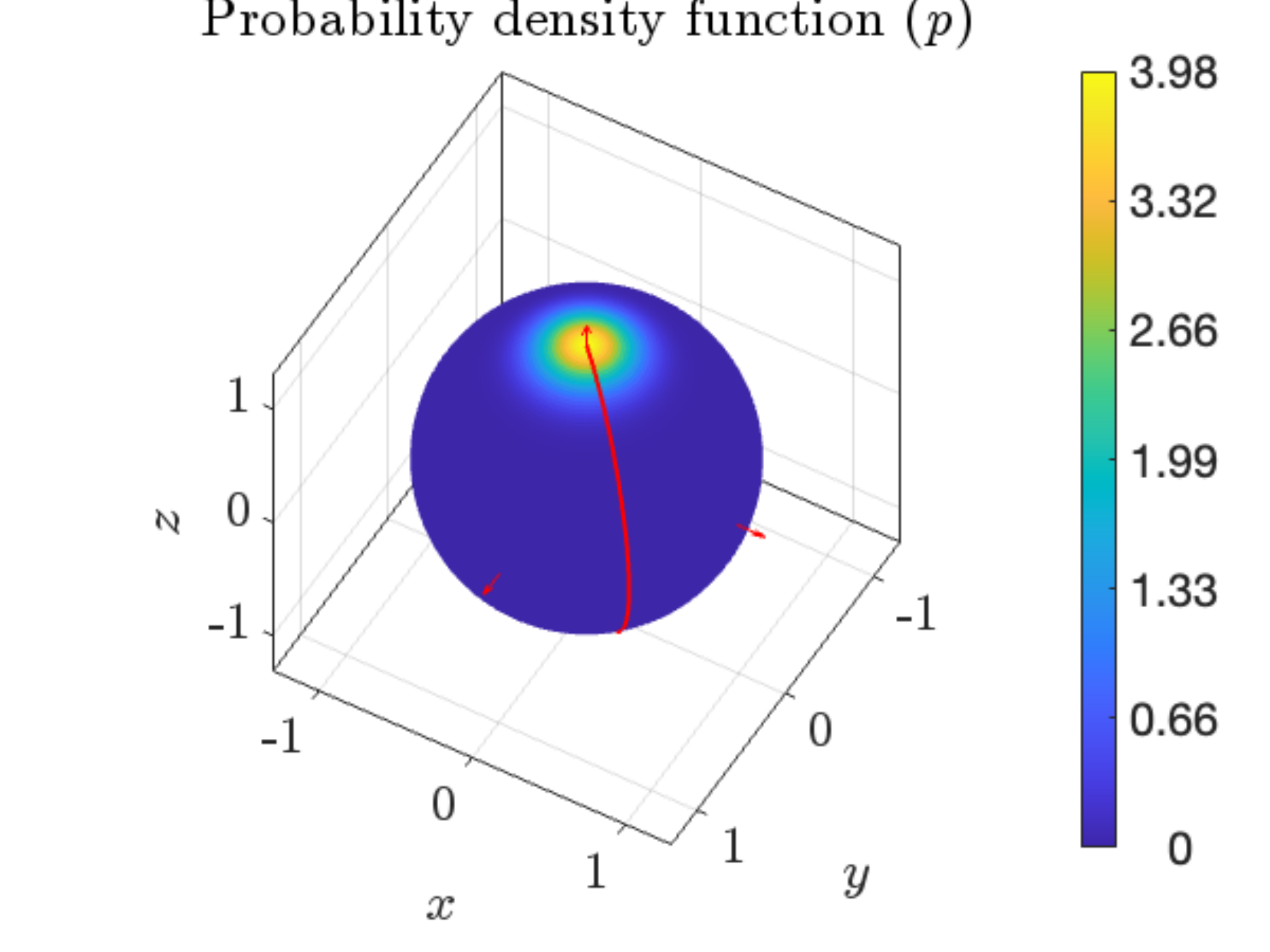}%
	\includegraphics[width=0.33\linewidth]{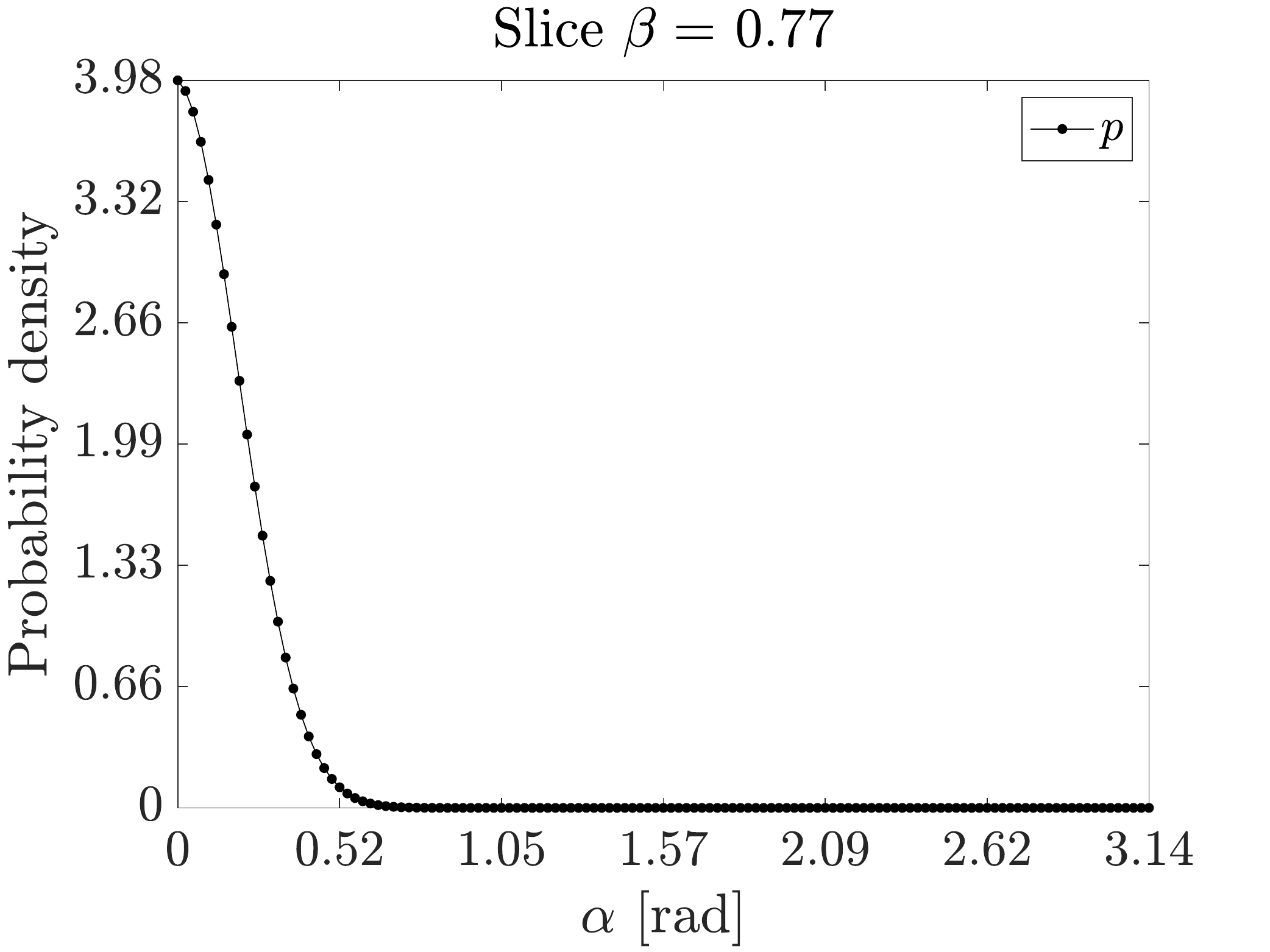}
	\caption{Prescribed distributions in Example \#1; $128^2$ sample points.}
	\label{fig:example_1-1}
\end{figure*}

\noindent The prescribed distributions are shown in Fig.~\ref{fig:example_1-1}, where we opted to plot $p$ on the unit sphere to facilitate comparisons to Fig.~\ref{fig:coneVector}.
Note that the most likely locations for the cone vector are close to the $z$-axis, i.e., relatively small-angle scattering, and we can be confident that $\alpha < \pi/2$.
In addition to the false-color plots, we have also `sliced' each distribution along the red lines with corresponding plots to the right of the accompanying false-color plots.
Note that we choose $f$ and $h$ such that (energy conservation)
\begin{equation}
	\int_{\S} f(x,y) \, \dd x \dd y = \int_{\U} h(\gamma,\nu) \sin(\gamma) \, \dd\gamma \dd\nu = 1,
\end{equation}
where $\U$ was found in the manner explained below.
This is the origin of the multiplicative factor $1/1.8202$ in $h$.

\begin{figure*}[htb!]
	\centering
	\includegraphics[width=0.33\linewidth]{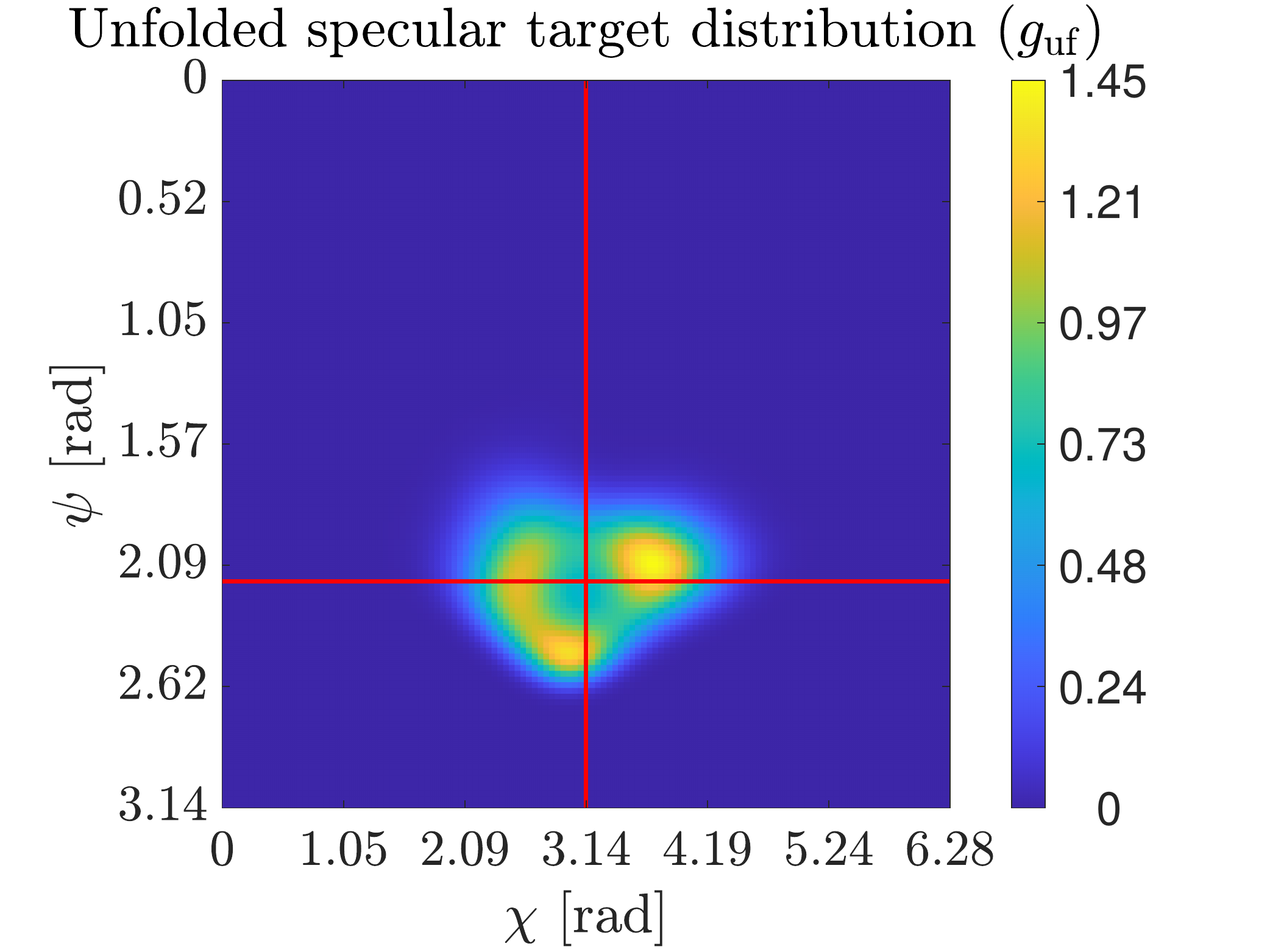}%
	\includegraphics[width=0.33\linewidth]{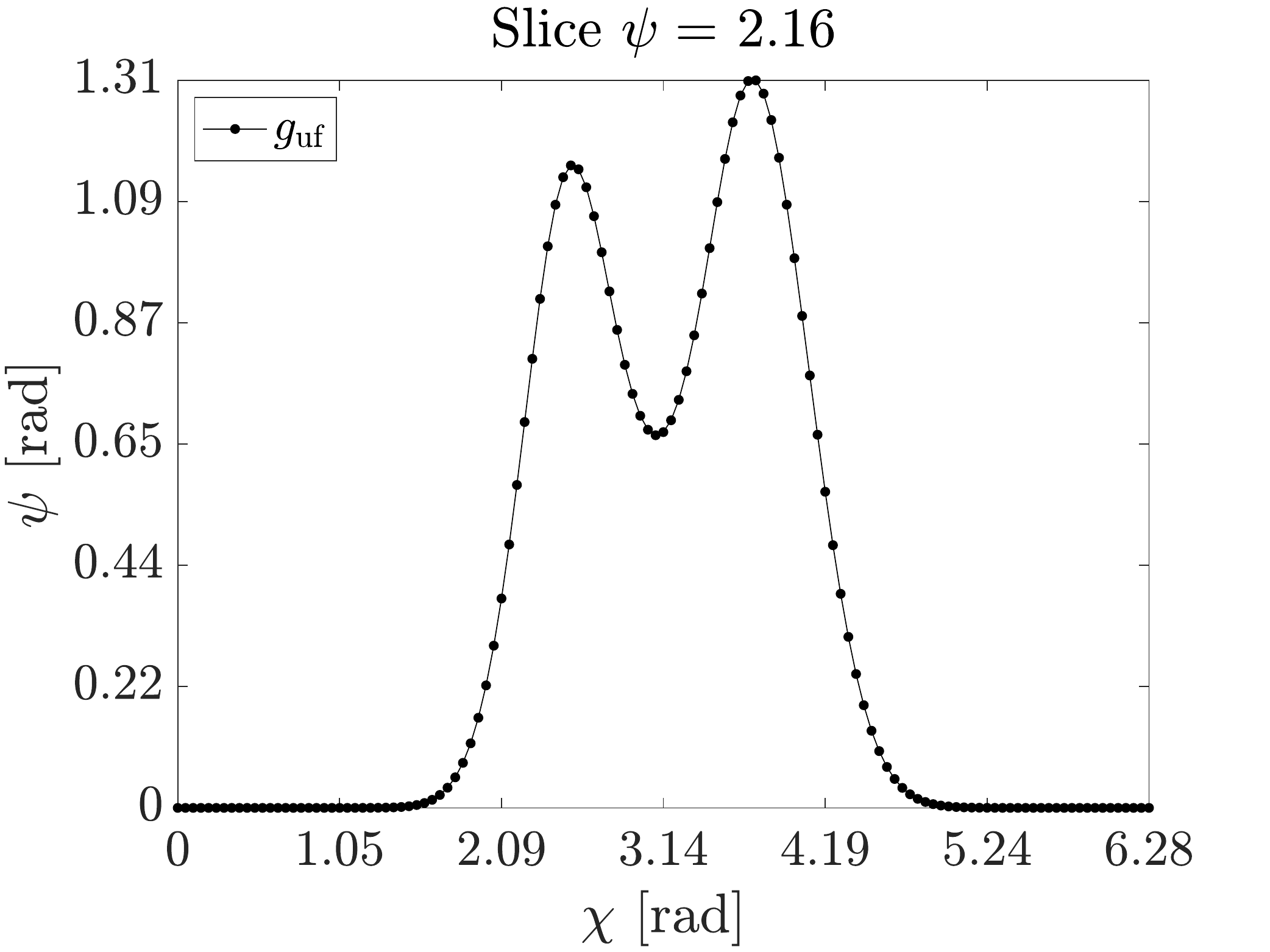}%
	\includegraphics[width=0.33\linewidth]{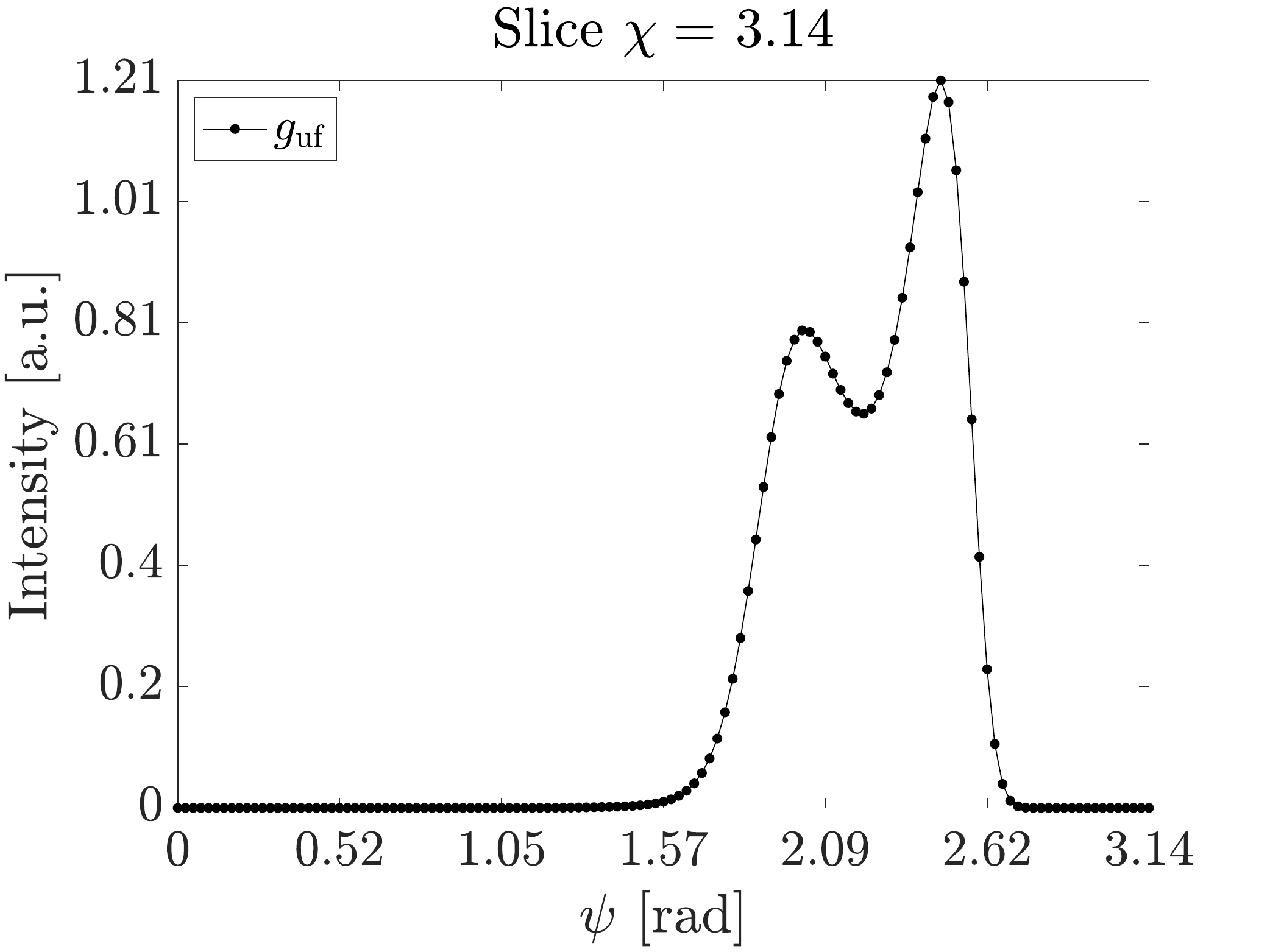}
	\caption{Unfolded specular target distribution $g_\uf$ in Example \#1; $128^2$ sample points, 1000 Richardson-Lucy iterations.}
	\label{fig:example_1-2a}
\end{figure*}

\begin{figure*}[htb!]
	\centering
	\includegraphics[width=0.33\linewidth]{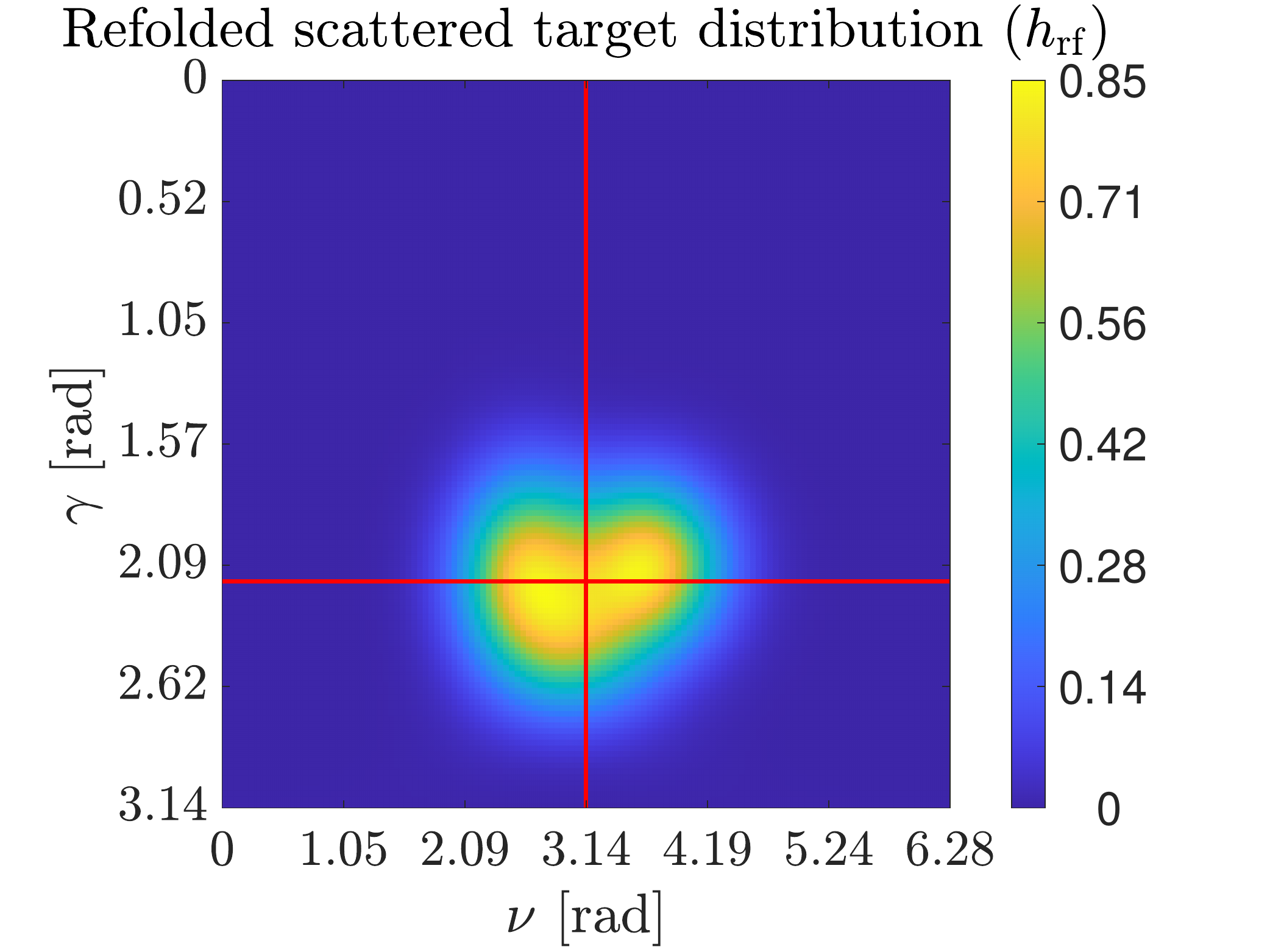}%
	\includegraphics[width=0.33\linewidth]{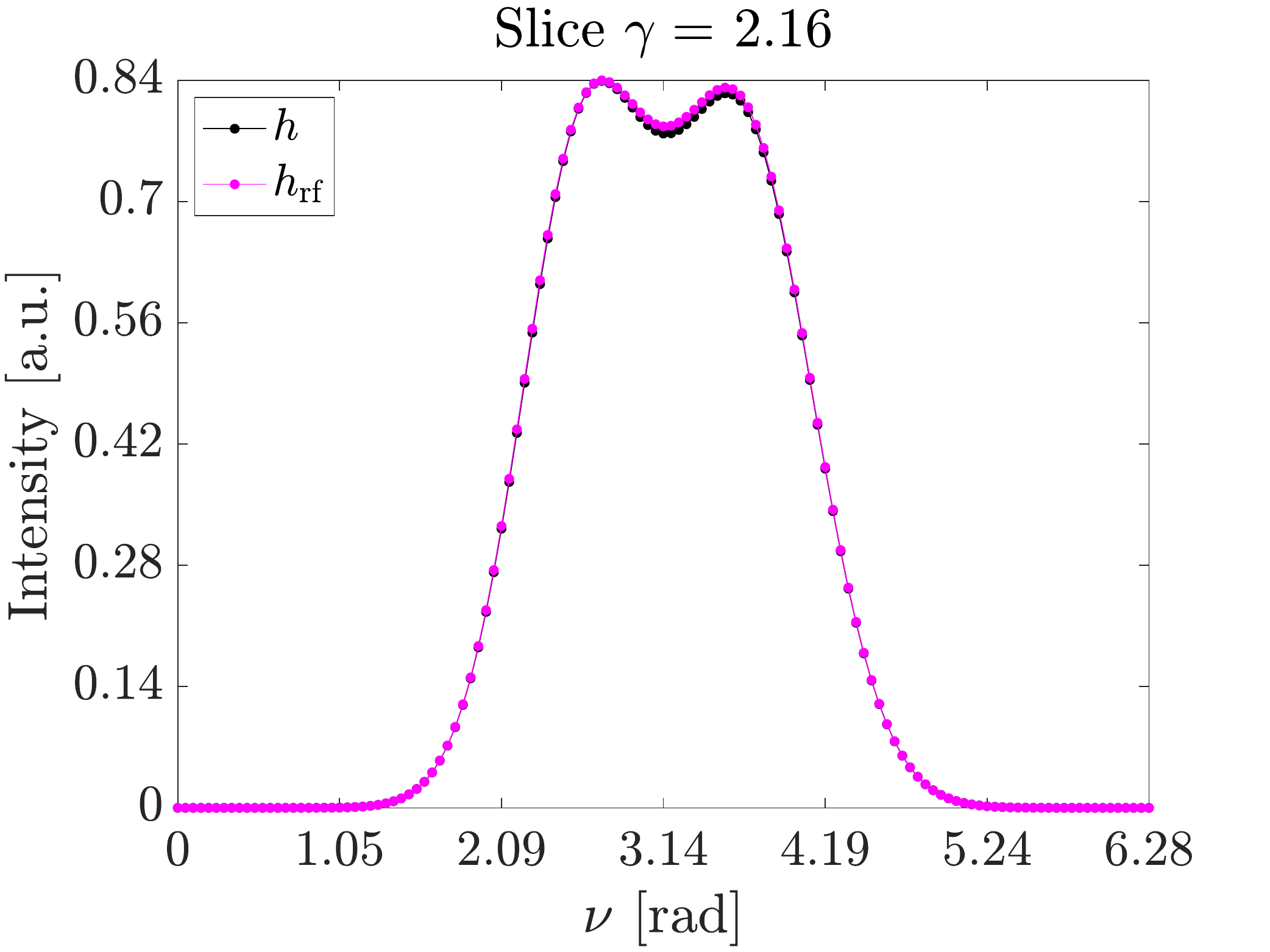}%
	\includegraphics[width=0.33\linewidth]{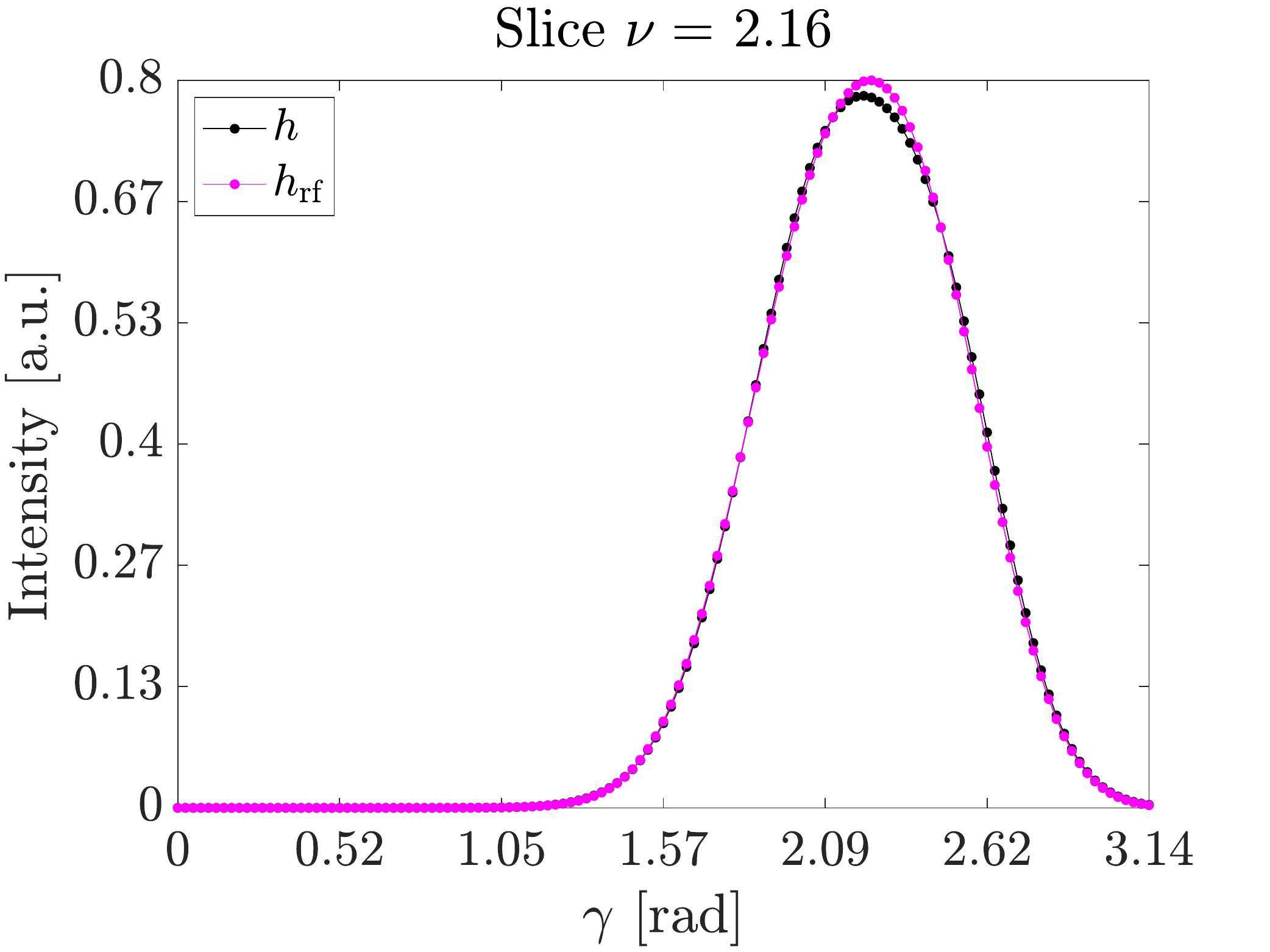}
	\caption{Refolded scattered target distribution $h_\rf$ in Example \#1; $128^2$ sample points.}
	\label{fig:example_1-2b}
\end{figure*}

Next, we computed the `unfolded' distribution $g_\uf$ using $1000$ Richardson-Lucy iterations, i.e., by applying Eq.~\eqref{eq:RL_unfolding} $1000$ times.
This yielded the distributions in Fig.~\ref{fig:example_1-2a}.
Note that we have nothing to directly compare $g_\uf$ to since the exact solution is unknown for this problem.
Thus, we computed the so-called `refolded' $\h_\rf := \p \g_\uf$, representing the scattered distribution that would occur from a reflector designed using $g_\uf$.
This is shown in Fig.~\ref{fig:example_1-2b}.
As we can see, $h$ and $h_\rf$ are very similar, meaning $g_\uf$ is a good representation of the `true' $g$ --- at least in the sense that the predicted scattered distribution from the reflector will be close to the prescribed target distribution.
This indicates that the Richardson-Lucy deconvolution method works well for the more general problem of unfolding our Fredholm integral.

\begin{figure*}[htb!]
	\centering
	\includegraphics[width=0.33\linewidth]{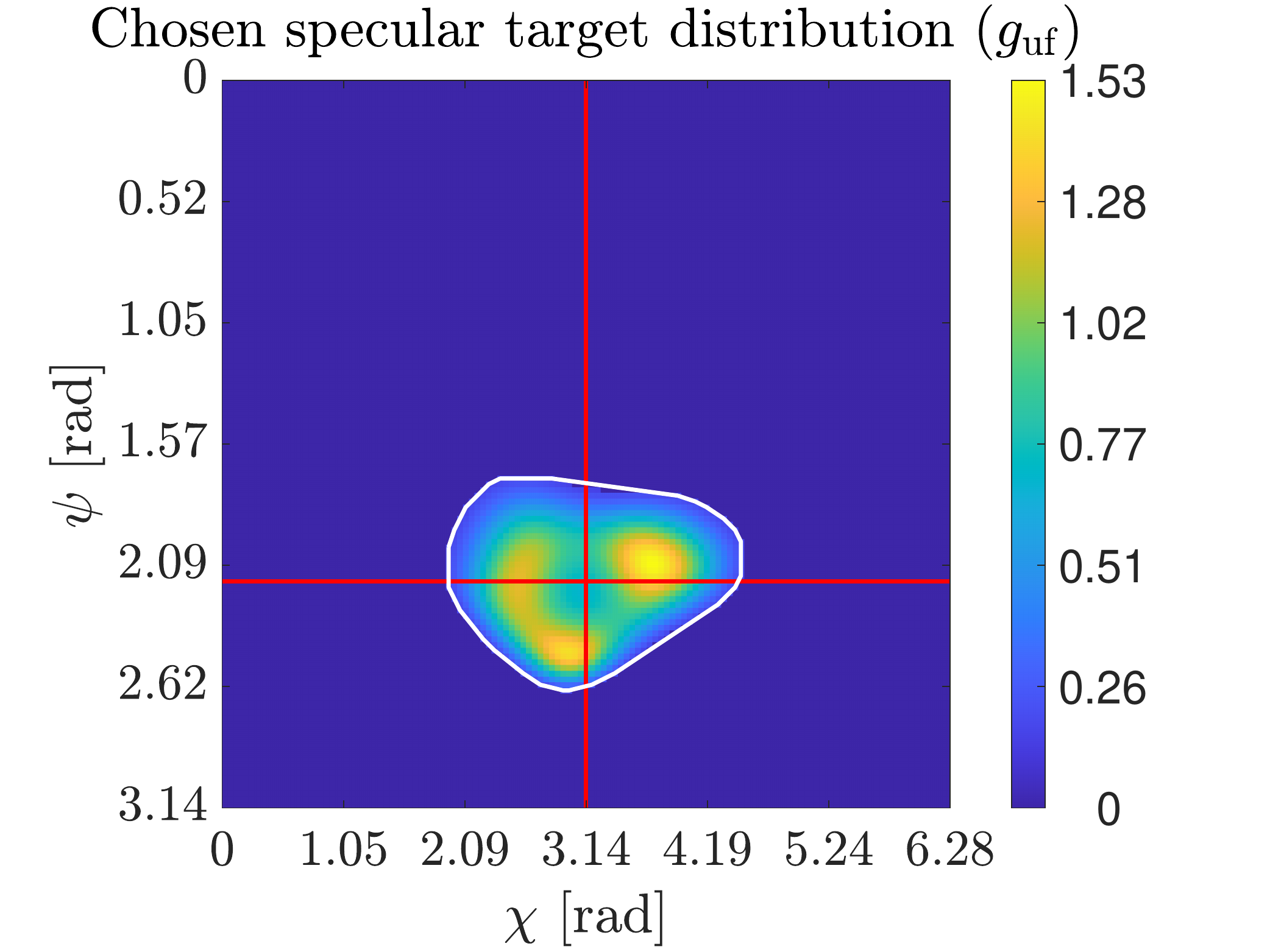}%
	\includegraphics[width=0.33\linewidth]{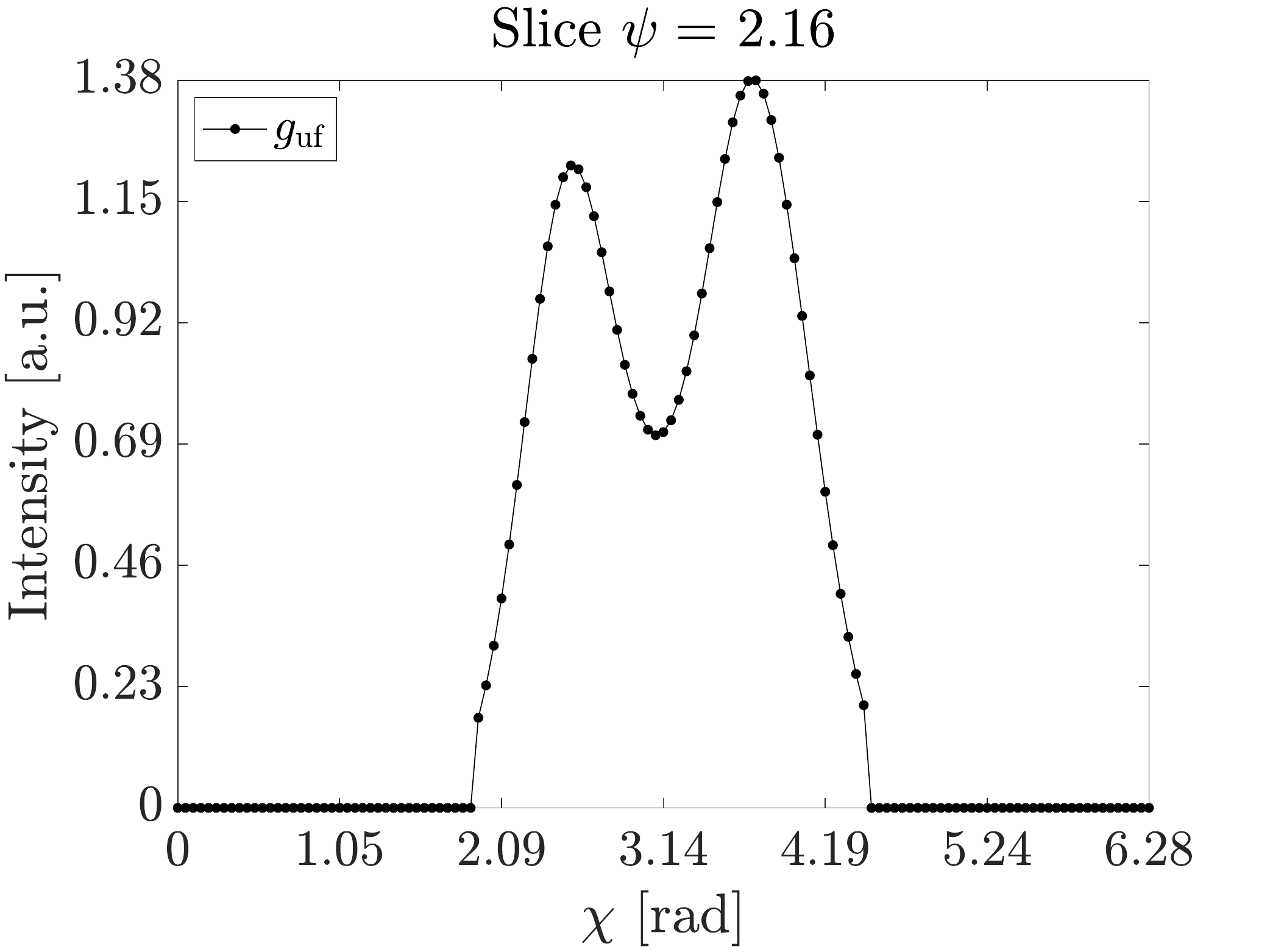}%
	\includegraphics[width=0.33\linewidth]{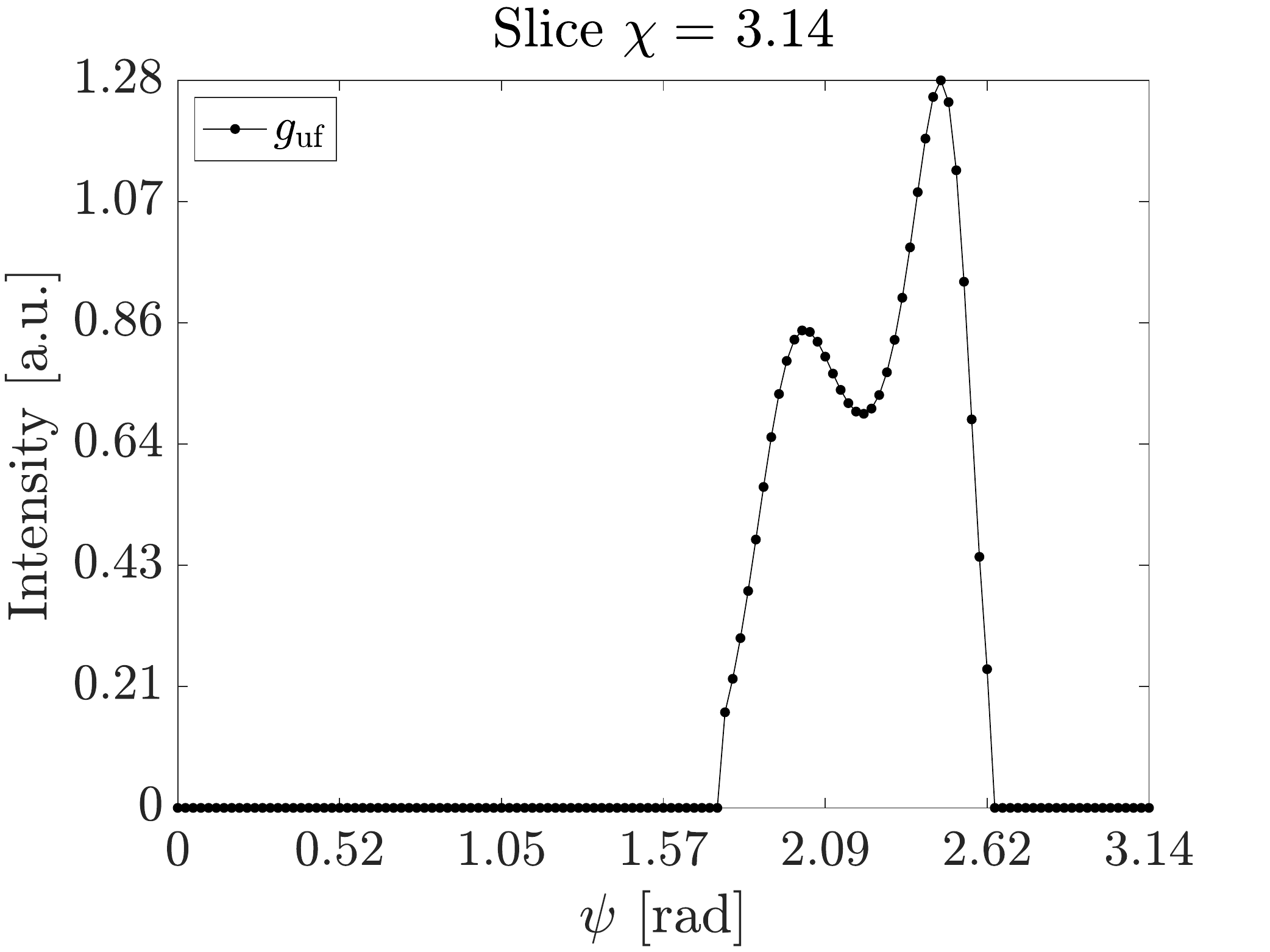}
	\caption{The specular target together with the support of $g_\uf$ in Example \#1; $128^2$ sample points.}
	\label{fig:example_1-3a}
\end{figure*}

Next, we computed the specular reflector that achieves $g_\uf$ in the far field, given $f$ as the source.
Because of the way the least-squares algorithm works, we must specify a boundary in the target domain and not have values too close to zero of $\tilde{g}_\uf$ within this boundary --- recall Eq.~\eqref{eq:MongeAmpere}.
Since our initial scattered target distribution consisted of overlapping Gaussians, the support of $g_\uf$ is not finite.
We thus chose some $\epsilon$ as a cutoff and found the boundary outlining the specular target domain by 
\begin{equation}
	\T = \big\{(\psi,\chi)\ |\ g_\uf(\psi,\chi) \geq \epsilon\big\}.
\end{equation}
The result of this process is shown in Fig.~\ref{fig:example_1-3a}, where $\epsilon = 0.1 \max(g_\uf)$, and we renormalized $g_\uf$ after introducing the boundary to maintain energy conservation.
The white outline shows the boundary of the target domain $\T$, i.e., the support of $g_\uf$.

\begin{figure*}[htb!]
	\centering
	\includegraphics[width=0.33\linewidth]{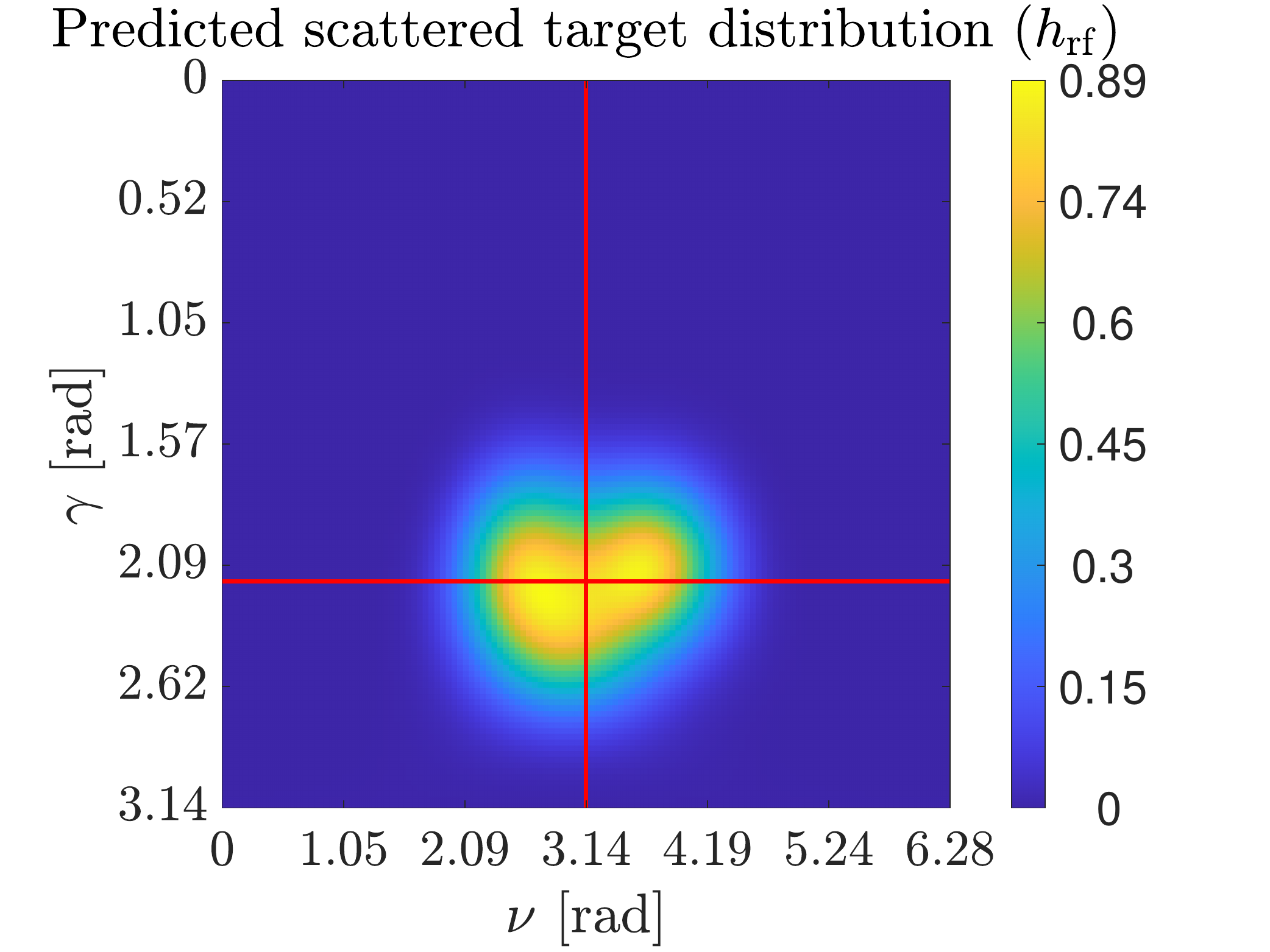}%
	\includegraphics[width=0.33\linewidth]{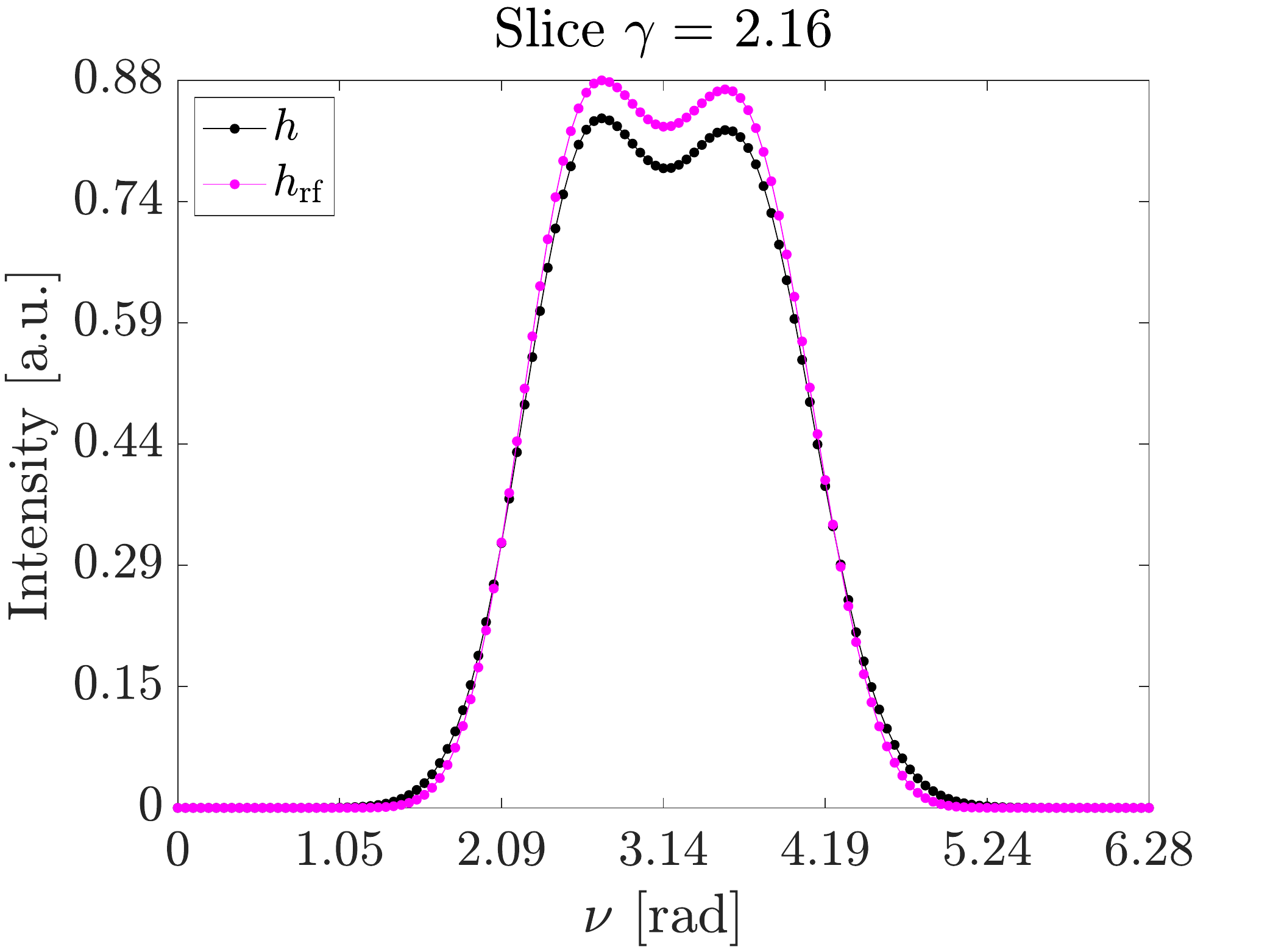}%
	\includegraphics[width=0.33\linewidth]{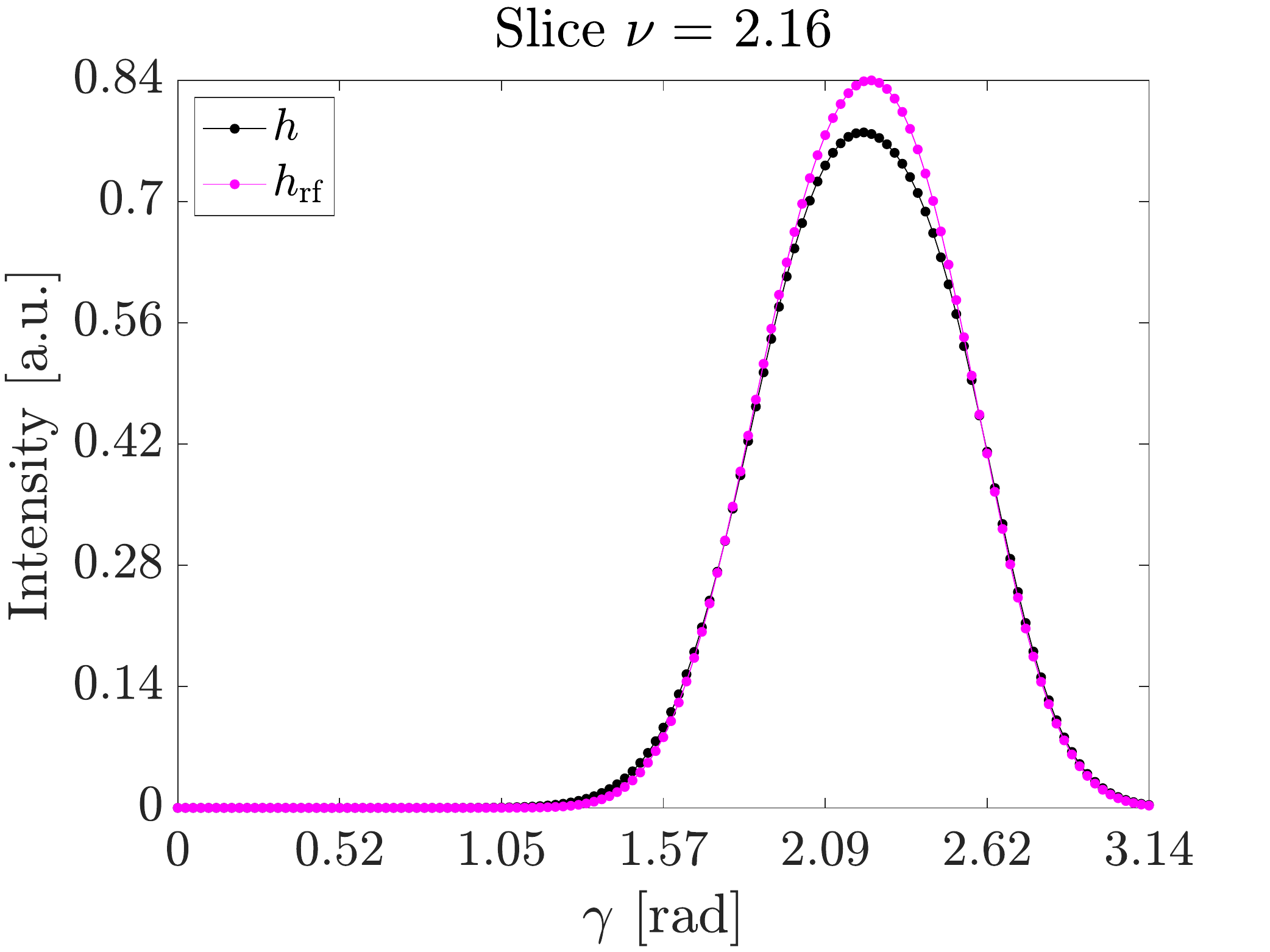}
	\caption{The final predicted scattered distribution in Example \#1; $128^2$ sample points.}
	\label{fig:example_1-3b}
\end{figure*}

Since we altered $g_\uf$ by giving it a finite domain and renormalizing, we must naturally update our predicted scattered light distribution accordingly, i.e., apply the Fredholm integral equation again, with the new $g_\uf$ as the specular target distribution.
The result is shown in Fig.~\ref{fig:example_1-3b}, where we can see that the final predicted $h_\rf$ is very similar to the original $h$, but with slightly increased values close to the center and somewhat decreased values further out due to the cut tails of the specular target $g_\uf$.

\begin{figure*}[htb!]
	\centering
	\includegraphics[width=0.33\linewidth]{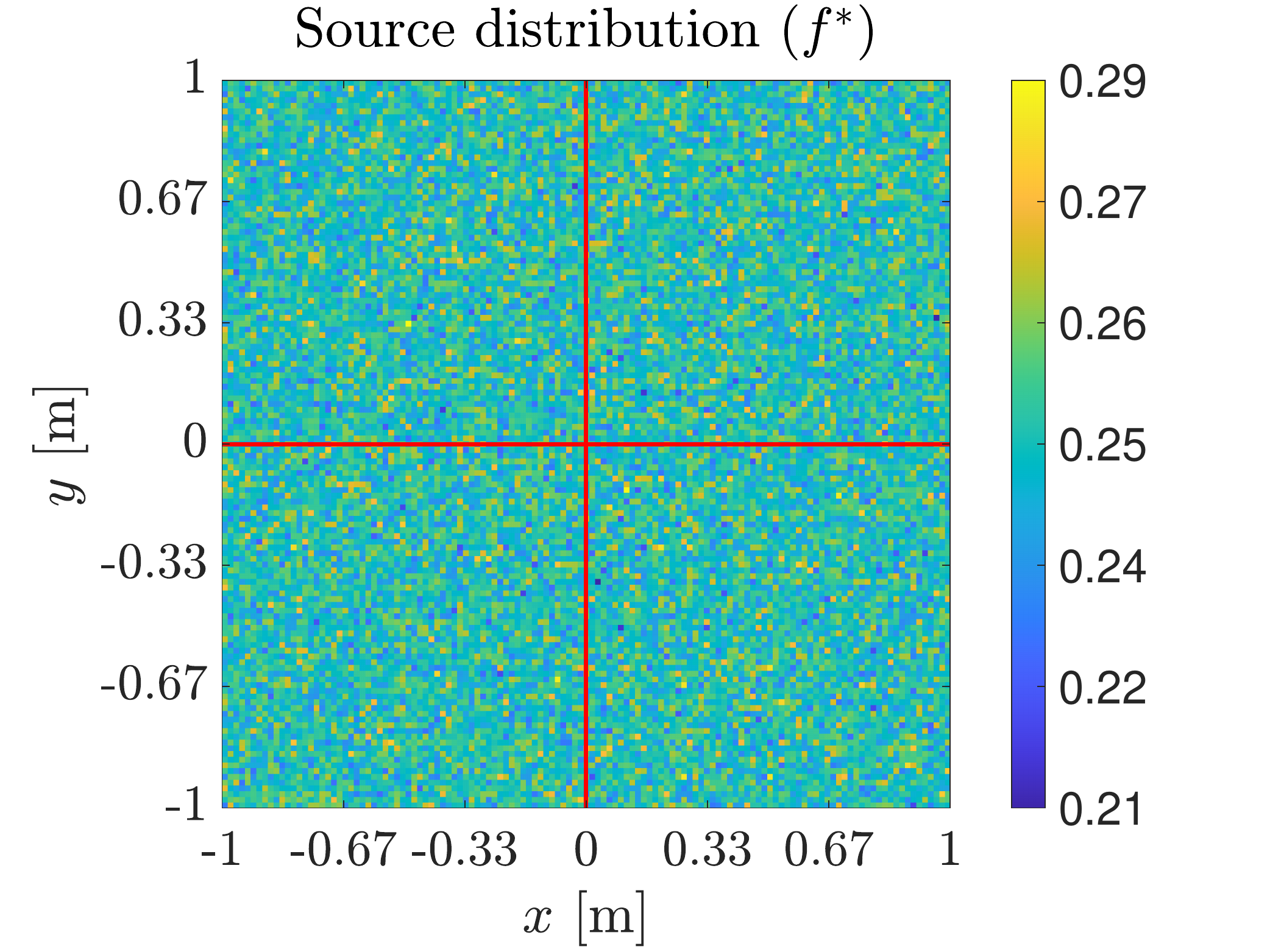}%
	\includegraphics[width=0.33\linewidth]{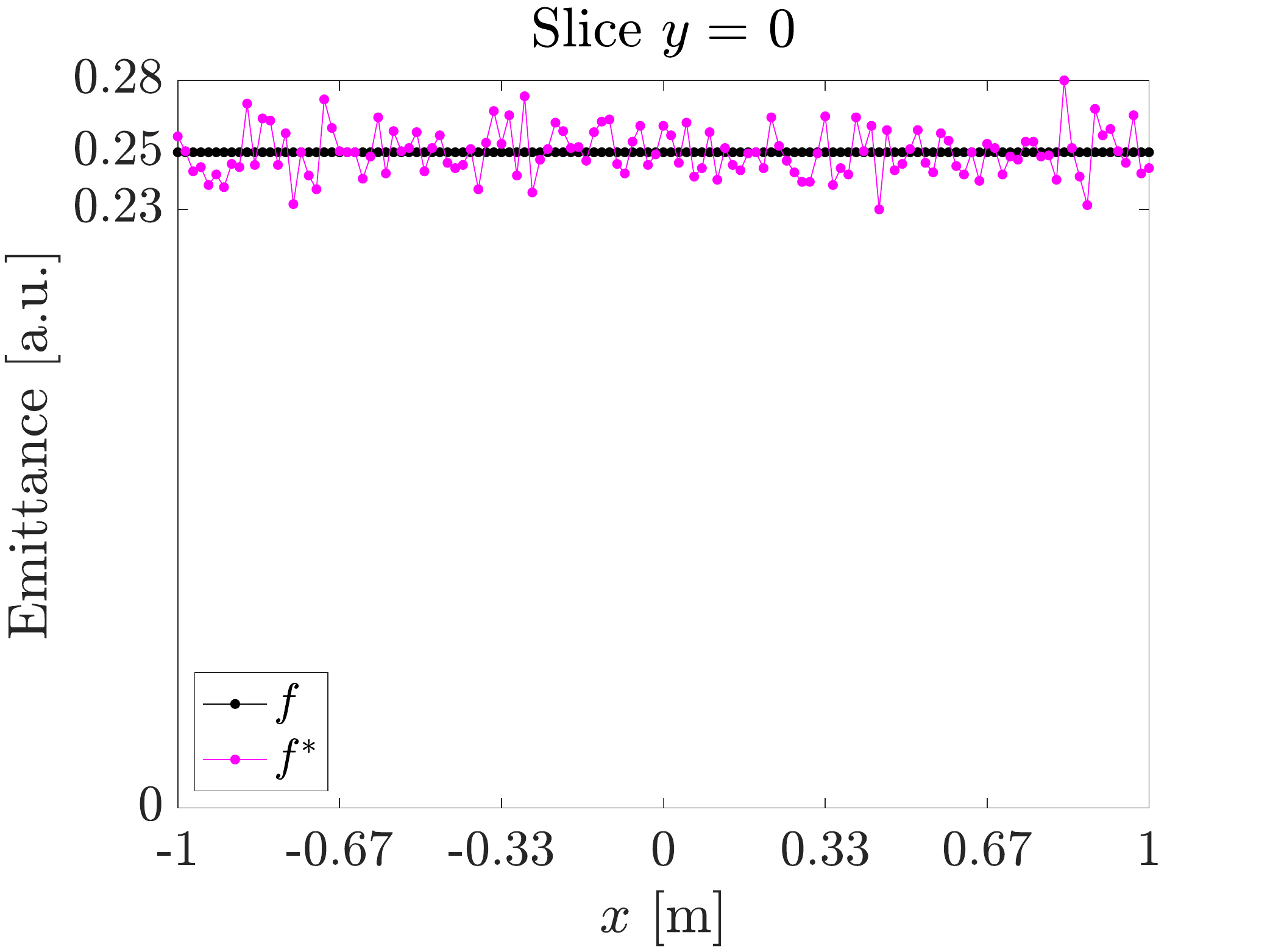}%
	\includegraphics[width=0.33\linewidth]{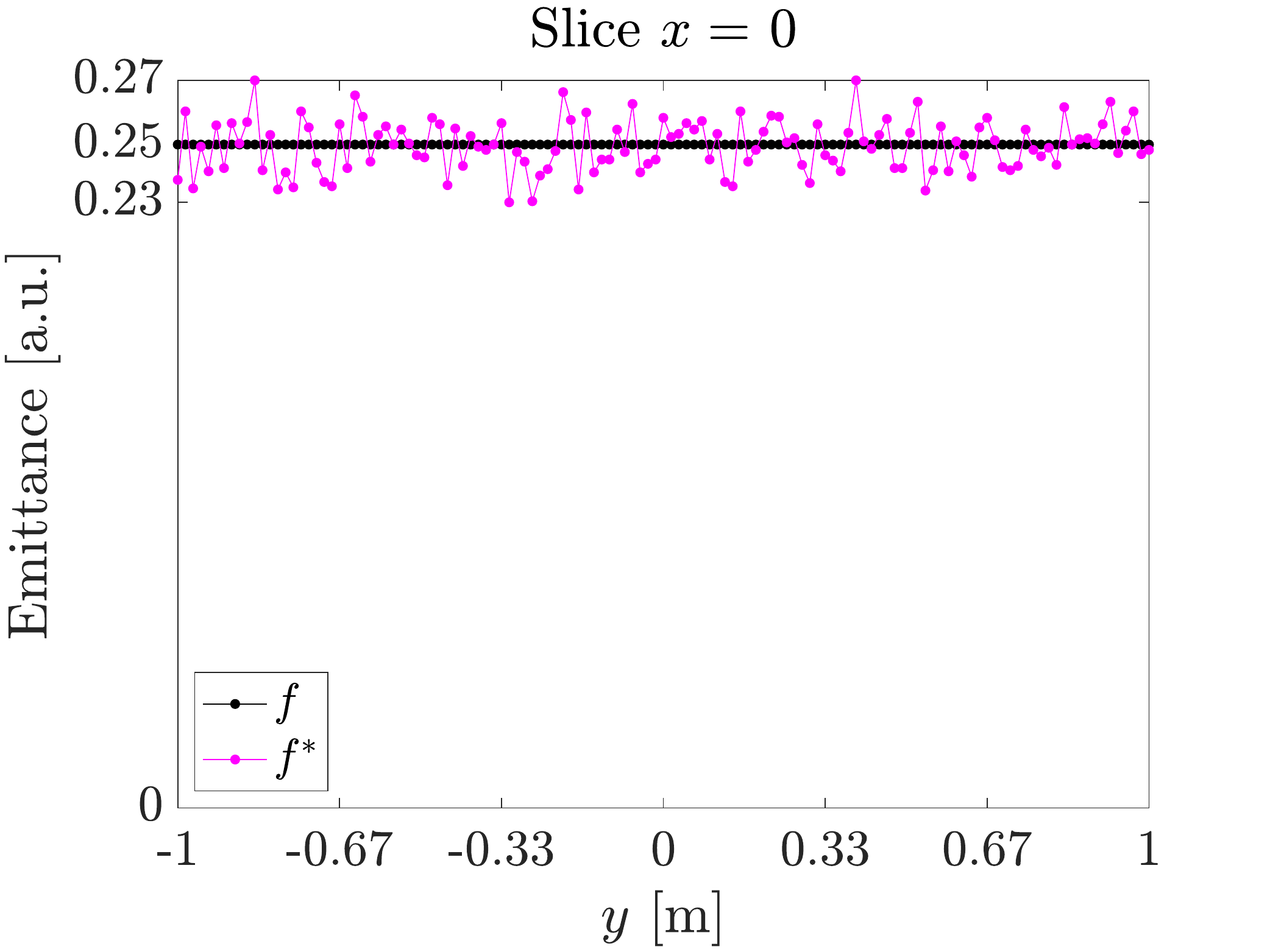}\\[10pt]
	\includegraphics[width=0.33\linewidth]{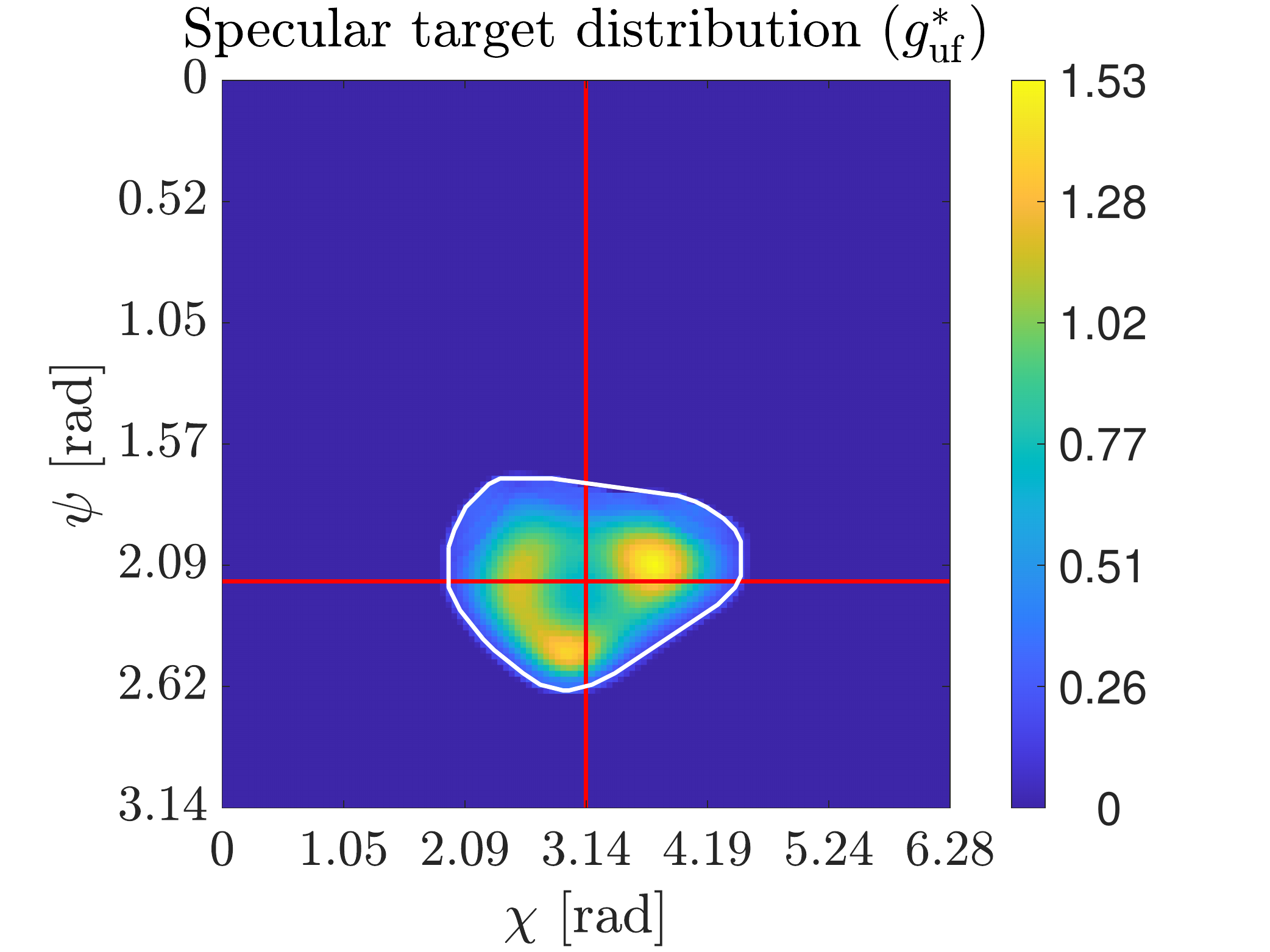}%
	\includegraphics[width=0.33\linewidth]{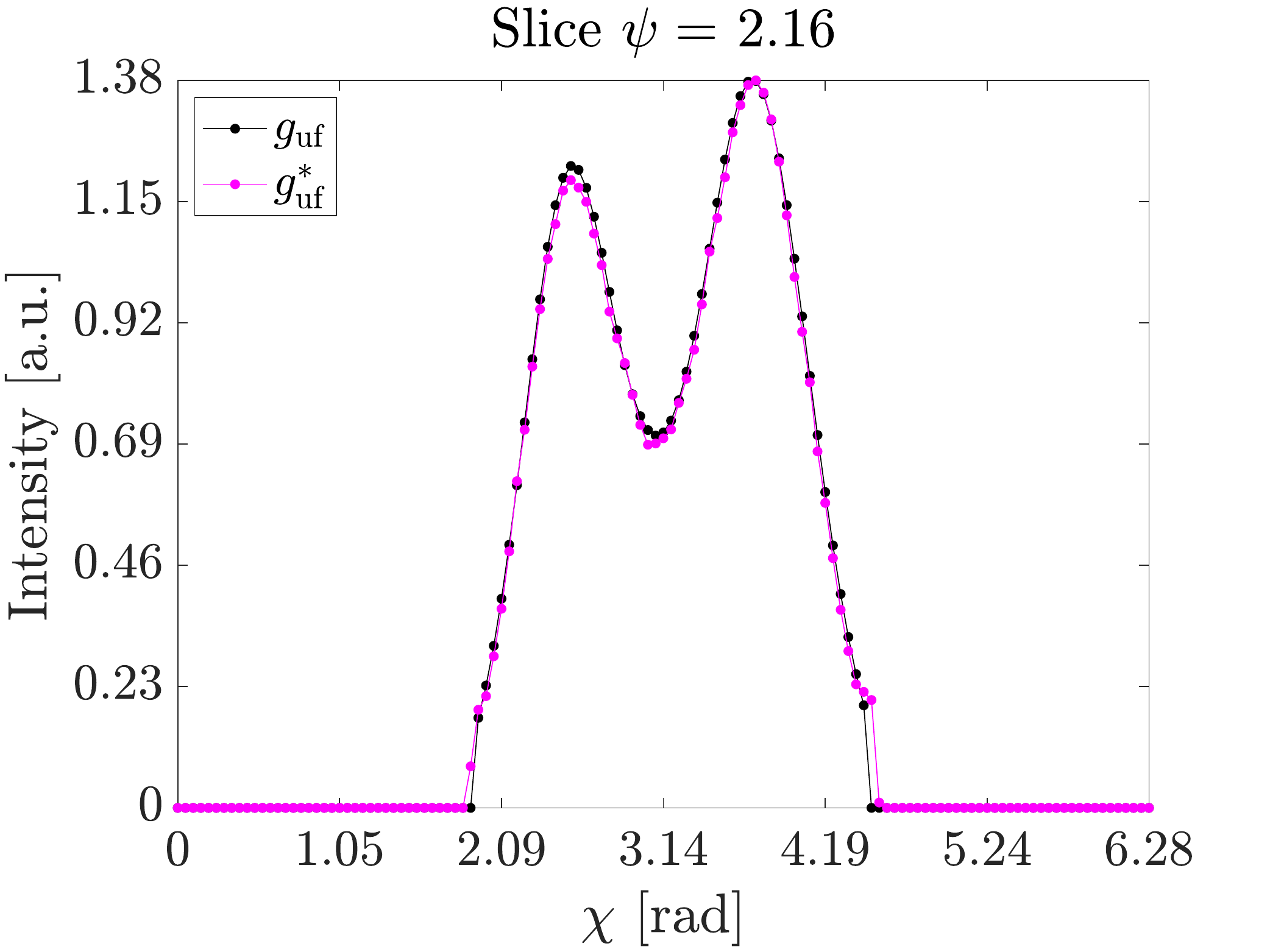}%
	\includegraphics[width=0.33\linewidth]{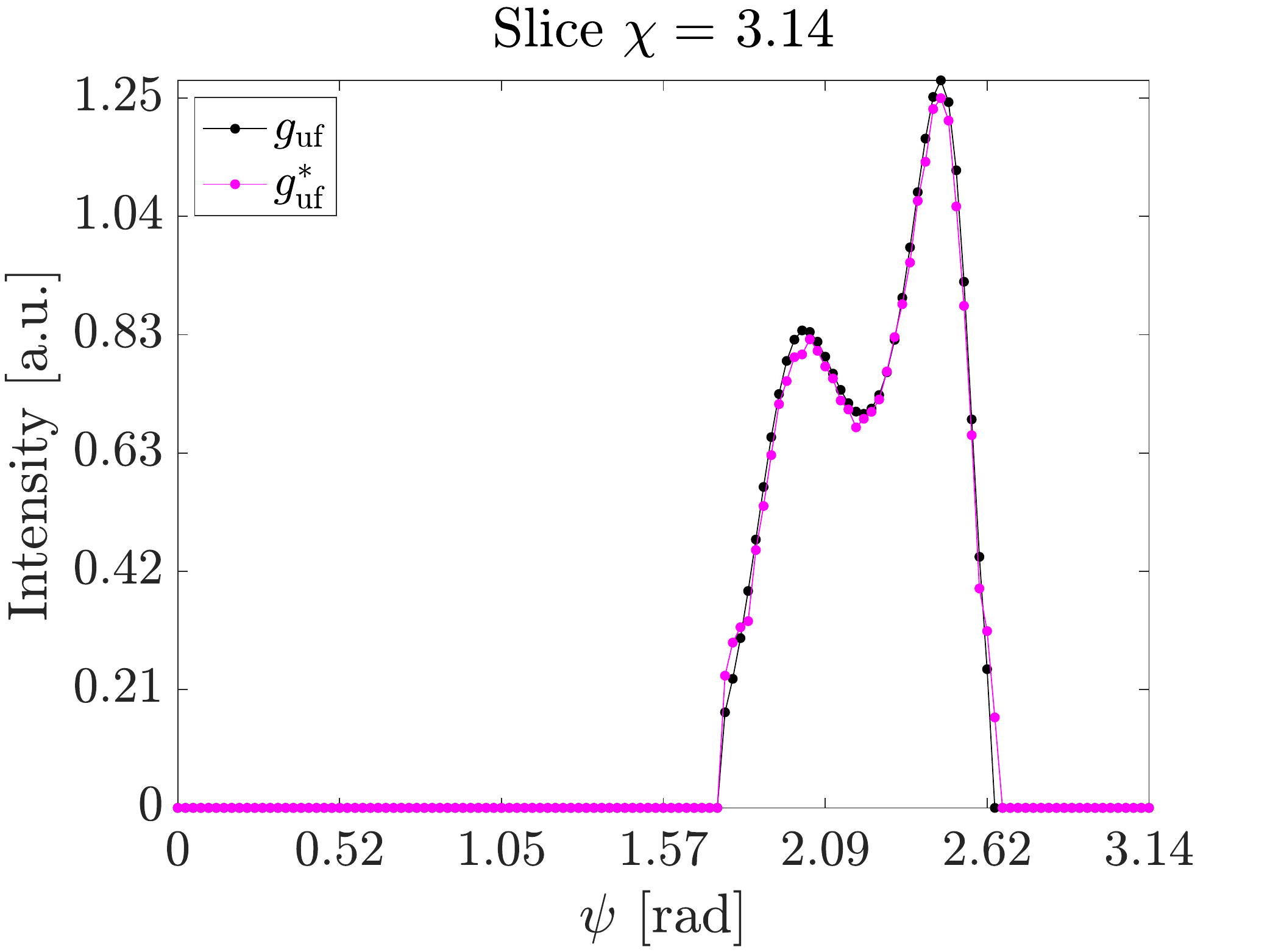}\\[10pt]
	\includegraphics[width=0.33\linewidth]{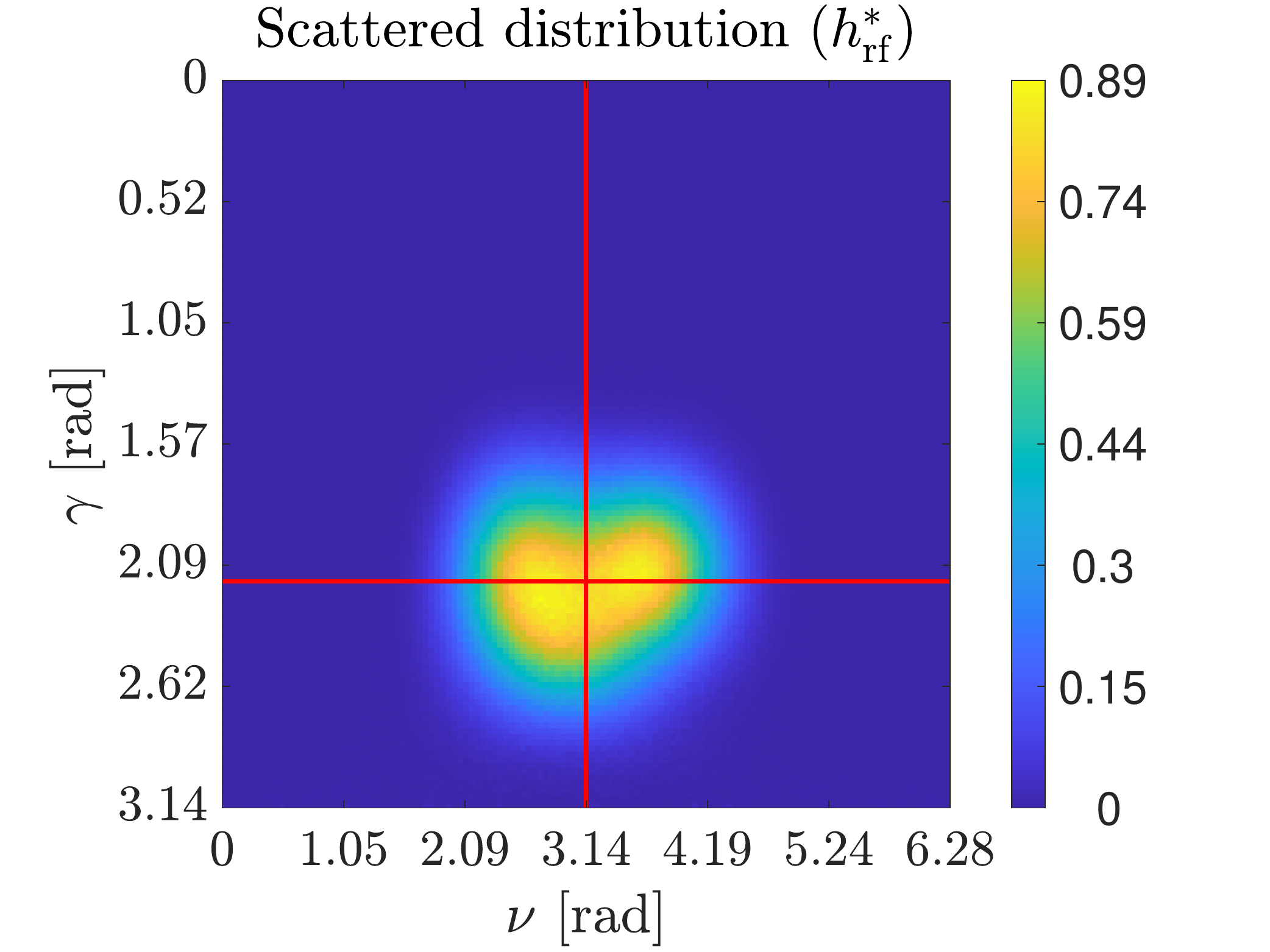}%
	\includegraphics[width=0.33\linewidth]{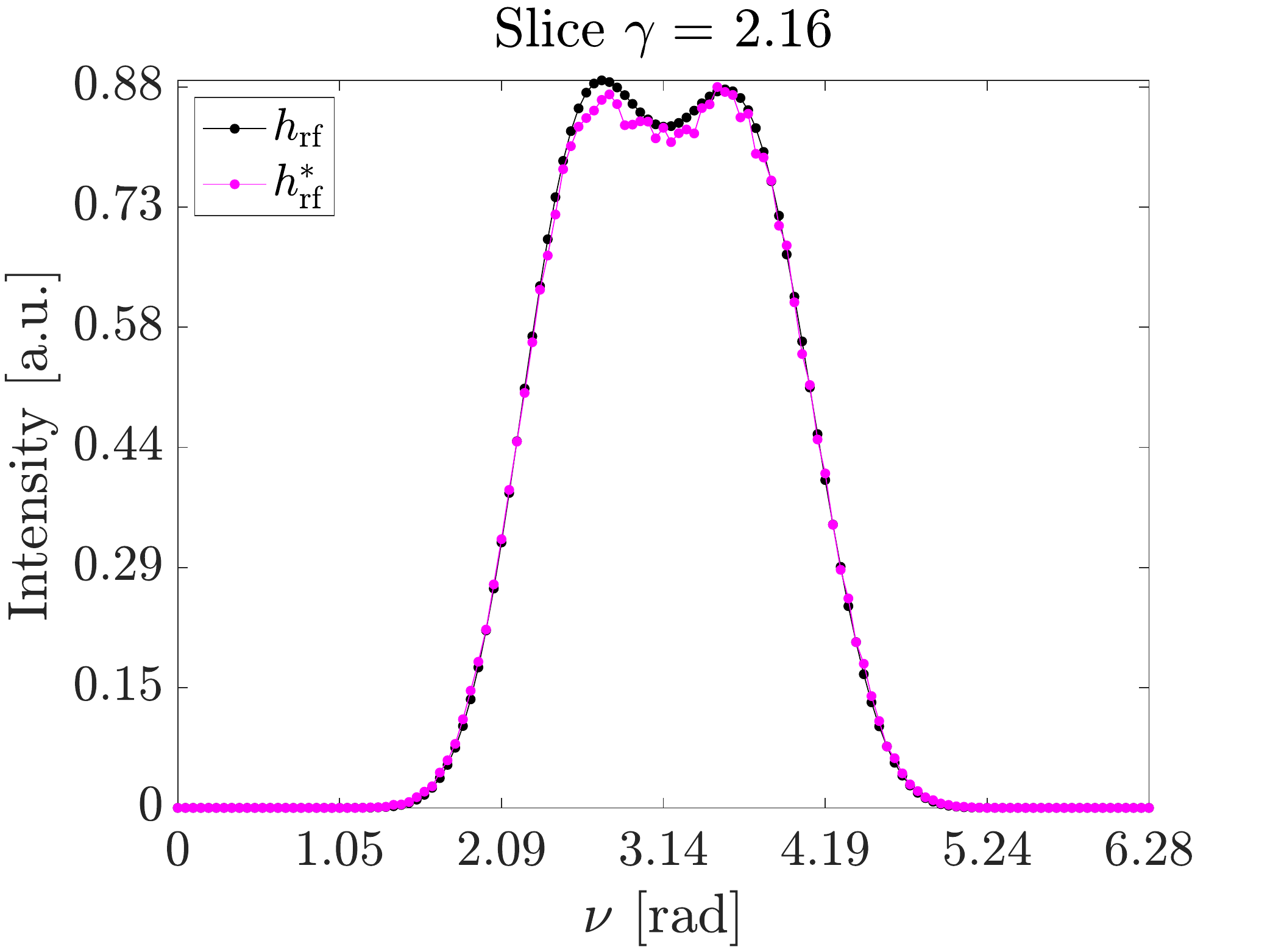}%
	\includegraphics[width=0.33\linewidth]{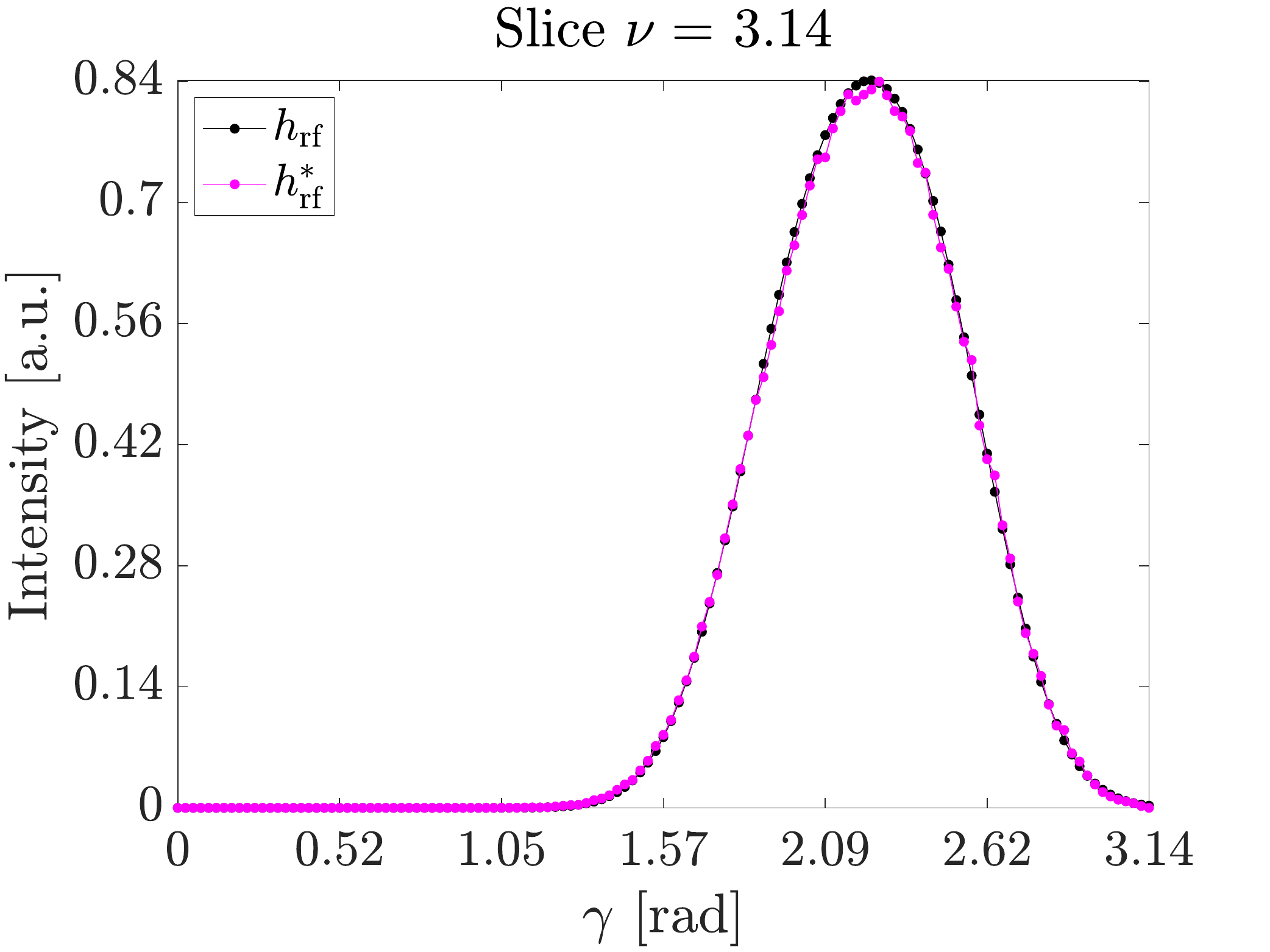}\\[10pt]
	\includegraphics[width=0.25\linewidth]{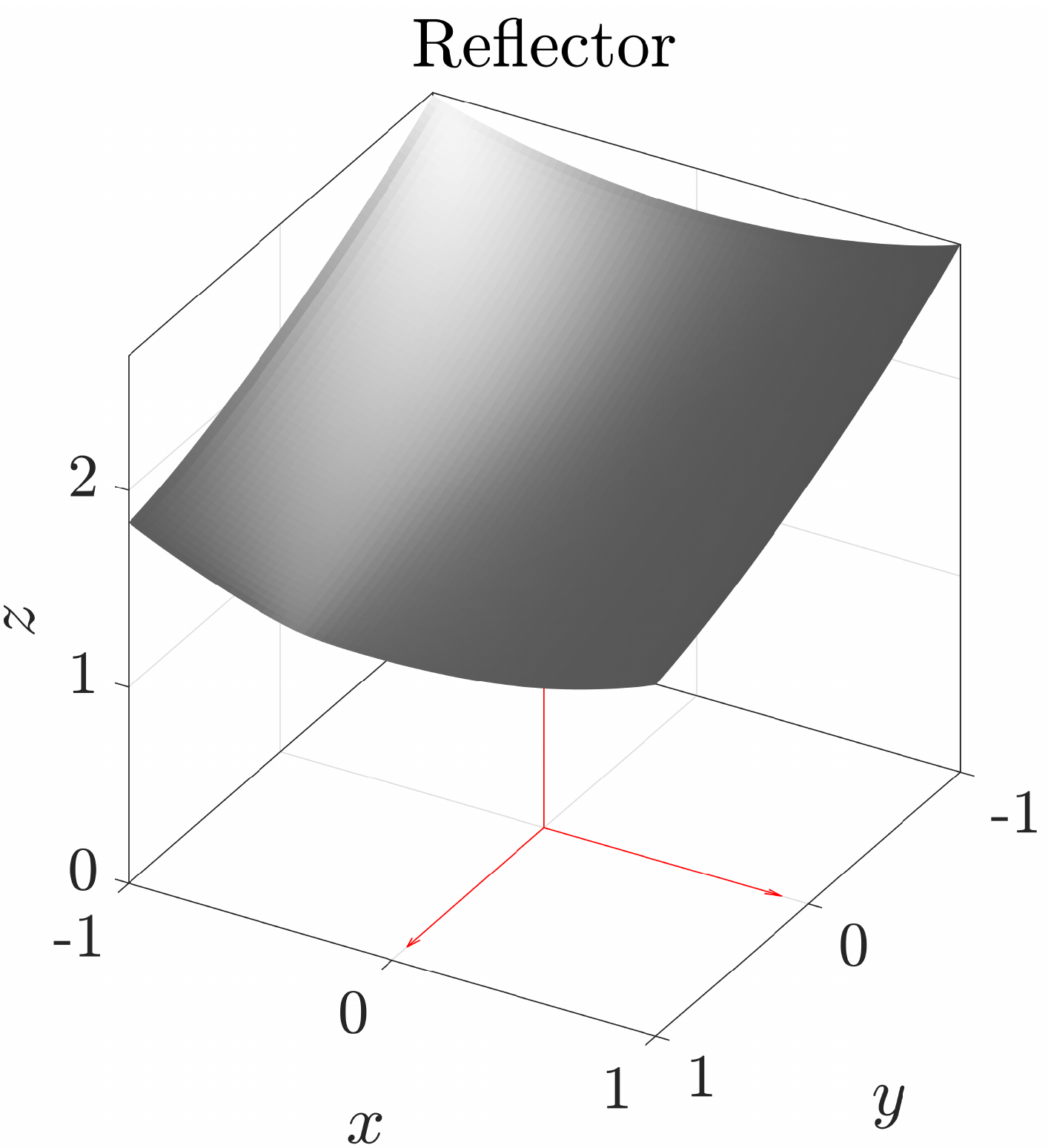}\hspace{5pt}%
	\includegraphics[width=0.33\linewidth]{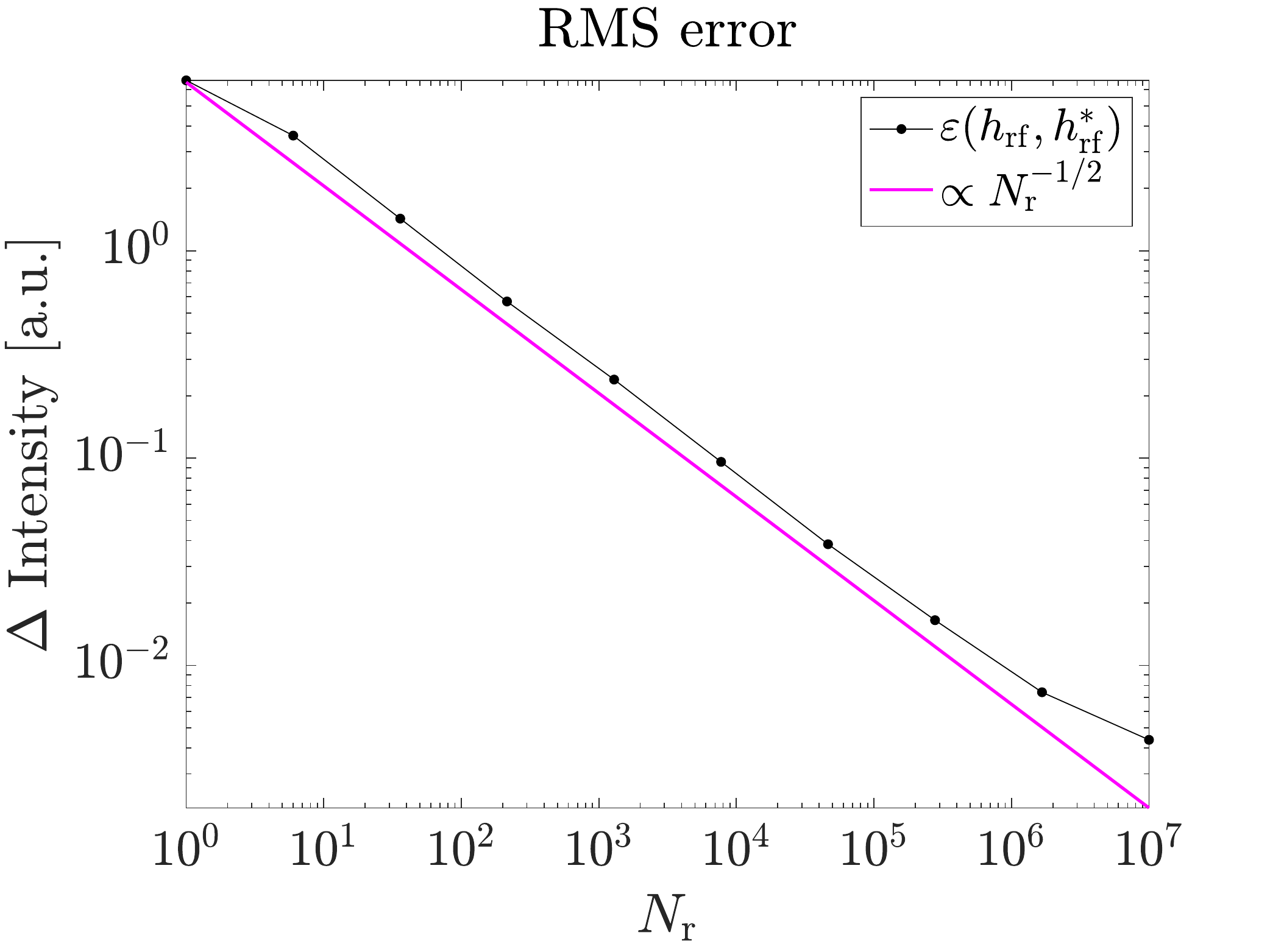}%
	\caption{Raytraced distributions and the reflector in Example \#1; $128^2$ sample points, $10^7$ rays traced.}
	\label{fig:example_1-4}
\end{figure*}

Next, the reflector was computed using the least-squares solver, and then the custom raytracer was employed to verify the shape of the surface and our scattering model.
The results of the ray trace after $10^7$ rays are shown in Fig.~\ref{fig:example_1-4}, where we see that the source sampling is homogeneous and that the scattered light distribution is very close to our prediction, as is also confirmed by the RMS error essentially following the expected $N_r^{-1/2}$ trend, where $N_r$ is the number of rays traced \cite[p.~9]{filosaPhaseSpaceRay2018}.
The specular distribution does deviate from the prescribed target in some places, which is best demonstrated by the slices --- see especially the data points close to the boundary of $\T$ and those close to the peaks.
These discrepancies come from the numerical least-squares solver used to compute the reflector surface. 
They are presumably the source of the slight deviation from the theoretical $N_r^{-1/2}$ slope of the RMS error, too.
The results could perhaps be improved by using a finer grid.
Crucially, however, the discrepancies do not significantly affect the scattered light, which is the main topic of concern here --- it is still clear that our predictions for the scattered light align very well with what is obtained from the raytracer.

 \clearpage
\subsection{Example \#2: Varying amounts of scattering}
In our second example, we wanted to visualize and quantify the differences in the reflector shape due to varying the amount of surface scattering.
The problem is outlined below, where we again enforced energy conservation such that the integrals over the source and target distributions were unity.
The surface scattering function $p$ can be found in Eq.~\eqref{eq:pAlpha}, with varying $\sigma$, i.e., varying amounts of scattering in the system.
Here, $\sigma = 0$ signifies a smooth, specular reflector.
\begin{equation*}
	\begin{split}
		&\text{Source domain:}\ \S = [-1,1] \times [-1,1]\\
		&\text{Target domain:}\ \text{see text}\\
		&\text{Source distribution:}\ f(x,y) = \frac{1}{4}\\
		&\text{Scattered target distribution:}\\
		&h(\gamma,\nu) = \frac{1}{0.685389} \, \N\bigg(\gamma,\nu;\frac{3\pi}{4},\pi,0.25,0.75\bigg)\\
		&\text{Surface scattering function:}\ p(\alpha;\sigma),\ \sigma \in \{0, 0.025, 0.05, 0.075, 0.1\}
	\end{split}
\end{equation*}

\begin{figure*}[htbp!]
	\centering
	\includegraphics[width=0.25\linewidth]{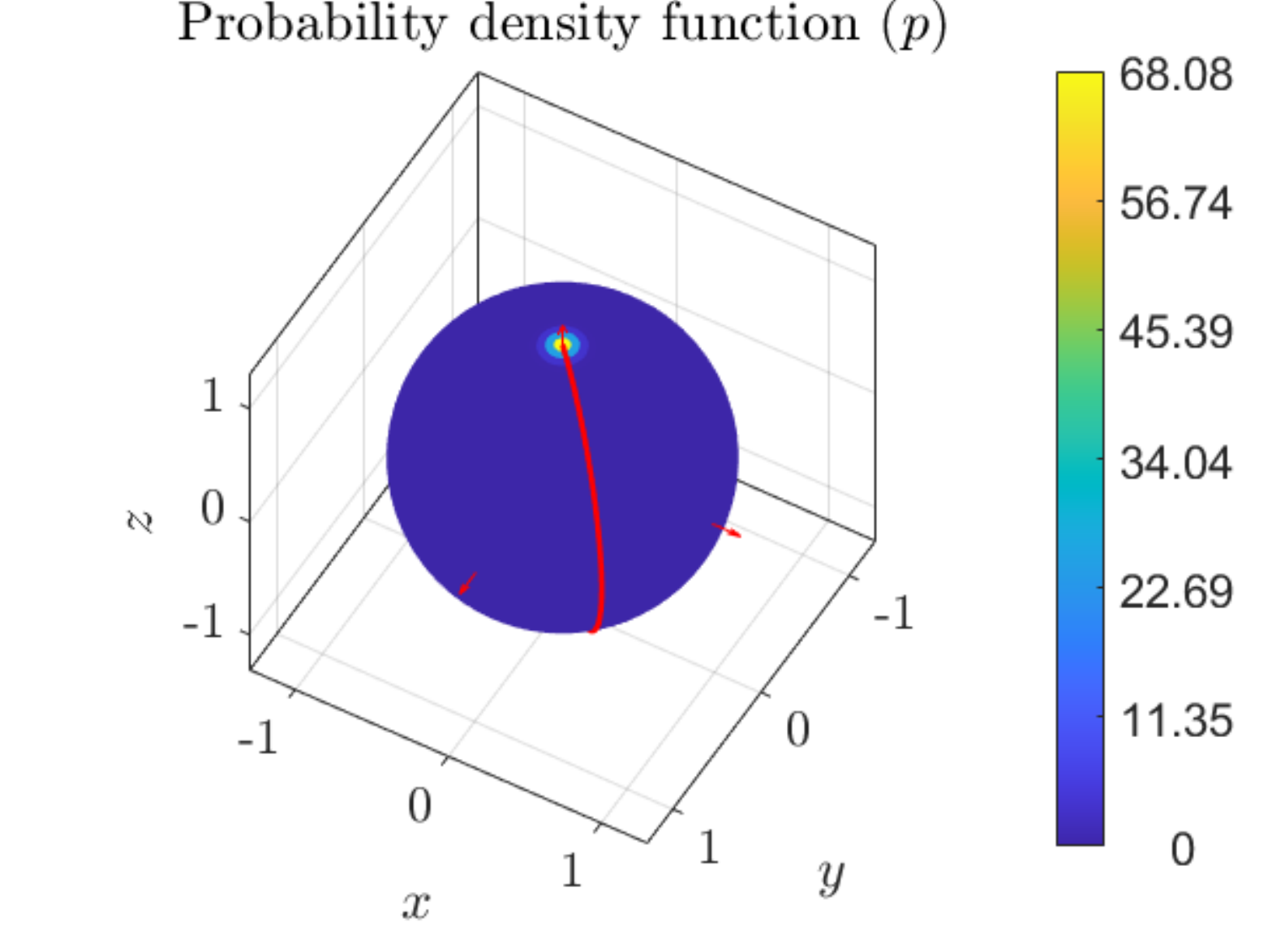}%
	\includegraphics[width=0.25\linewidth]{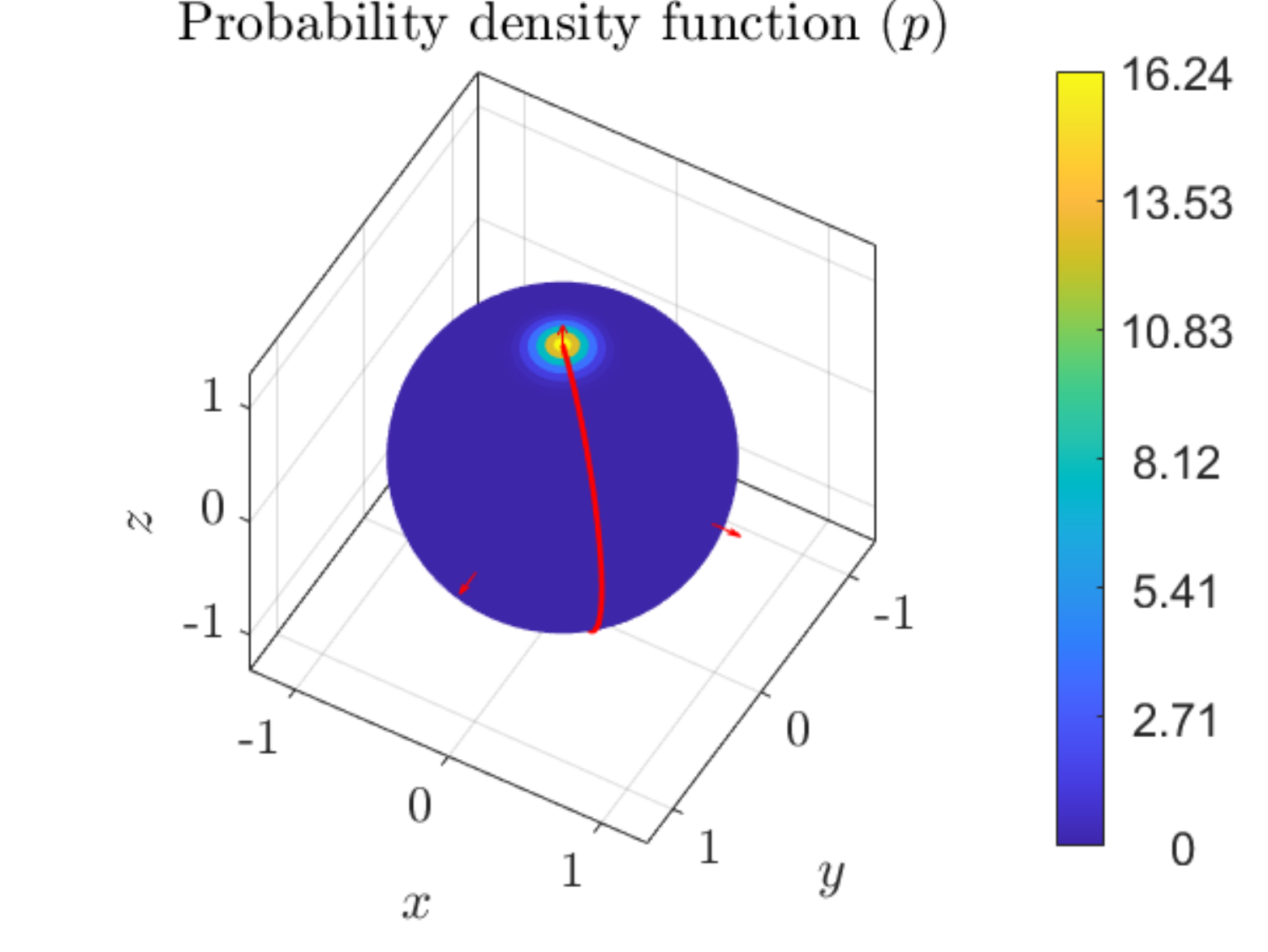}%
	\includegraphics[width=0.25\linewidth]{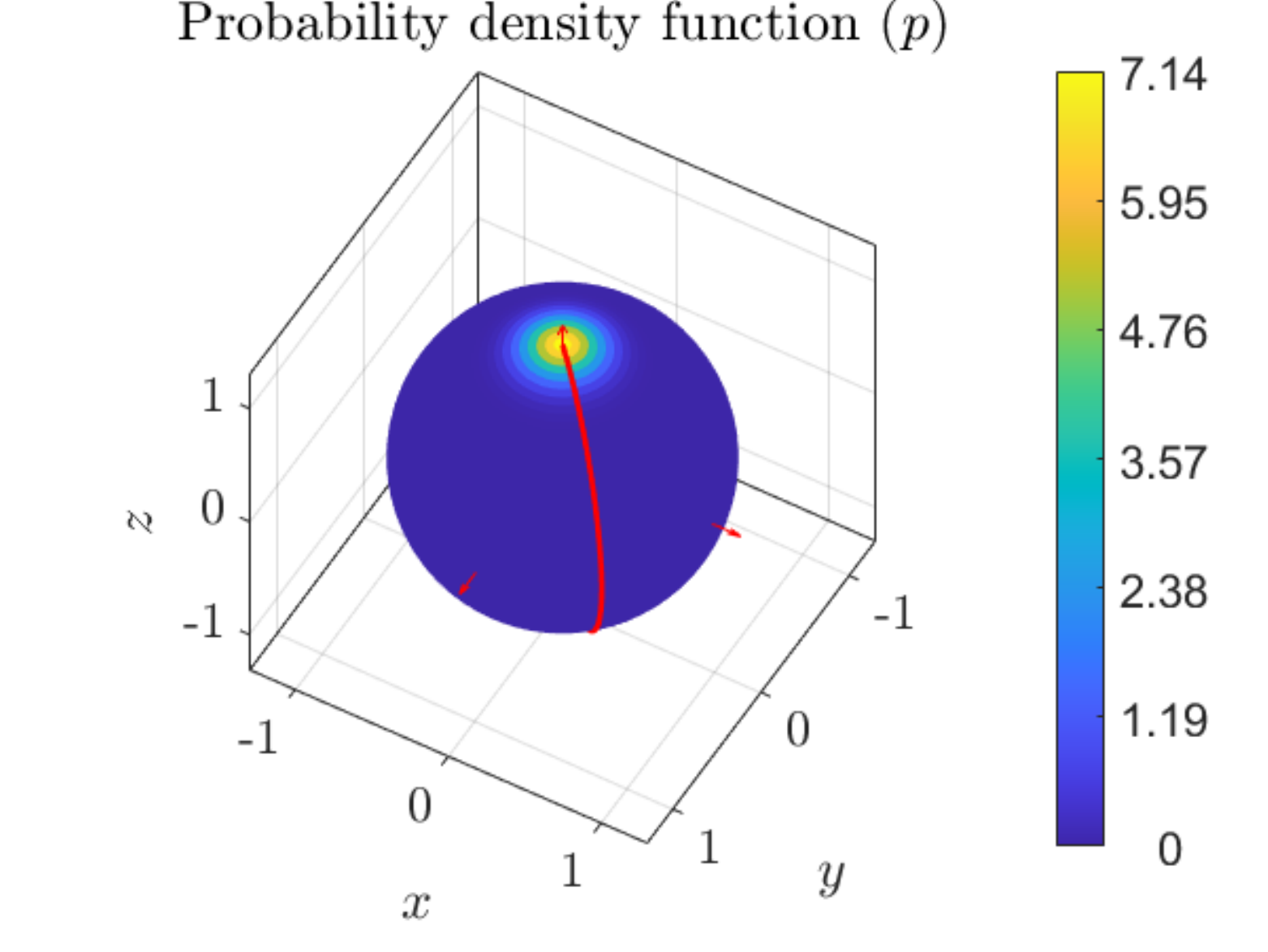}%
	\includegraphics[width=0.25\linewidth]{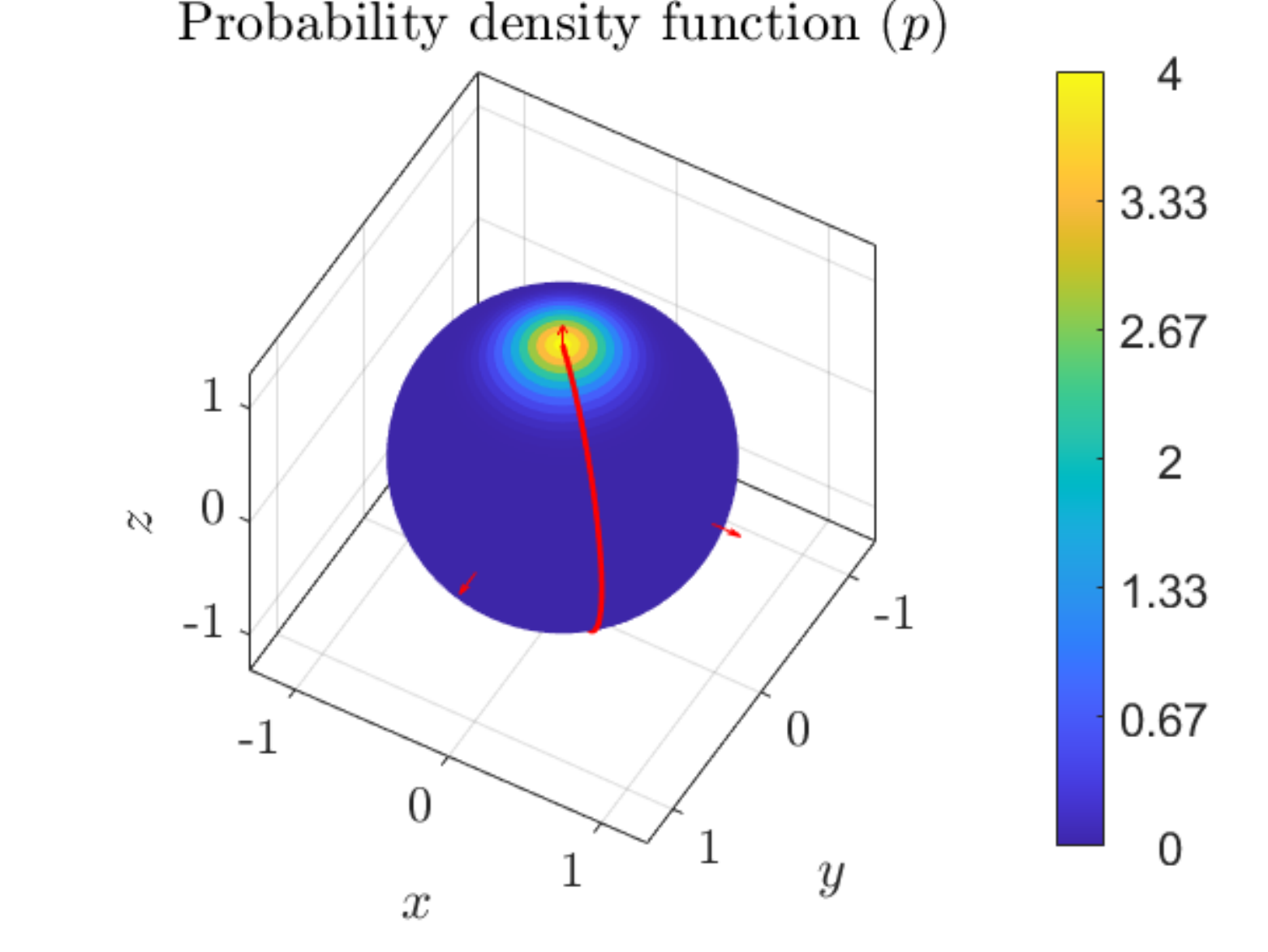}\\[10pt]
	\includegraphics[width=0.25\linewidth]{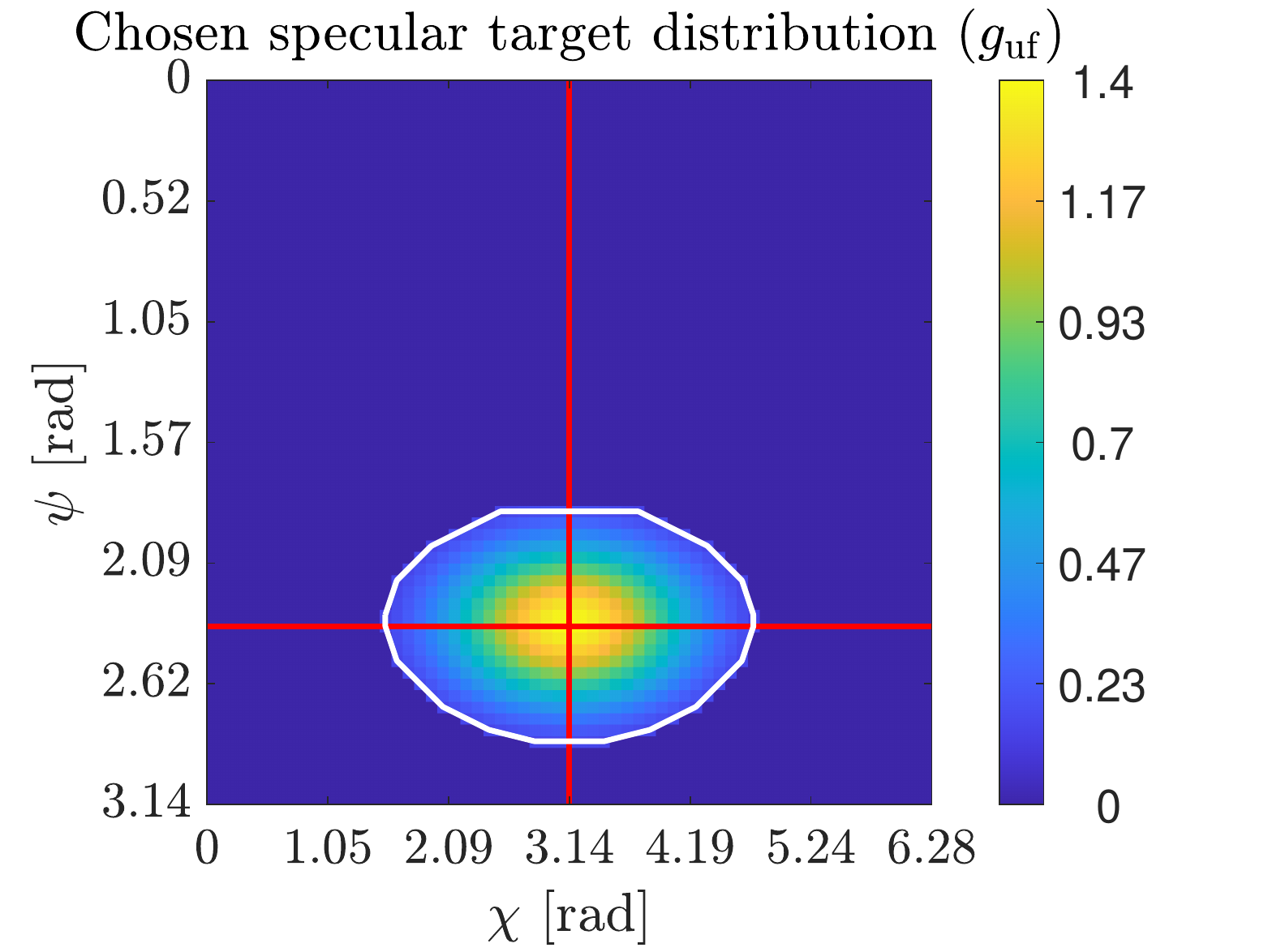}%
	\includegraphics[width=0.25\linewidth]{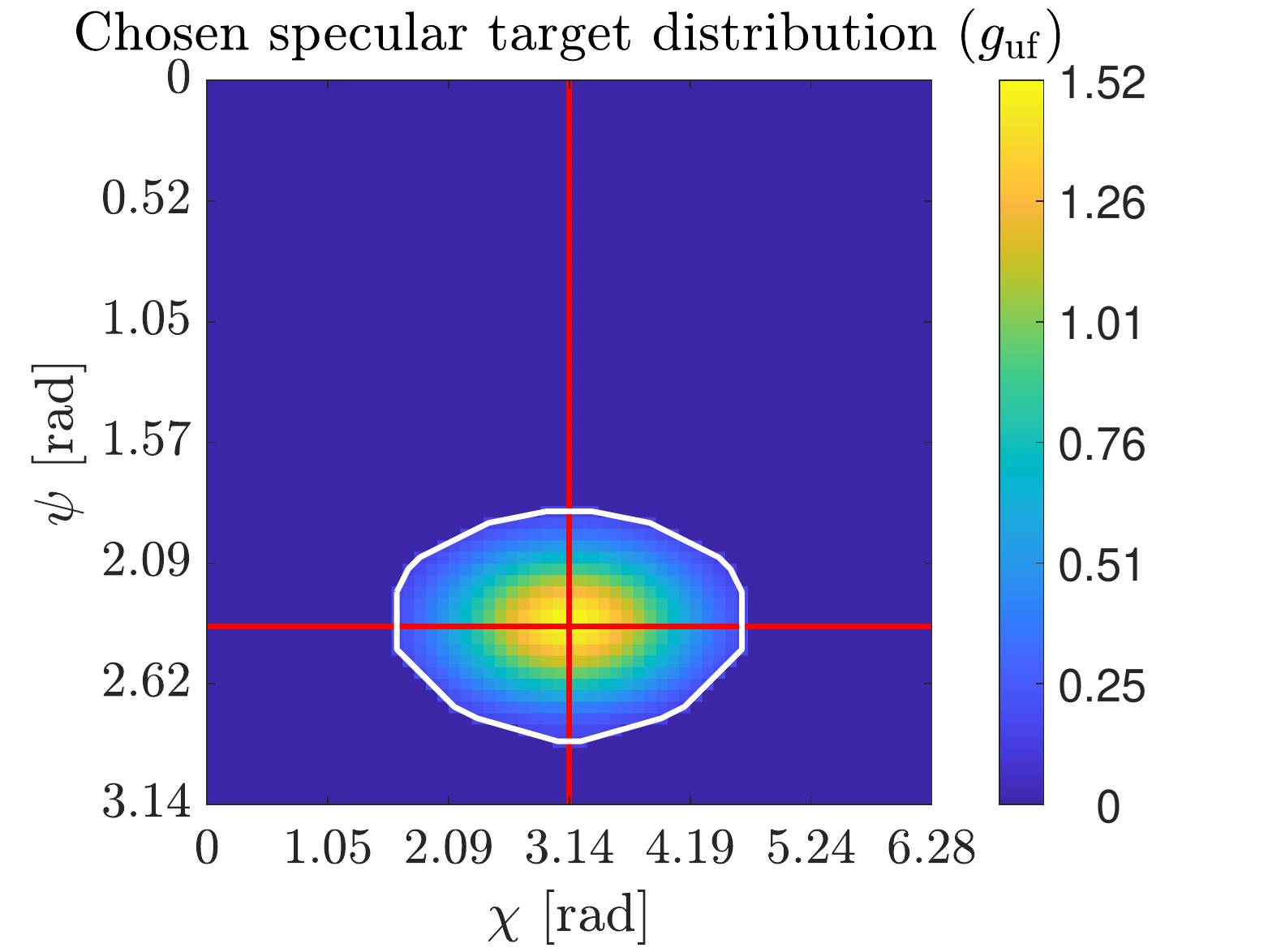}%
	\includegraphics[width=0.25\linewidth]{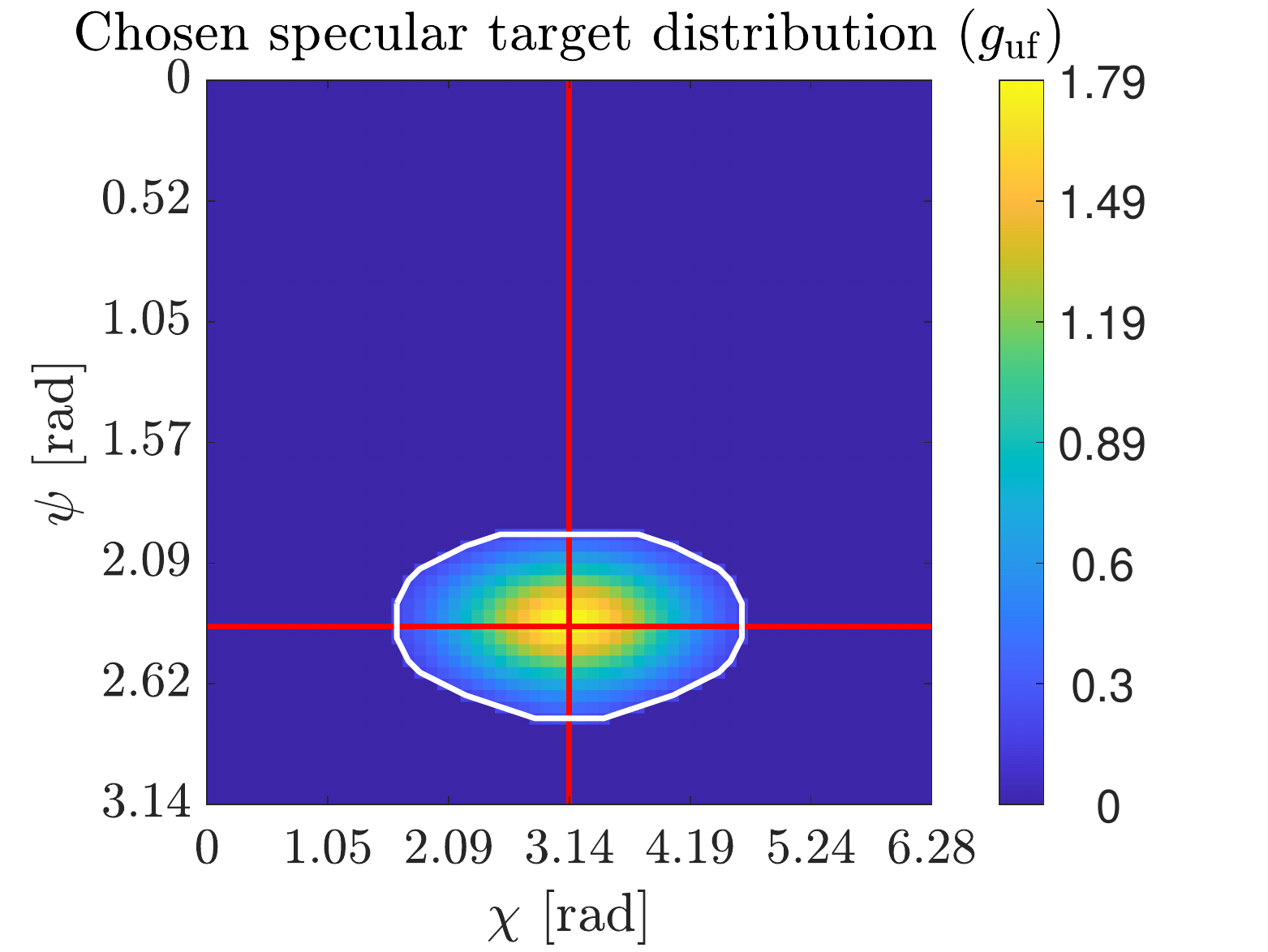}%
	\includegraphics[width=0.25\linewidth]{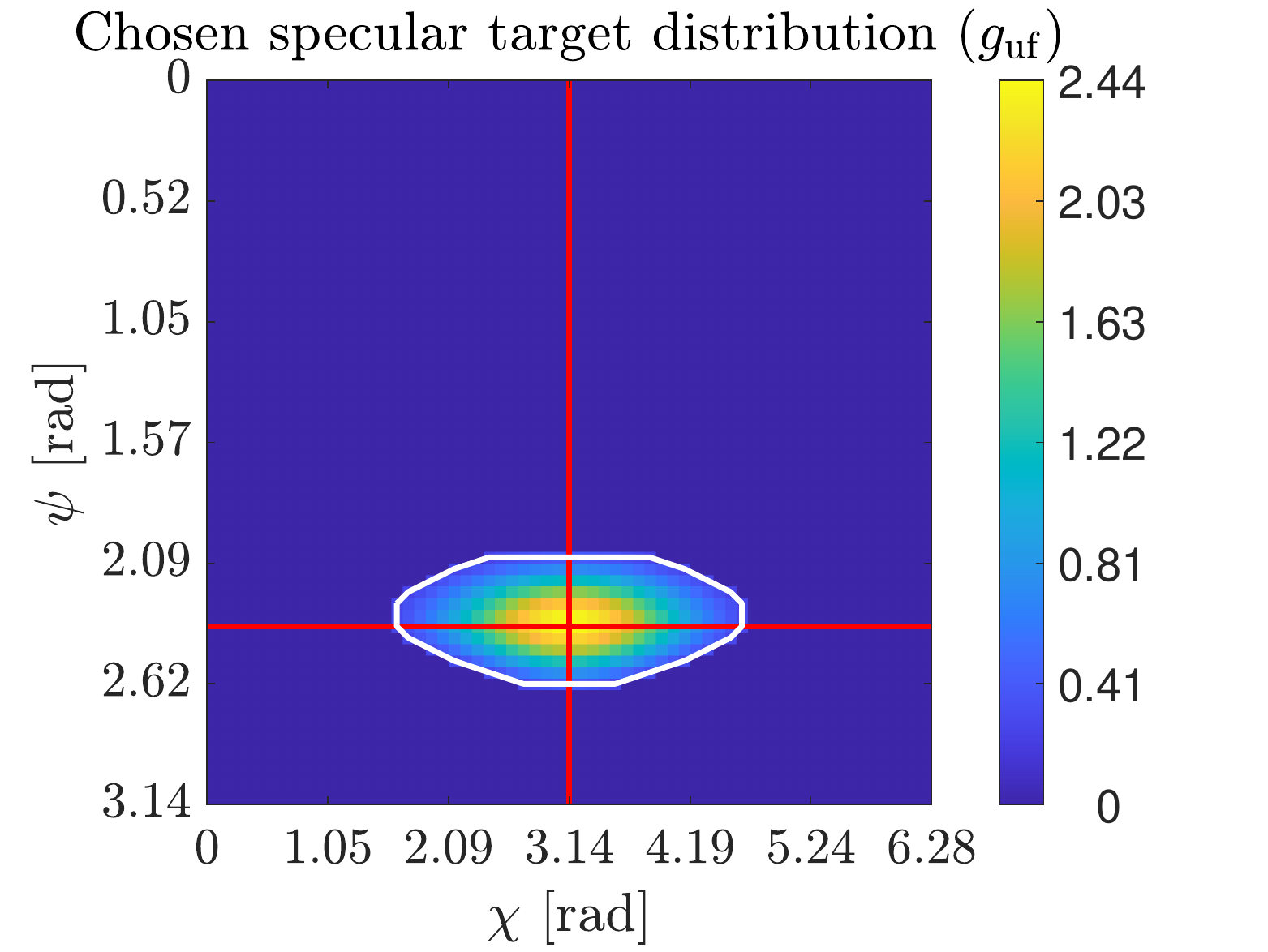}%
	\caption{The probability density functions and specular target distributions in Example \#2; $\sigma \in \{0.025, 0.05, 0.075, 0.1\}$ from left to right; $64^2$ sample points.}
	\label{fig:example_2-1}
\end{figure*}

\noindent The probability density functions $p$ with nonzero values of $\sigma$ and the associated specular target distributions are shown in Fig.~\ref{fig:example_2-1}, together with the boundary of $\T$, found by fixing $\epsilon = 0.1 \max(g_\uf)$.
Successively increasing $\sigma$ shows a `sharpening' of the target; the target domain shrinks, and the maximum value increases.

\begin{figure*}[htbp!]
	\centering
	\includegraphics[width=0.3\linewidth]{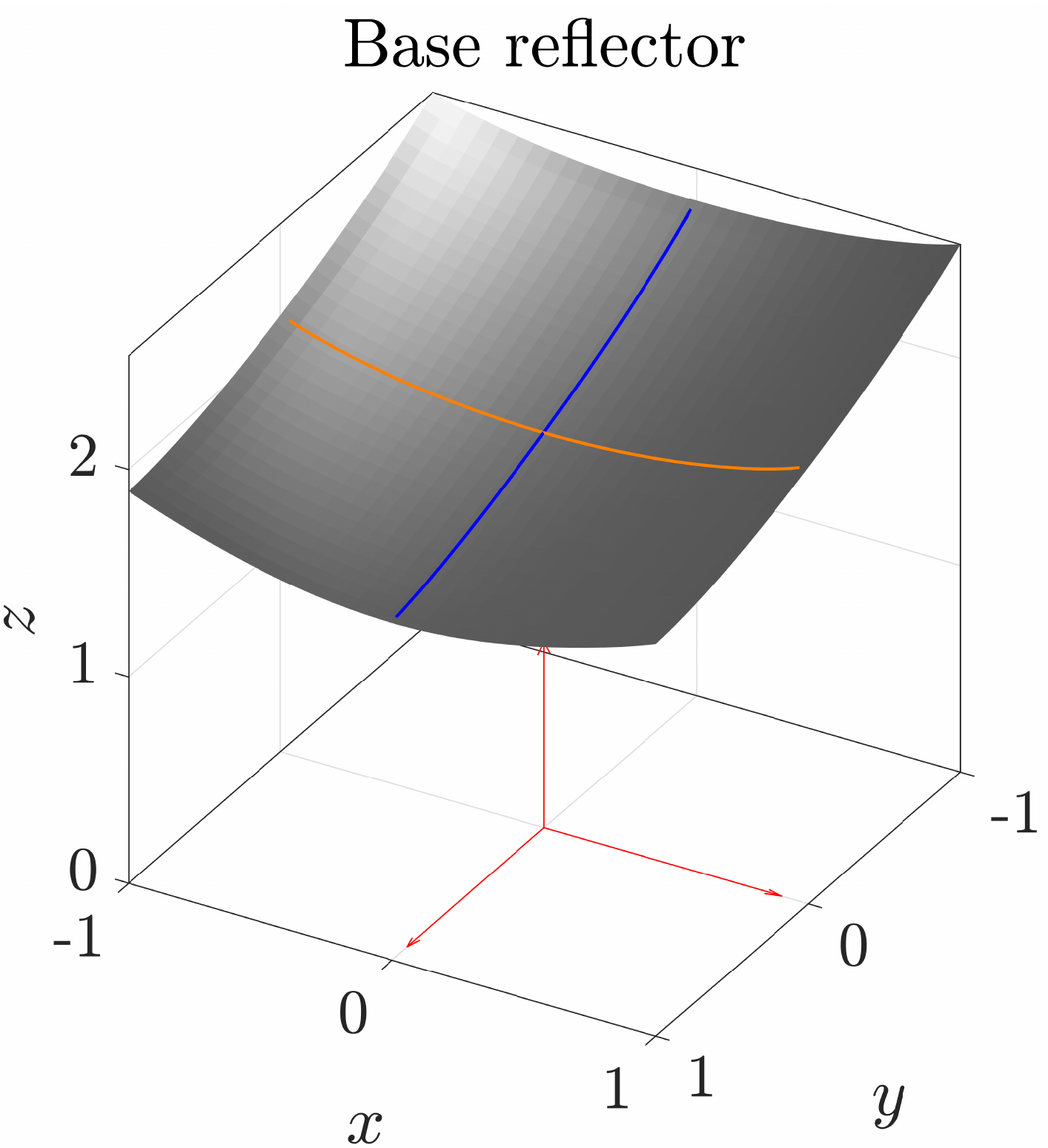}\hfill
	\includegraphics[width=0.33\linewidth]{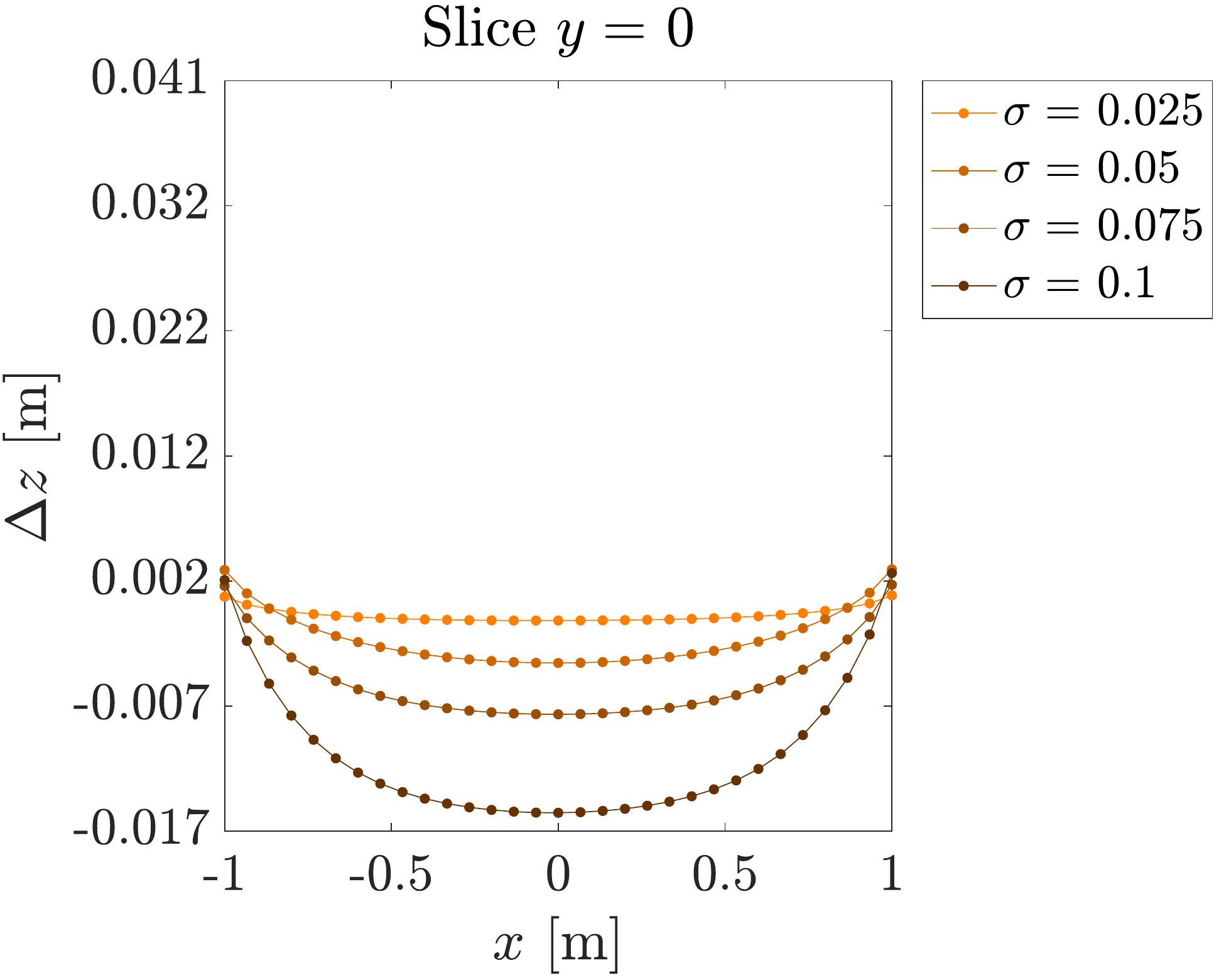}%
	\includegraphics[width=0.33\linewidth]{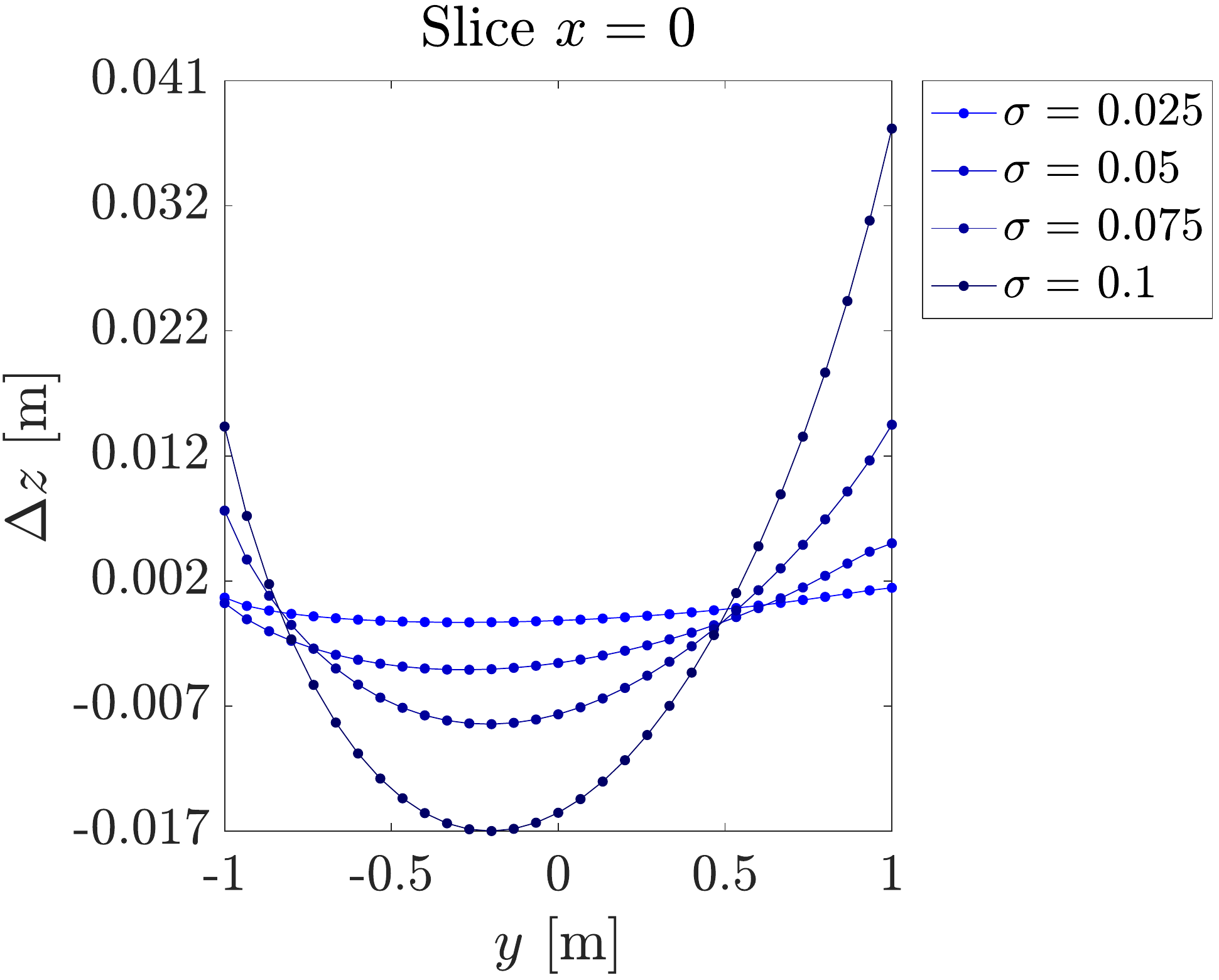}\\[10pt]
	\includegraphics[width=0.3\linewidth]{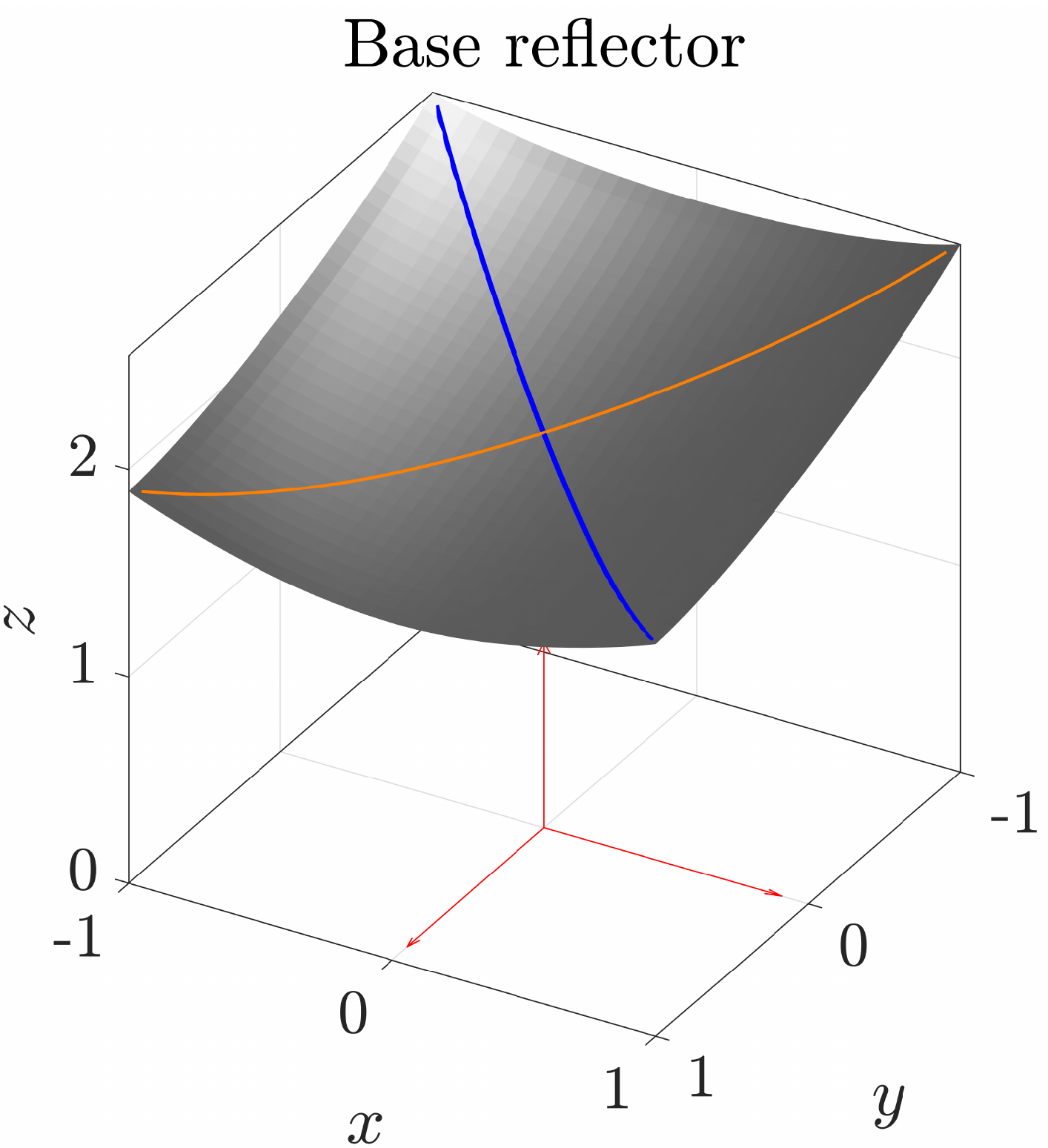}\hfill
	\includegraphics[width=0.33\linewidth]{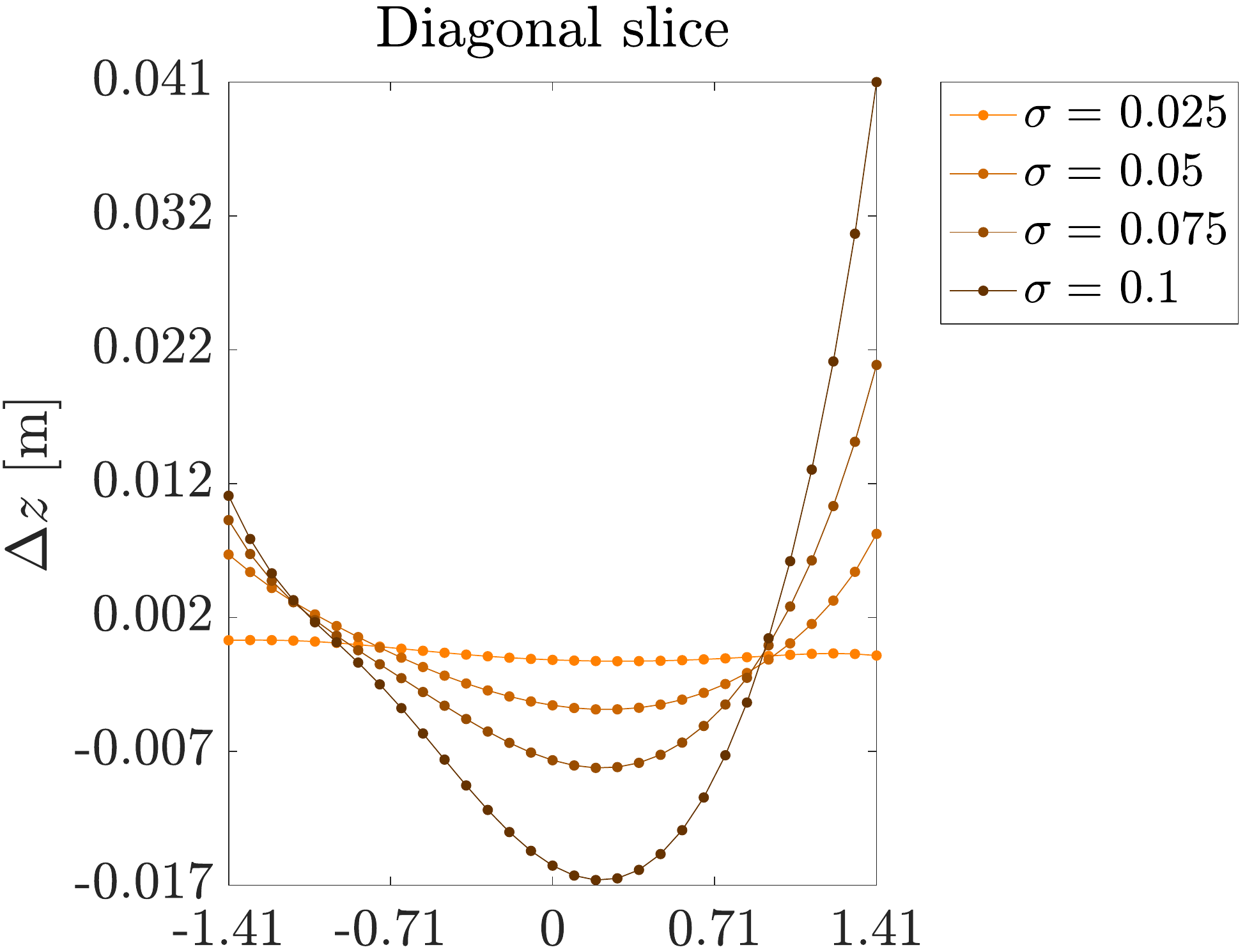}%
	\includegraphics[width=0.33\linewidth]{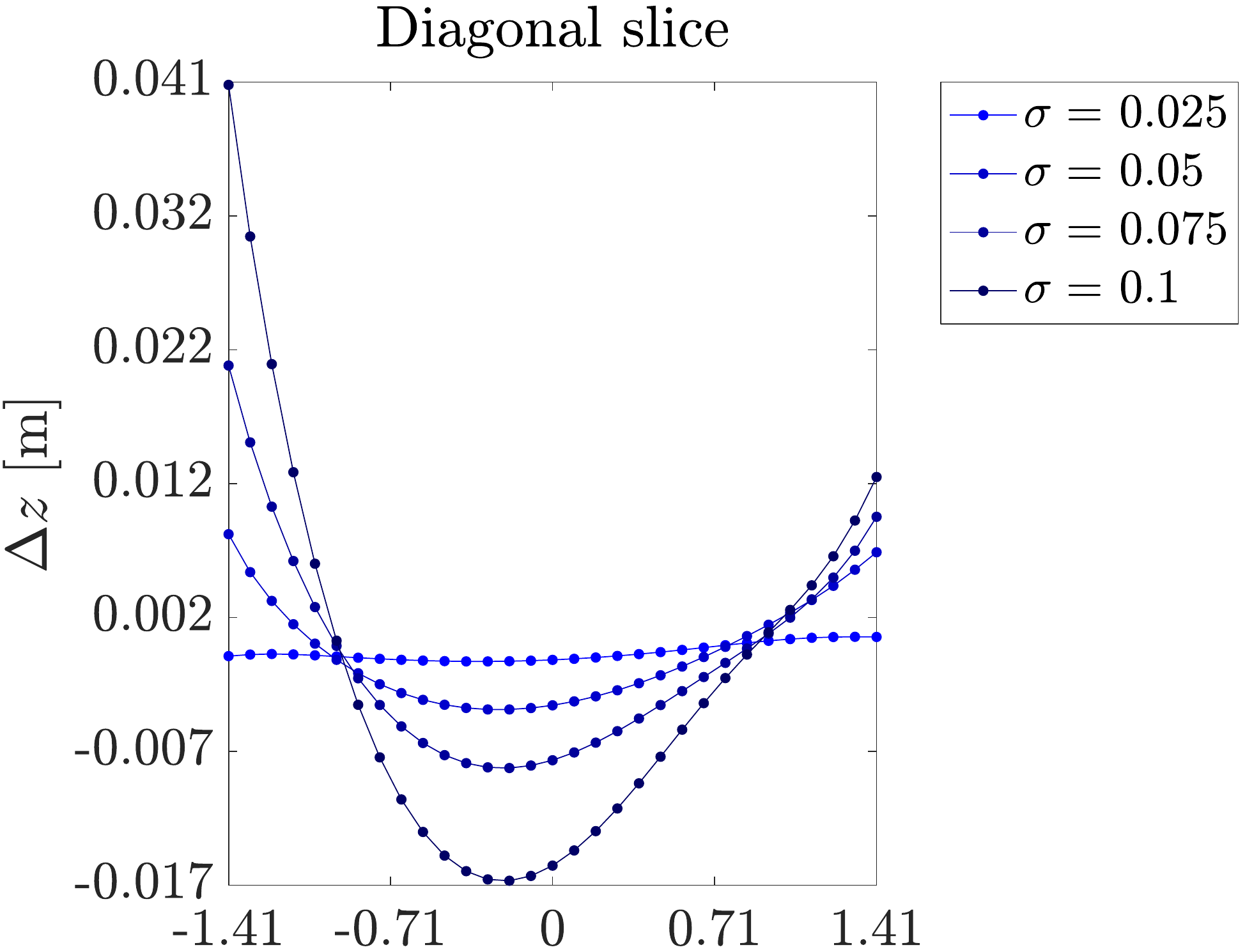}
	\caption{Slices along the indicated lines for reflectors with varying amounts of surface scattering, all fulfilling the problem in Example \#2; $32^2$ sample points.}
	\label{fig:example_2-3}
\end{figure*}

Fig.~\ref{fig:example_2-3} shows the effect scattering has on the reflector, where the \textit{base reflector} was taken as the specular reflector achieving $h$ given $f$, i.e., with $\sigma = 0$ so that $p$ is a delta function and $g \equiv h$.
Successive reflectors have increasing values of $\sigma$, associated with more and more scattering up to $\sigma = 0.1$.
As expected, more scattering requires more modification of the reflector versus the base one, and we see variations in height up to a few percent of the size of the reflectors, which is consistent with previous observations we made in the two-dimensional case \cite{kronbergTwodimensionalFreeformReflector2023}.
As noted there, variations of this order of magnitude are typically considered manufacturable.

\clearpage
\section{Conclusions}
We have developed a novel approach to computing freeform reflectors with scattering surfaces inspired by optimal transport.
Our method involves using a density function with specific properties, which results in a Fredholm integral equation of the first kind.
This equation provides information about scattered light in the far field based on a probability density function (PDF) that defines the surface's scattering properties and the specular distribution.
We can create a virtual specular target distribution by unfolding the Fredholm integral.
This virtual distribution can then be used as a target when solving the inverse problem of illumination optics, i.e., computing the reflector.
This approach ensures that the prescribed target is achieved when considering surface scattering. 
 
As a result, the process of designing freeform reflectors with scattering surfaces becomes a two-step process.
We first modify the target distribution by unfolding the Fredholm integral equation that governs scattering in our model. Then, we use existing methods to compute the associated specular reflector.

Our future goal is to expand on the approach we introduced in \cite{kronbergTwodimensionalFreeformReflector2023} by applying it to three dimensions.
In our previous work, we utilized microfacets to model the rough surface that causes light scattering.
Furthermore, one way to enhance the applicability of our model is by removing the current limitation of isotropic surfaces.\\

\hrule
\begin{changemargin}{1cm}{1cm}
	\noindent\textsc{\textbf{Funding:}} This work was partially supported by the Dutch Research Council (\textit{Dutch:} Nederlandse Organisatie voor Wetenschappelijk Onderzoek (NWO)) through grant P15-36.\\
	\textsc{\textbf{Disclosures:}} The authors declare no conflicts of interest.\\
	\textsc{\textbf{Data availability:}} Data underlying the results presented in this paper are not publicly available at this time but may be obtained from the authors upon request.
\end{changemargin}
\hrule

\pagestyle{ref}
\addcontentsline{toc}{section}{References}
\printbibliography

\clearpage
\section{Appendix: Finding $p(\alpha;\sigma)$}\label{sec:appendix}
\noindent Suppose we pick $\alpha$ and $\beta$ from a rotationally symmetric Gaussian centered around the origin in the stereographic plane.
Recalling the two-dimensional normal distribution in Eq.~\eqref{eq:2DGaussian}, we get that the PDF in stereographic coordinates $\q := (q_1, q_2)^\intercal$ is
\begin{equation}
	p_\ster(\q;\sigma) = \frac{1}{2\pi \sigma^2} \, \mathrm{exp}\bigg(-\frac{\norm{\q}^2}{2 \sigma^2}\bigg),
\end{equation}
where $\sigma$ is the standard deviation in both $q_1$ and $q_2$.
Since $p_\ster$ is a PDF, it follows that, for all $\sigma \in \mathbb{R}$,
\begin{equation}\label{eq:appendix:energyConservation_q}
	\int_{\mathbb{R}^2} p_\ster(\q;\sigma) \, \dd \q = 1.
\end{equation}

The analogous stereographic and inverse stereographic mappings from the south pole to those in Eqs.~\eqref{eq:sterNP} and \eqref{eq:sterNPInverse} are
\begin{equation}\label{eq:appendix:sterSP}
	\q(\uc) =
	\begin{pmatrix}
		q_1\\
		q_2
	\end{pmatrix}
	=
	\frac{1}{1+c_3}
	\begin{pmatrix}
		c_1\\
		c_2
	\end{pmatrix}
	=
	\frac{\sin(\alpha)}{1+\cos(\alpha)}
	\begin{pmatrix}
		\cos(\beta)\\
		\sin(\beta)
	\end{pmatrix},
\end{equation}
and
\begin{equation}
	\uc(\q) = \frac{1}{1 + \norm{\q}^2}
	\begin{pmatrix}
		2 q_1\\
		2 q_2\\
		1 - \norm{\q}^2
	\end{pmatrix},
\end{equation}
where $\q$ is the 2-tuple stereographic representation of $\uc$.

Transforming Eq.~\eqref{eq:appendix:energyConservation_q} to angular coordinates gives
\begin{equation}
	\int_0^{2\pi} \int_0^\pi p_\ster(\q(\alpha,\beta);\sigma) \, \abs{\pdv{\q(\alpha,\beta)}{(\alpha,\beta)}} \, \dd \alpha \dd \beta = 1.
\end{equation}
The Jacobian, $\partial \q/\partial(\alpha,\beta)$, can readily be evaluated using Eq.~\eqref{eq:appendix:sterSP}:
\begin{equation}
	\pdv{\q(\alpha,\beta)}{(\alpha,\beta)} = \frac{\tan(\alpha/2)}{1 + \cos(\alpha)}.
\end{equation}

Finally, letting
\begin{equation}
	p(\alpha;\sigma) := p_\ster(\q(\alpha,\beta);\sigma) \frac{\tan(\alpha/2)}{1 + \cos(\alpha)} \frac{1}{\sin(\alpha)},
\end{equation}
so that
\begin{equation}
	\int_0^{2\pi} \int_0^\pi p(\alpha;\sigma) \sin(\alpha) \, \dd\alpha \dd\beta = 1,
\end{equation}
yields the expression for $p(\alpha;\sigma)$ in Eq.~\eqref{eq:pAlpha}, since the $\beta$ dependence in $p_\ster(\q(\alpha,\beta);\sigma)$ drops out.

\end{document}